\documentclass[useAMS,usenatbib]{mn2e}
\usepackage[english]{babel}
\usepackage[ansinew]{inputenc}
\usepackage[T1]{fontenc}
\usepackage{amsmath}
\usepackage{amssymb}
\usepackage[dvips]{graphicx}
\usepackage{nameref,everysel,ragged2e}
\usepackage{epsfig}
\usepackage{multirow}
\usepackage{color}

\definecolor{turquoise}{rgb}{0.1,0.5,0.5}
\newcommand{\dbar}[1]{\bar{\bar{#1}}}

\title[Modeling $\xi(r_p, \pi)$]
{A model of the anisotropic correlation function $\xi(r_p, \pi)$ in redshift space including redshift errors}

\author[H.~A. Schlagenhaufer et al.]
{Holger A.~Schlagenhaufer$^{1,2}$\thanks{schlagen@mpe.mpg.de},
  Stefanie Phleps$^2$, and Ariel G.~S{\'a}nchez$^2$\\
  $^1$Universit\"atssternwarte M\"unchen, Ludwig-Maximillians Universit\"at, Scheinerstr. 1,
      D-81679 M\"unchen, Germany\\
  $^2$Max-Planck-Institut f\"ur Extraterrestrische Physik, Giessenbachstra{\ss}e, 
      D-85748 Garching, Germany\\}
\date{Released 2011 Xxxxx XX}

\pagerange{\pageref{firstpage}--\pageref{lastpage}} \pubyear{2002}

\begin{document}
\label{firstpage}
\maketitle

\begin{abstract}
With the advent of very large volume, wide-angle photometric redshift surveys like e.g. Pan-STARRS, DES, or PAU, which aim at using the spatial distribution of galaxies as a means to constrain the equation of state parameter of dark energy, $w_{DE}$, it has become extremely important to understand the influence of redshift inaccuracies on the measurement. We have developed a new model for the anisotropic two point large-scale ($r\ga 64 h^{-1}$\,Mpc) correlation function $\xi(r_p,\pi)$, in which nonlinear structure growth and nonlinear coherent infall velocities are taken into account, and photometric redshift errors can easily be incorporated. In order to test its validity and investigate the effects of photometric redshifts, we compare our model with the correlation function computed from a suite of 50 large-volume, moderate-resolution numerical $N$-body simulation boxes, where we can perform the analysis not only in real- and redshift space, but also simulate the influence of a gaussian redshift error distribution with an absolute rms of $\sigma_z= 0.015$, $0.03$, $0.06$, and $0.12$, respectively. We conclude that for the given volume ($V_{box}=2.4 h^{-3}$\,Gpc$^3$) and number density ($\bar{n}\approx1.25\times 10^{-4} $) of objects the full shape of $\xi(r_p,\pi)$ is modeled accurately enough to use it to derive unbiased constraints on the equation of state parameter of dark energy $w_{DE}$ and the linear bias $b$, even in the presence of redshift errors of the order of $\sigma_z$ = 0.06.
\end{abstract}

\begin{keywords}
cosmology -- large scale structure -- correlation function
\end{keywords}

\section{Introduction}

The accelerated expansion of the Universe was first detected by \citet{1998AJ....116.1009R} and \citet{1999ApJ...517..565P} from the analysis of supernova type Ia observations. Over subsequent years a variety of independent data sets have confirmed this finding \citep{2005MNRAS.362..505C,2009ApJS..180..330K,2009ApJ...700.1097H,2009ApJS..185...32K,2009MNRAS.400.1643S,2010MNRAS.401.2148P,2011ApJS..192...18K}. The origin of the accelerated expansion of the Universe is one the most challenging open problems in cosmology. Several theories have been proposed to provide an explanation for this phenomenon  \citep[see e.g.][for a review]{2008GReGr..40..301D}. One of the most promising solutions is the inclusion of a homogeneously and isotropically distributed fluid with a repulsive gravitational force to the energy-momentum tensor in the Einstein equations. This so-called {\it dark energy} can be characterized by its equation of state parameter $w_{\rm DE}= p_{\rm DE}/\rho_{\rm DE}$, the ratio of the pressure and energy density of this component. Current observations are consistent with the simplest solution, i.e. a cosmological constant, for which $w_{\rm DE}= -1.$ \citep[see e.g.][]{2009MNRAS.400.1643S,2010MNRAS.401.2148P,2011MNRAS.416.3017B,2011MNRAS.418.1707B,2011ApJS..192...18K,2012arXiv1202.0092M,2012arXiv1203.6616S}, although plenty of room still exists for alternative scenarios.

Several observational probes have been proposed as means to obtain new clues about the nature of dark energy. Among them, the analysis of Baryonic Acoustic Oscillations (BAOs) is considered as one of the most promising alternatives \citep{arXiv:astro-ph/0609591}. BAOs are the fossil signal of the acoustic waves that propagated through the photon-baryon fluid until the epoch of recombination, when electrons and protons formed neutral hydrogen and the speed of sound dropped rapidly to zero. The maximum distance the sound waves were able to travel is called sound horizon. This scale is imprinted on the large-scale matter density fluctuations. In the two point correlation function $\xi(r)$ the signature of the BAOs shows up as a single broad bump at around $110\,h^{-1}{\rm Mpc}$ \citep{2004ApJ...615..573M}, in the power spectrum as a series of small wiggles \citep{1998ApJ...496..605,2012MNRAS.421.2656M}. Since galaxies form in the high-density peaks of the matter density field, BAOs are also present in the distribution of galaxies at later times, and have indeed been detected \citep{2005ApJ...633..560E,2005MNRAS.362..505C,2009MNRAS.396.1119C,2011MNRAS.416.3017B,2011MNRAS.418.1707B} locally and at intermediate redshift. Since they originate in the early Universe where dark energy does not play a role and their propagation is described by well understood plasma physics, the BAO signal can be used as a standard ruler. By measuring the apparent extent of the acoustic scale in the directions parallel and perpendicular to the line-of-sight it is possible to recover the redshift evolution of the Hubble parameter $H$ and the angular diameter distance $D_{\rm A}$ through a simple geometrical relation \citep{2003ApJ...594..665B}. 

Due to the low amplitude of the BAO signal, their analysis requires the observation of large volumes, which due to the expenditure of time spectroscopy causes, are often only accessible with photometric surveys -- which in turn comes at the price of large redshift inaccuracies. Besides this technical difficulty, which we will address in this paper, galaxy clustering differs from the linear theory predictions in a number of ways that need to be taken into account if these measurements are to be used to obtain constraints on cosmological parameters. Nonlinear growth of structure leads to coupling between Fourier modes, changing the shape of the power spectrum and correlation function. The measured clustering statistics are also affected by the gravitationally induced peculiar motions of galaxies which introduce a distortion when the distance to each galaxy is inferred from its observed redshift. Besides this, galaxies are biased tracers of the underlying dark matter density field. As a result, the correlation function of the galaxies could be a modified version of that of the mass. Besides their effect on the broad-band shape of the correlation function and power spectrum, these effects alter the shape and location of the BAO signature \citep{2008MNRAS.383..755A,2008MNRAS.390.1470S,2008PhRvD..77d3525S} and might jeopardize the success of any analysis if they are not modelled accurately.

Redshift-space distortions originating from the peculiar velocities of the galaxies are unfortunately not the only distortions of the line-of-sight component of the correlation function. The measurement of redshifts are not free of errors, which add to the distortion of the clustering signal. These errors are particularly important in the case of photometric redshifts where they can be as large as $\sigma_z$ = 0.03 -- 0.04, smearing out the clustering signal along the line-of-sight, and leading to a significant reduction of its amplitude at a given scale. Examples of large photometric redshift surveys with the aim to measure the equation of state of dark energy using galaxy clustering are the Panoramic Survey Telescope And Rapid Response System Pan-STARRS \citep{2004AAS...204.9702C},  the upcoming Dark Energy Survey DES \citep{2010AAS...21547009T}, or PAU ({\bf P}hysics of the {\bf A}ccelerating {\bf U}niverse, see \citet{Benitez09}).

As described above, in order to obtain constraints on both $D_{\rm A}(z)$ and $H(z)$ it is necessary to measure the full two-dimensional clustering pattern by splitting the comoving distance between the galaxies into a component perpendicular, $r_p$, and parallel, $\pi$, to the line-of-sight. Despite this, most of the theoretical analyses and observations to date have focused on angle-averaged statistics which are sensitive instead to the parameter combination $D_{\rm A}^2(z)/H(z)$. In this paper we model the full $\xi(r_p,\pi)$ by taking nonlinear structure growth, nonlinear coherent infall velocities, and redshift errors into account. The paper is structured as follows: In Section \ref{method} the analytic model and the tests performed to compare it with the results of N-body simulations are discussed in detail. The performance of the model for real-, redshift- and redshift-error space is described in Section \ref{discussion}, where it is also compared to previous work. Finally, in Section \ref{conclusions} the most important results of this paper are summarized.

\section{Method}
\label{method}
\subsection{The L-BASICC II simulations}
\label{L-BASICC}

The validity of the model we are going to describe in the following is tested by comparing it to the anisotropic correlation function measured in a suite of 50 numerical large-volume, medium-resolution $N$-body simulations, the L-BASICC II simulations \citep{2008MNRAS.387..921A,2008MNRAS.390.1470S}; we use the snapshots at redshift $z=0.5$. The L-BASICC II simulations correspond to fifty different realizations of a flat $\Lambda$CDM cosmology with $\Omega_{\rm M} = 0.237$, $n_{\rm s}= 0.954$, and $\sigma_8 = 0.77$, which is in close agreement with the latest constraints on cosmological parameters from CMB and LSS measurements (e.g. \citealp{2006MNRAS.366..189S,Spergel07,2009MNRAS.400.1643S,2011ApJS..192...18K}). Each simulation box covers a volume of ($1340\,h^{-1}{\rm Mpc})^3 = 2.4\,h^{-3}{\rm Gpc}^{3}$ with a particle mass of $M_{dm} = 1.75\times 10^{12}\,h^{-1} M_{\odot}$. Haloes were identified by a Friend-of-Friends halo finder as gravitationally bound systems with at least ten dark matter particles. Hence, the lowest mass halo in the catalogue has a mass of $M_{\rm min}=1.75\times 10^{13} \,h^{-1}M_{\odot}$. A detailed description of the L-BASICC II simulations is given in \citet{2008MNRAS.387..921A}. 

\subsection{The model}
\label{model}

This section describes our model of the anisotropic two-point correlation function $\xi(r_p,\pi)$ for real, redshift and redshift error space. Non-linear structure growth, described by third-order perturbation theory \citep{1994ApJ...431..495J}, the non-linear Kaiser effect (\citealp{1987MNRAS.227....1K,2004PhRvD..70h3007S}) and redshift errors are taken into account. The inclusion of all of these effects results in an accurate description of $\xi(r_p, \pi)$, helping to get more reliable estimates of the cosmological parameters. 

\subsubsection{Non-linear evolution}

The evolved structure of the Universe at lower redshifts is not described with sufficient accuracy by linear perturbation theory. In $\xi(r)$ the BAO peak is broadened and shifted towards smaller scales \citep{2008MNRAS.383..755A,2008PhRvD..77d3525S,2008PhRvD..77b3533C,2008MNRAS.390.1470S}. An incorrect modelling of this feature will produce biased estimates of the cosmological parameters. It is particularly important to model the nonlinearities correctly if the full shape of the correlation function is taken into account in the analysis. Nonetheless, non-linearities must also be included even if only the BAO peak is used as a standard ruler, as these introduce a small shift of its position which can lead to an up to a five times larger shift in the estimated value of $w_{{\mathrm DE}}$ \citep{2005MNRAS.363.1329B,2005ApJ...633..560E,2008MNRAS.383..755A}. Including smaller scales (where nonlinearities are even stronger) in the analysis makes an accurate treatment of nonlinear clustering growth a necessity. 

Several methods have been developed in order to describe the nonlinear growth of structure, e.g. third-order perturbation theory \citep{1994ApJ...431..495J,Jeong_06,Jeong_09}, renormalized perturbation theory \citep{2006PhRvD..73f3519C,crocce_RPT2,mcdonald_renormbias,matarrese_rengroup1, bernardeau_08,matarrese_rengroup2} and Lagrangian perturbation theory \citep{2008PhRvD..77f3530M, mats_LPT2,pietroni_flowtime,taruya_closure,smith09_SZ,taruya_09,Elia_10}. In our analysis we model the non-linear evolution of density fluctuations by means of third-order perturbation theory. The main input for the evaluation of the third-order perturbation theory $P(k)$ is the linear theory power spectrum, which we generate using the {\sc CAMB} code\footnote{http://camb.info/sources/} \citep{2000ApJ...538..473L}. The numerical routines taken from http://gyudon.as.utexas.edu/\~{}komatsu/CRL/index.html are then used for the further processing of the linear CAMB output. The third-order perturbation theory results in terms proportional to the squared linear power spectrum which are added to the linear power spectrum; this sum is denoted $2^{nd}$ order power spectrum \citep{1994ApJ...431..495J}. The output obtained by these routines are the $2^{nd}$ order power spectra for the density (Equation (A15) and (A16) in \citet{2007PASJ...59.1049N}) and the velocity (Equation (A17) and (A18) in \citet{2007PASJ...59.1049N}) field as well as the density - velocity (Equation (A19) and (A20) in \citet{2007PASJ...59.1049N}) cross power spectrum. These are needed for the calculation of the non-linear Kaiser effect \citep{2004PhRvD..70h3007S}. In \citet{1994ApJ...431..495J} only the density - density case is derived, see Equations (17) and (18).

\subsubsection{From P(k) to $\xi(r_p,\pi)$ }

The non-linear correlation function is obtained by Fourier transforming the third-order perturbation theory power spectrum: $\xi(r)=\frac{1}{2\pi^2}\int{P(k)}\frac{\sin kr}{kr}k^2dk$. Then, assuming small angles (large redshifts), such that $r^2=r_p^2+\pi^2$, we can split the distance between galaxies into a component perpendicular ($r_p$) and parallel ($\pi$) to the line-of-sight, and calculate the full two-dimensional real-space correlation function $\xi_{\rm rs}(r_p,\pi)$.

\begin{figure}
\centerline{\psfig{figure=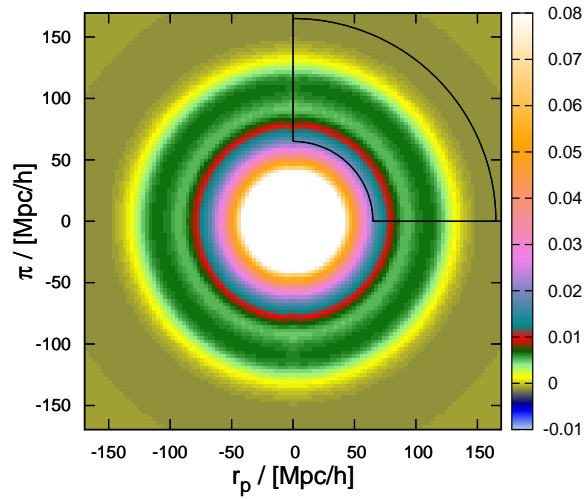,angle=0,clip=t,width=12.cm}}
\caption{Our model of the two-point correlation function of dark matter haloes in real space, $\xi_{rs}(r_p , \pi)$, evaluated at $z = 0.5$. The BAO peak is now a ring at about $r=\sqrt{r_p^2+\pi^2} \sim 110 h^{-1}$\,Mpc, here corresponding to the dark green feature. The black lines indicate the range in which we fit the model: Distances smaller than $|r|\leq 64 h^{-1}$\,Mpc and larger than $|r|\geq 165 h^{-1}$\,Mpc will not be taken into account in the analysis. Only one quadrant is used in the fit, since the information in the other three quadrants is redundant. The colour bar on the right shows the values of $\xi_{rs}(r_p , \pi)$.}
\label{DMxirppi}
\end{figure}

So far only the clustering of dark matter was considered. In real space the clustering of dark matter and collapsed objects (haloes or galaxies) can be related by the bias parameter $b$, which in the two-point statistics to first order approximation can be assumed to be a constant factor  \citep{1994ApJ...425....1F} at the large scales we are interested in, 
\begin{equation}
\xi_{halo/galaxy}(r_p,\pi) = b^{2}\,\xi_{dm}(r_p, \pi)~,
\label{biasdefinition}
\end{equation}
where $\xi_{dm}(r_p, \pi)$ and $\xi_{halo/galaxy}(r_p, \pi)$ are the dark matter and halo/galaxy $\xi_{rs}(r_p, \pi)$, respectively. As can be seen in Figure \ref{DMxirppi}, which shows the prediction of the dark matter haloes real-space $\xi_{rs}(r_p,\pi) = \xi_{halo/galaxy}(r_p,\pi)$  evaluated at $z=0.5$, for the cosmology of the L-BASICC II simulations, the BAO peak in $\xi(r)$ is now distributed over a ring with a radius of $r\sim 110\,h^{-1}{\rm Mpc}$. In redshift space this simple equation does not hold, since the bias parameter also changes the size and shape of the redshift space distortions (mainly through the quadrupole and hexadecapole contribution of the non-linear Kaiser effect, see Equation (\ref{quadrupole}) and (\ref{hexadecapole}) below), and is thus not a simple multiplicative factor anymore. A precise modelling of the redshift space distortions is of vital importance; we will describe how we incorporate them in our model in detail in the following section.

\subsubsection{From real to redshift space}\label{zdistortions}

If the exact positions of the objects (galaxies, haloes) were known, $\xi_{rs}(r_p, \pi)$ would describe their clustering. However, peculiar velocities of these objects lead to a blurring of their exact positions. The peculiar velocities are induced by the surrounding density field as galaxies move in the local gravitational potential. There are two different effects, which have a different influence on the correlation function of galaxies: on small scales the random motions of galaxies within their host dark matter halo make structures appear elongated along the line-of-sight, an effect known as Fingers-of-God \citep[FoG][]{1972MNRAS.156P...1J}, and lead to a reduction of the clustering signal. We ignore the FoGs in this analysis -- first of all we are interested in very large scales, where their effect is negligible, secondly we test our model against the L-BASICC II halo catalogues, which due to the lack of substructure (or galaxies) do not display any FoGs.

On large scales coherent infall of objects onto large structures dominates the redshift-space distortions. This effect was first described by \citet{1987MNRAS.227....1K}. The formulation is based on the distant observer approximation, meaning that the structure is distorted in a plane-parallel fashion, and therefore only the line-of-sight component is affected. The Kaiser effect in its original, linear form includes the linear power spectrum; for details we refer to \citep{1992ApJ...385L...5H}. In order to obtain unbiased and more accurate constraints on the cosmological parameters, we have to go a step further \citep{Cole94}, and model the non-linear Kaiser effect \citep{2004PhRvD..70h3007S} on the anisotropic two-point correlation function in redshift space $\xi_{zs}(r_p, \pi)$. Following the theoretical derivation of the linear Kaiser effect for $\xi_{zs}(r_p, \pi)$ in \citet{1992ApJ...385L...5H} and \citet{1993ApJ...417...19H} the anisotropic redshift space correlation function can be expressed as
\begin{equation}
\xi_{zs}(r_p, \pi) =  \xi_0(r) P_0(\mu) + \xi_2(r) P_2(\mu) + \xi_4(r) P_4(\mu),
\label{sum}
\end{equation}
where $P_i(\mu)$ are the Legendre polynomials and the multipoles $\xi_i(r)$ are given by

\begin{eqnarray}
\xi_0(r) & = &  b^2 \xi_{\delta\delta}(r) + \frac {2} {3}f b \xi_{\delta\theta}(r) + \frac{1} {5}f^{2} \xi_{\theta\theta}(r), 
\label{monopole}\\
\xi_2(r) & = &  \frac {4} {3}f b \lbrack \xi_{\delta\theta}(r) - \bar\xi_{\delta\theta}(r) \rbrack + \frac{4} {7}f^{2} \lbrack \xi_{\theta\theta}(r) - \bar\xi_{\theta\theta}(r) \rbrack, 
\label{quadrupole}\\
\xi_4(r) & = & \frac{8} {35}f^{2}\cdot \lbrack \xi_{\theta\theta}(r) + \frac{5} {2} \bar\xi_{\theta\theta}(r) - \frac{7} {2} \dbar\xi_{\theta\theta}(r) \rbrack~,
\label{hexadecapole}
\end{eqnarray}
where $\xi_{\delta\delta}(r)$, $\xi_{\delta\theta}(r)$ and $\xi_{\theta\theta}(r)$ are the dark matter density-density, density-velocity and velocity-velocity correlation functions, respectively, Fourier transformed from the corresponding power spectra, and $f=\frac{d\ln D}{d \ln a}$ is the logarithmic derivative of the growth factor $D$ with respect to the scale factor of the Universe, which we compute as
\begin{equation}
f(z) =  \left( \frac{\Omega_M \cdot \left( 1 + z \right)^{3}}{\Omega_M \left( 1 + z \right)^{3} + \Omega_{\Lambda} \left( 1 + z \right)^{3\cdot \left( 1 + w_{{\mathrm DE}} \right)}} \right)^{\gamma} ~.
\label{dimles_growth}
\end{equation}

To allow for the dark energy to be different from a cosmological constant (in which case $\gamma=0.55$), we use $\gamma = 0.55 + 0.05 \cdot (1 + w_{{\rm DE}}(z))$ for $w_{{\rm DE}}(z)>-1$ and $\gamma = 0.55 + 0.02 \cdot (1 + w_{{\rm DE}}(z))$ for $w_{{\rm DE}}(z)<-1$ \citep{2007arXiv0709.1113L} in all our calculations. 

The barred correlation functions in Equations (\ref{monopole})--(\ref{hexadecapole}) are given by 
\begin{eqnarray}
\bar\xi_{\delta\theta}(r) & \equiv & 3r^{-3}\int_{0}^{r} \xi_{\delta\theta}(r^{\prime})r^{\prime 2}dr^{\prime}  \\
\bar\xi_{\theta\theta}(r) & \equiv & 3r^{-3}\int_{0}^{r} \xi_{\theta\theta}(r^{\prime})r^{\prime 2}dr^{\prime}  \\
\dbar\xi_{\theta\theta}(r) & \equiv & 5r^{-5}\int_{0}^{r} \xi_{\theta\theta}(r^{\prime})r^{\prime 4}dr^{\prime} ~.
\label{Integrals}
\end{eqnarray}

\begin{figure}
\centerline{\psfig{figure=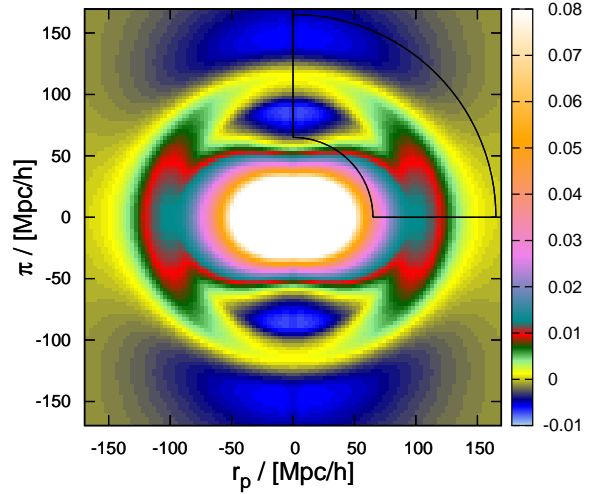,angle=0,clip=t,width=12.cm}}
\caption{$\xi_{zs}(r_p,\pi)$ of dark matter haloes with nonlinear Kaiser effect at redshift $z$ = 0.5. As in Figure \ref{DMxirppi} the black lines indicate the range in which we fit the model.}\label{kaisereffect}
\end{figure}

If the linear power spectrum would be used instead of the 2$^{nd}$ order power spectra, the derived formulas above would result in the linear Kaiser effect. In Figure \ref{kaisereffect} the contours of $\xi_{zs}(r_p,\pi)$ modeled including the nonlinear Kaiser effect are shown. Redshift-space distortions destroy spherical symmetry and change the shape and location of the BAO ring. The structure, affected by the coherent infall of objects, looks squashed along the line-of-sight, which also increases the clustering signal compared to $\xi_{rs}(r_p,\pi)$. In addition to the change in the amplitude, redshift-space distortions also contain information  about the cosmological parameters and bias, as both growth rate and linear bias enter into the multipoles which describe them (Equations \ref{monopole}--\ref{hexadecapole}).

\subsubsection{From redshift to redshift error space}\label{redshifterrorspace}

So far the model is able to describe redshift space clustering as long as the redshifts (which translate into distances between the observer and the objects) are determined without large redshift errors, as in spectroscopic surveys, where they are negligible. However, collecting a large number of spectroscopic redshifts is  time consuming, especially if the aim is to observe a large volume at high redshift. A faster alternative is to estimate redshifts from deep photometric data, which makes it necessary to include the effect of large redshift errors on $\xi_{zs}(r_p,\pi)$ in the model. How this can be done will be described in the next subsection.

Traditionally photometric redshifts are derived from observed fluxes in five or more broad to medium band filters \citep{1962IAUS...15..390B}. The probability distribution of the fitted redshifts depends on the spectral type, magnitude and redshift of the observed objects, the filter set and the fitting scheme (e.g. a neural network or a comparison with a library of template spectra, which are used to perform synthetic photometry). However, several authors have found the redshift error distribution in realistic surveys to be very close to gaussian. For example, \citet{Cunha2009} calculated the full probability distribution function for $\sim 78$ million SDSS DR7 galaxies using photometric observables and weighted sampling from a spectroscopic subsample of the data, and \citet{Saglia2012} estimated photometric redshifts for objects found in the Medium-Deep Fields of Pan-STARRS1 using available spectroscopic surveys (including SDSS spectra) as training and/or verification sets. From a direct comparison between spectroscopic and photometric redshifts obtained for the same objects, they find the width of the distribution of all galaxies in the sample to be $\sigma_z\la 5\%$, with $\sim 1$\% extreme outliers (defined as the fraction of objects for which $|z_{phot}-z_{spec}|> 0.15 \times (1 + z_{spec})$ ), and $\sigma_z/(1+z)=0.024$ and an insignificant fraction of only $0.4$\%  outliers for the object types which will eventually be used to measure BAOs in Pan-STARRS, namely luminous red galaxies (LRGs) at $z\la 0.5$.

Hence, as a first order approximation we adopt here a single gaussian peak to simulate the influence of photometric redshift errors on the measurement of $\xi(r_p,\pi)$. Since the uncertainty of the redshift estimation mainly results in a distortion of the distance along the line-of-sight, redshift errors have a similar effect as peculiar velocities, and to first order they do not depend on the local density field.

\begin{figure}
\centerline{\psfig{figure=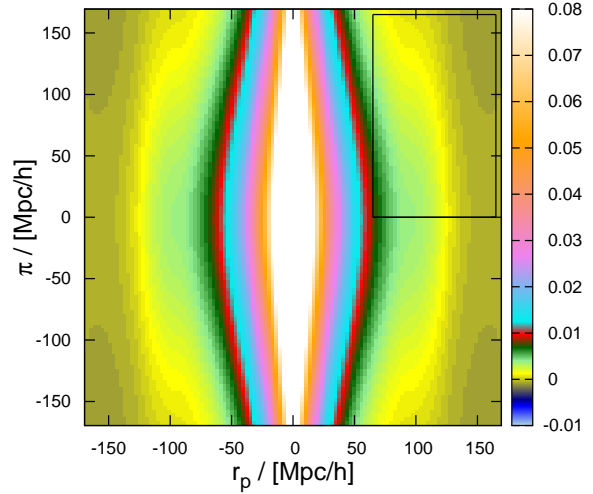,angle=0,clip=t,width=12.cm}}
\caption{The model of $\xi_{zerr}(r_p,\pi)$ at $z = 0.5$ convolved with a pairwise redshift error distribution, which assumes gaussian redshift errors with an rms of $0.03$. Because the clustering signal is now smeared out along the line-of-sight, in order to capture (to zeroth order) the same information as in the fit of the correlation function in real and redshift space, we replace the spherical shell in which we fit the model (see Figures (\ref{DMxirppi}) and (\ref{kaisereffect})) by a cylinder (as indicated by the black rectangular box).}\label{redshifterrorpicture}
\end{figure}

The main effect of photometric redshift errors is that the line-of-sight component of the distance to a galaxy is smeared out by an amount $\delta x_{||}$. This section follows the description in \citet{2005MNRAS.363.1329B} and \citet{2007AA...468..113P} for modeling the redshift errors where the redshift errors are assumed to be gaussian distributed such that the spatial displacement $\delta x$ is given by

\begin{equation}
f(\delta x_{||}) \propto \exp{\left[-0.5\left(\frac{\delta x_{||}}{\sigma_x}\right)^2\right]}~,
\label{smear}
\end{equation}

where $\sigma_x$ is the comoving distance corresponding to the rms $\sigma_z$ of the redshift error probability distribution function at the considered  redshift . Equation (\ref{smear}) describes the probability distribution function for one single galaxy, but the correlation function is a pairwise statistic, hence in order to simulate the impact on the model $\xi_{zerr}(r_p,\pi)$ we have to compute the {\it pairwise} error distribution function. For two galaxies with errors $\delta x_{||,i}$ and $\delta x_{||,j}$ the rms of their pairwise error would be 
\begin{equation}
\delta x_{||,ij}=\sqrt{\delta x_{||,i}^2+\delta x_{||,j}^2}~.
\end{equation}
If $N_{g}$ galaxies are contained in the survey then there are $N_{p}$ = $\frac{N_{g} (N_{g} - 1)}{2}$ pairs. Every galaxy is contained in $N_{g} - 1$ pairs, and the overall convolving function $f_p(\pi)$ is a sum of the Gaussians corresponding to all pairs $ij$: 
\begin{equation}
f_p(\delta \pi) = \frac{1}{N_{p}} \sum_{n=1}^{N_{p}} \exp{\left[-0.5\left(\frac{\pi}{\delta x_{||,ij,n}}\right)^2\right]}~.
\label{smearing function}
\end{equation}
We estimate this by randomly drawing $N_{g}$ values $\delta x_{||,n}$ from Equation (\ref{smear}), from which we calculate all possible pairwise errors $\delta x_{||,ij,n}$. Equation (\ref{smearing function}) is the probability distribution of the smearing of the correlation signal along the line-of-sight due to the redshift errors. By convolving $\xi_{zs}(r_p,\pi)$ with this function, we model the effect of photometric redshifts on the anisotropic correlation function. The resulting $\xi_{zerr}(r_p ,\pi)$ for redshift errors with an rms of $\sigma_z= 0.03$ is shown in Figure \ref{redshifterrorpicture}. The redshift errors have the same effect on large scales as the FoG have on small scales, only the order of magnitude is much larger. The clustering signal is smeared out and the structure looks extremely elongated, the BAO ring is almost completely washed out and can barely be identified; with increasing size of the redshift errors the signal to noise of the BAO ring decreases rapidly, until the signal completely vanishes and cannot be used anymore to constrain any cosmological parameters. 

\subsection{Constraining cosmological parameters }\label{analysis}

We will now demonstrate the capability of the model of $\xi(r_p,\pi)$ described in the previous section to constrain the equation of state parameter of dark energy $w_{{\mathrm DE}}$ and the linear bias $b$, by fitting the correlation function computed from the L-BASICC II simulations. The anisotropic correlation function is  calculated using the estimator of \citet{1993ApJ...412...64L} in each single box up to a distance of $\pm 300.0 h^{-1}$\,Mpc, and in bins of $3.0h^{-1}$\,Mpc in both $r_p$ and $\pi$, for real, redshift and redshift error space, respectively. In the latter case ``redshift errors'', that is, small offsets, are added to the coordinate which has been designated the line-of-sight. These offsets have been randomly drawn from a gaussian error distribution function with a given rms. From the 50 estimates of $\xi(r_p,\pi)$ the mean is calculated.

The model $\xi(r_p,\pi)$ is also evaluated from ($-300 h^{-1}$\,Mpc,$-300 h^{-1}$\,Mpc) to ($300 h^{-1}$\,Mpc, $300 h^{-1}$ Mpc), in steps of ($0.5 h^{-1}$ Mpc,$0.5 h^{-1}$\,Mpc). In order to facilitate an accurate comparison with the simulation, which has been calculated with a six times larger bin size, we average over the model bins when fitting $w_{{\mathrm DE}}$ and $b$. While keeping $H_0$, $\Omega_m$ and $\Omega_\Lambda$ fixed at the values determined by the L-BASICC II simulation, we vary $w_{{\mathrm DE}}$ and $b$, which both change the amplitude of the correlation function and the shape of the redshift space distortions (see Section \ref{zdistortions}). 

When we calculate the correlation function of the simulation boxes, we have to assume a fiducial cosmology -- for simplicity we chose the input parameters of the L-BASICC II, i.e. the correct cosmology (and thus the correct correlation function with the acoustic peak at the position predicted by theory). For any other choice of cosmological parameters the measured redshifts and positions on the sky translate differently into distances perpendicular and parallel to the line-of-sight, which leads to a distortion \citep{1979Natur.281..358A}. Therefore when we fit the model to the data, we have to rescale the distances accordingly:

\begin{equation}
r_p^{t} = \frac{D_{A}^{t}\left( z \right)}{D_{A}^{f}\left( z \right)}\cdot r_p^{f}
\label{cosmological test transverse size}
\end{equation}

\noindent and

\begin{equation}
\pi^{t} = \frac{H^{f} \left( z \right)}{H^{t} \left( z \right)} \cdot \pi^{f} ~,
\label{cosmological test along the line sight size}
\end{equation}

\noindent where the quantities for the fiducial cosmology of the model are superscripted with $f$ and the ``true'' cosmology (of the simulation) with $t$, $D_{A}(z)$ is the angular diameter distance and $H(z)$ is the evolution factor,

\begin{equation}
D_{A}\left( z \right) \equiv \frac{a \cdot r_p}{\theta} = \frac{c}{1 + z}\int_0^{z}\frac{dz}{H(z)}
\label{angular diameter distance}
\end{equation}

and

\begin{equation}
H\left( z \right) = H_0\cdot\sqrt{\left( \Omega_{M} \left( 1 + z \right)^{3} + \Omega_{\Lambda}\left( 1 + z \right)^{3\left(1 + w_{{\mathrm DE}}\right)} \right)} ~
\label{evolution part}
\end{equation}
where $a$ is the scale factor (the evolution of $a$ is given by the Friedmann equations and today's value is defined to be unity) and $\theta$ the angular extent on the sky. The model and the halo catalog correlation functions are compared within a spherical shell of $64 h^{-1}$\,Mpc $\leq \sqrt{r_p^{2} + \pi^{2}} \leq 165 h^{-1}$\,Mpc in both real and redshift space. In redshift error space the shape of that shell is distorted due to the smearing of the redshift errors, so in order to sample similar scales we replaced the spherical shell by a cylinder, as indicated by a rectangle in Figure \ref{redshifterrorpicture}. The corners of the rectangle are defined by $\left[64 h^{-1} Mpc, 0 h^{-1} Mpc\right]$ (left lower corner), $\left[64 h^{-1} Mpc, 165 h^{-1} Mpc\right]$ (left upper corner), $\left[165 h^{-1} Mpc, 0 h^{-1} Mpc\right]$ (right lower corner) and $\left[165 h^{-1} Mpc, 165 h^{-1} Mpc\right]$ (right upper corner).

The model was tested for two different cases, in one case the information contained in the amplitude was taken into account, and in the other case we analytically marginalized over the amplitude using the scheme described in \citet{2002PhRvD..66j3511L}. In the latter case only the shape of the model is examined. 

The fit is performed by means of a Monte Carlo Markov Chain (MCMC) after averaging and rescaling in order to find the best fitting values for $w_{{\mathrm DE}}$ and $b$, and to estimate their errors. Due to the limited number of realisations it is unfortunately not possible to calculate correct (invertible) covariance matrices (the full covariance matrix would take $100\times100$ pixels into account, hence at least $100\times100+1$ independent simulation boxes are required). If only the variance of the correlation function is used when fitting the {\it mean} $\xi(r_p,\pi)$, the resulting errors of $w_{{\mathrm DE}}$ and $b$ (in real and redshift space of the order of $\sim2$\%) are clearly underestimated. Instead we assume that the scatter of 50 fits to the {\it single} realisations can at least partly account for cosmic variance and the otherwise ignored contribution of the off-diagonal elements in the covariance matrix. Therefore the variance of the best-fitting values inferred in this way can be thought to represent a more realistic estimate of the errors of the fit parameters, while leaving the actual best-fitting values unchanged. Indeed, although the resulting values of  $w_{{\mathrm DE}}$ and $b$ reassuringly do not depend on the way of fitting, the size of the errors does. Throughout this paper we quote errors which have been calculated from the variance of the 50 values of $w_{{\mathrm DE}}$ and $b$, which are  about an order of magnitude larger than those inferred from the fit to the mean.  

We restrict the range in which we fit the parameters to $-1.6\leq w_{{\mathrm DE}} \leq-0.4$ and $1\leq b\leq 20$. 40000 steps are sufficient for the chain to converge towards the best fitting values and to explore the likelihood. Since the calculation of the third order perturbation theory power spectrum is time consuming, to speed up the analysis we run the MCMC on a grid, for which we construct a library of correlation functions $\xi(r)$ for $-1.6\leq w_{{\mathrm DE}} \leq -0.4$ in steps of $-0.001$. Once the random sampling process of the MCMC has chosen a new value for $w_{{\mathrm DE}}$, the appropriate $\xi(r)$ for the nearest value of $w_{{\mathrm DE}}^{lib}$ is selected from the library and used as starting point for the calculation of the model $\xi(r_p,\pi)$ (and $w_{{\mathrm DE}}$ set to $w_{{\mathrm DE}}^{lib}$).

\section{Results}
\label{discussion}

In this section the results of the fits of $\xi_{rs}(r_p,\pi)$ (real space), $\xi_{zs}(r_p,\pi)$ (redshift space) and $\xi_{zerr}(r_p,\pi)$ (redshift error space) will be discussed in detail. The analytic model will be tested by fitting the equation of state parameter of dark energy $w_{{\mathrm DE}}$ and the linear bias $b$ against the correlation function of the halo catalogue from the L-BASICC II simulations by following the description of Section \ref{analysis}. 

\subsection{Real Space}
\label{real space measurement}

In order to investigate the validity of the third order perturbation theory, we first test the model against the dark matter correlation function of the L-BASICC II simulations, where $b=1.0$. The best fitting parameters are found to be $w_{{\mathrm DE}} = -0.992\pm0.091$, $b = 0.998\pm0.073$, from which we conclude that the nonlinear structure growth is modeled  accurately enough to obtain unbiased estimates of these parameters.

If we now want to describe the clustering of collapsed objects like galaxies -- or dark matter haloes -- which are biased tracers of the dark matter density field, we have to include the bias (as defined in Equation \ref{biasdefinition}) in our calculation, as described in Section \ref{zdistortions}. Since in real space the linear bias is only a multiplicative factor which boosts the amplitude of the dark matter correlation function, but does not alter its shape, any information about the bias is contained in the amplitude at a given radius $r=\sqrt{r_p^2+\pi^2}$. Including the amplitude in the fit of  $\xi_{rs}(r_p, \pi)$ of the dark matter haloes in the 50 L-BASICC II boxes we obtain $w_{{\mathrm DE}}=-1.010 \pm0.117$ and $b=2.641\pm0.183$, if we disregard the amplitude, we cannot fit the bias, but the same value for $w_{{\mathrm DE}}$ is obtained when the amplitude information is disregarded. The measured value of $w_{{\mathrm DE}}$ is in good agreement with the fiducial value of the simulation. Since the simulation has only medium-resolution, it does not contain very small haloes (the lowest mass halo has ten particles, which corresponds to a minimum mass of $M_{\rm min}=1.75\times 10^{13} \,h^{-1}M_{\odot}$), hence a relatively large mean bias measured for all dark matter haloes is expected.

\begin{figure*}
\centerline{\psfig{figure=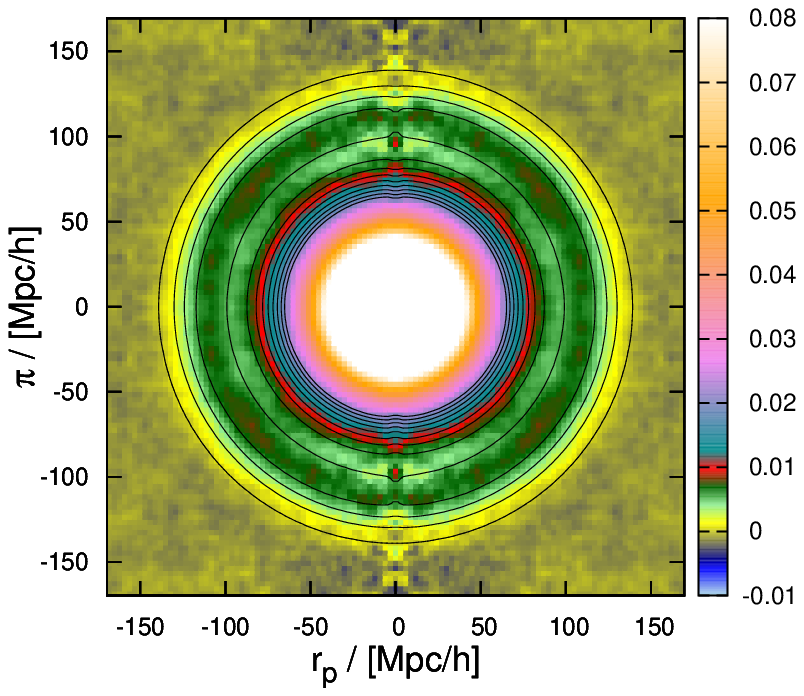,angle=0,clip=t,width=12.cm}
\psfig{figure=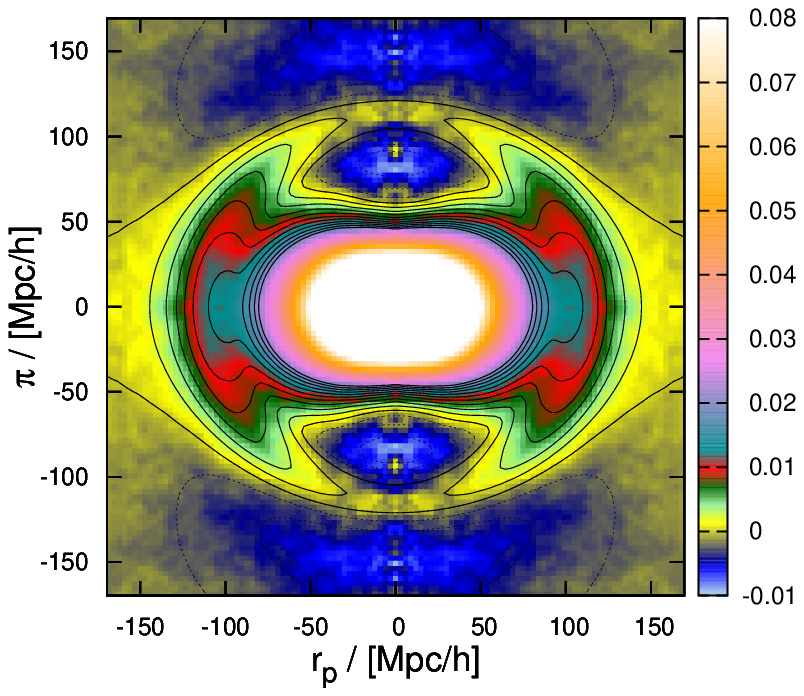,angle=0,clip=t,width=12.cm}}
\caption[ ]{Contours from the best fit model plotted over the correlation function calculated from the L-BASICC II dark matter halo catalogues  in real (left) and redshift space (right), respectively.}\label{xi_rp_pi_rs_zs}
\end{figure*}

\begin{figure*}
\centerline{\psfig{figure=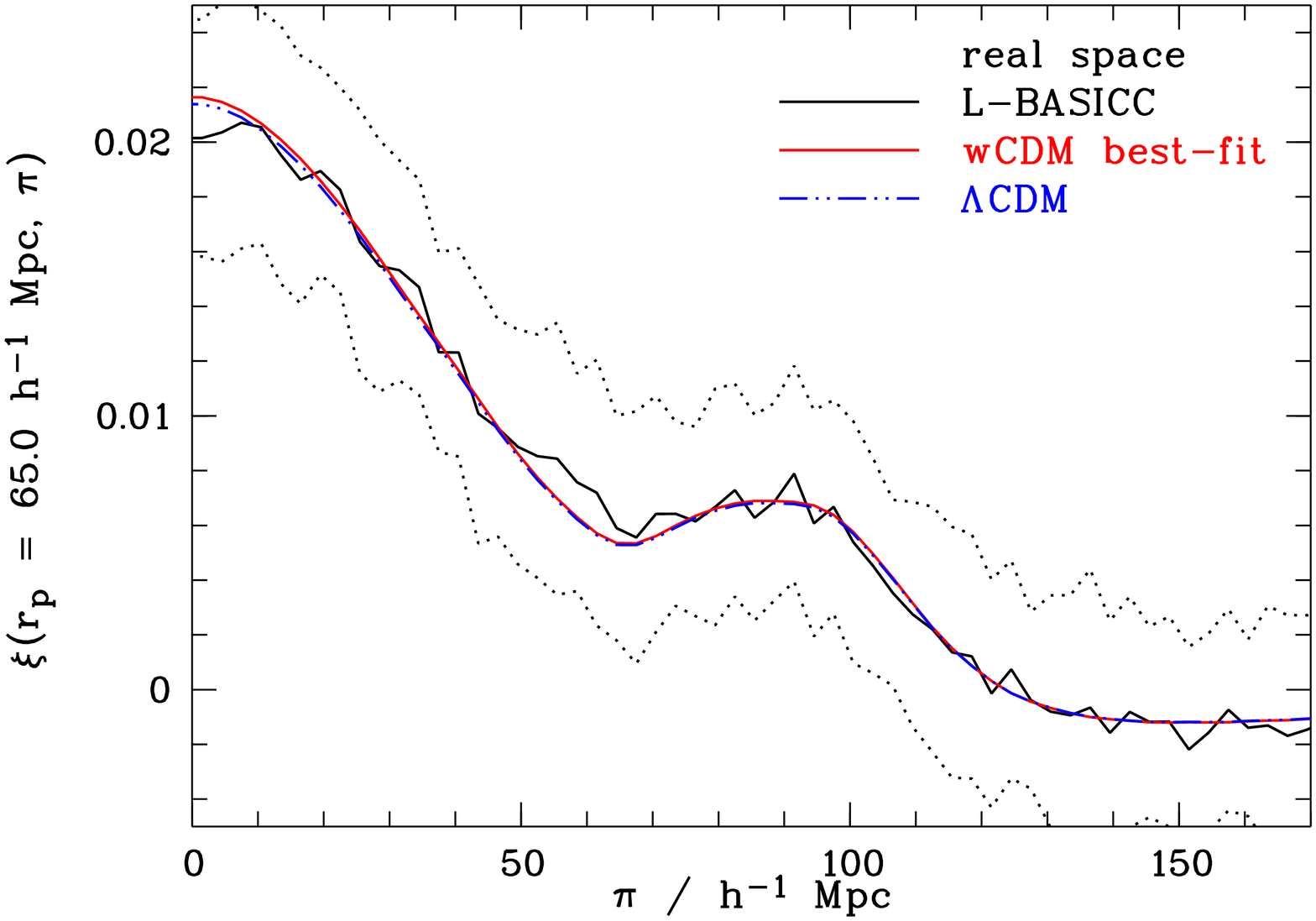,angle=270,clip=t,width=6.cm}
\psfig{figure=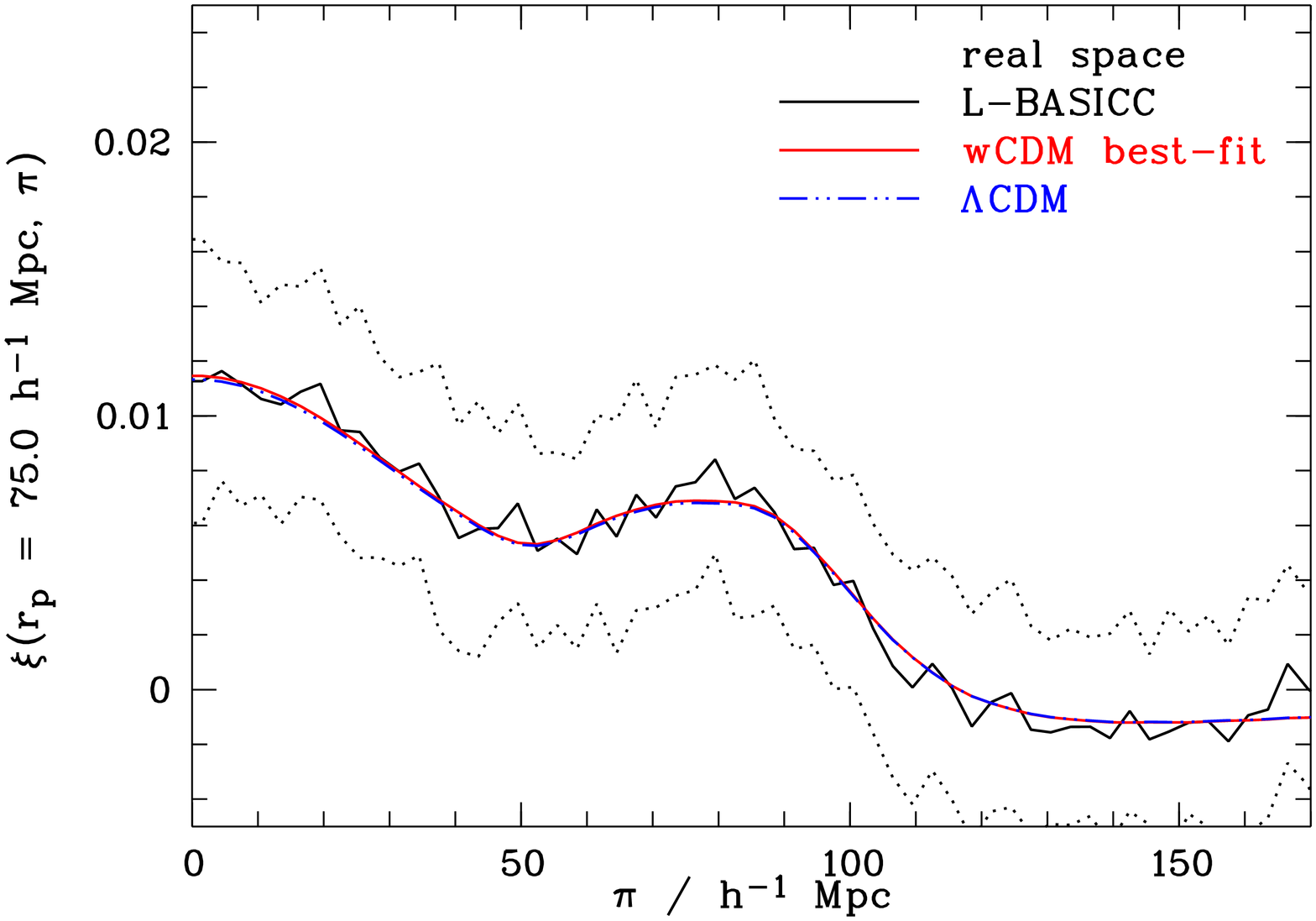,angle=270,clip=t,width=6.cm}
\psfig{figure=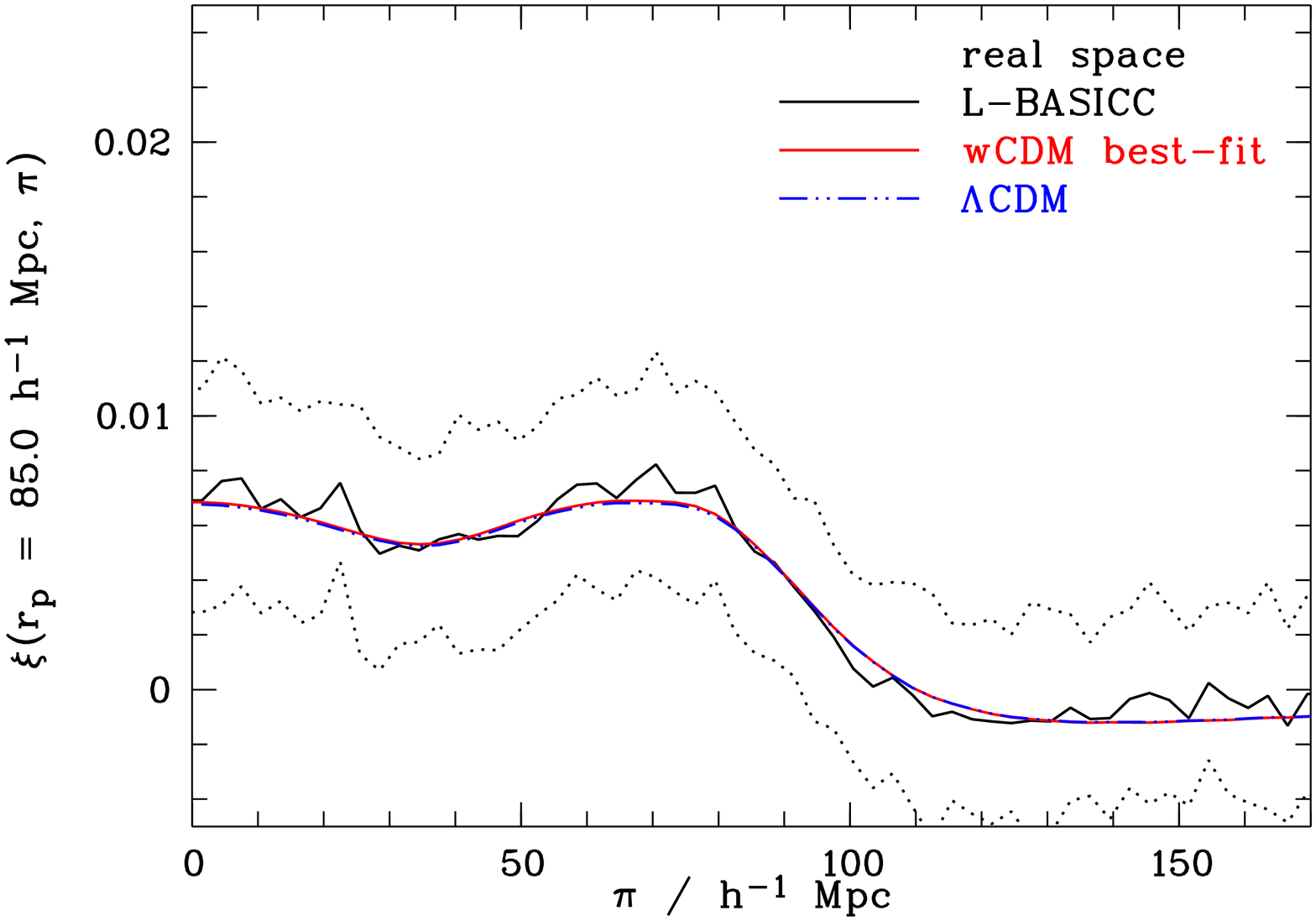,angle=270,clip=t,width=6.cm}}
\centerline{\psfig{figure=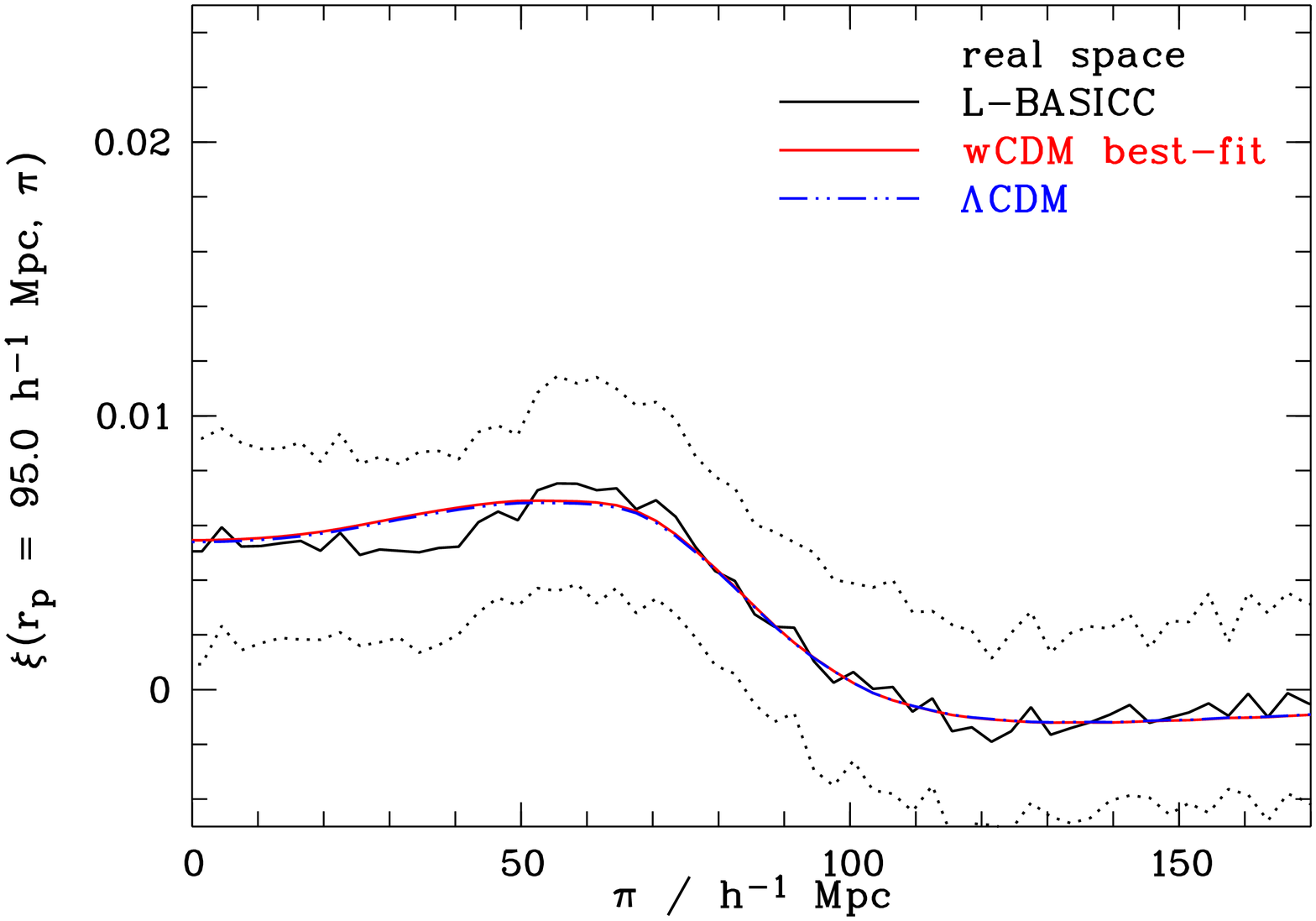,angle=270,clip=t,width=6.cm}
\psfig{figure=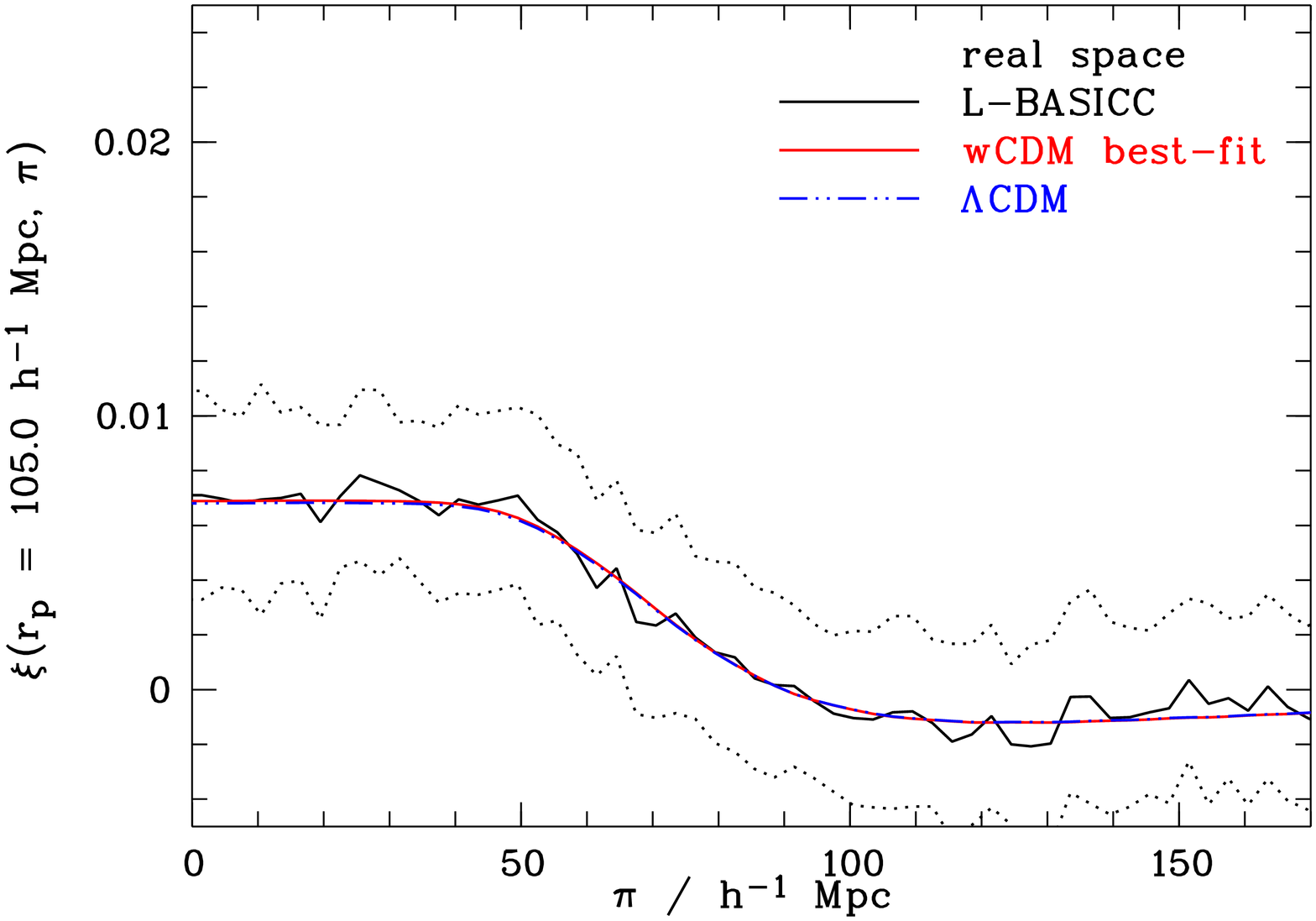,angle=270,clip=t,width=6.cm}
\psfig{figure=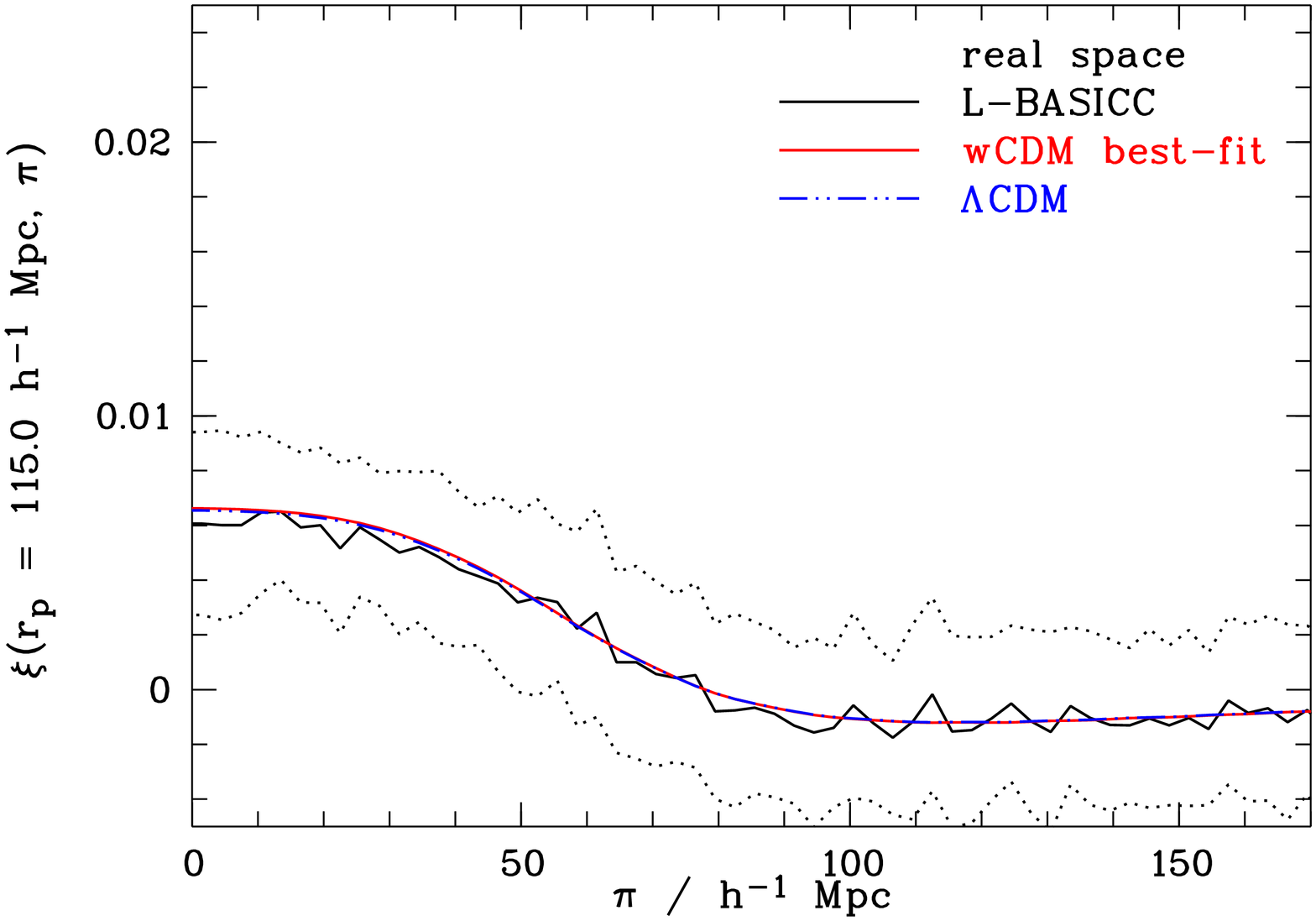,angle=270,clip=t,width=6.cm}}
\centerline{\psfig{figure=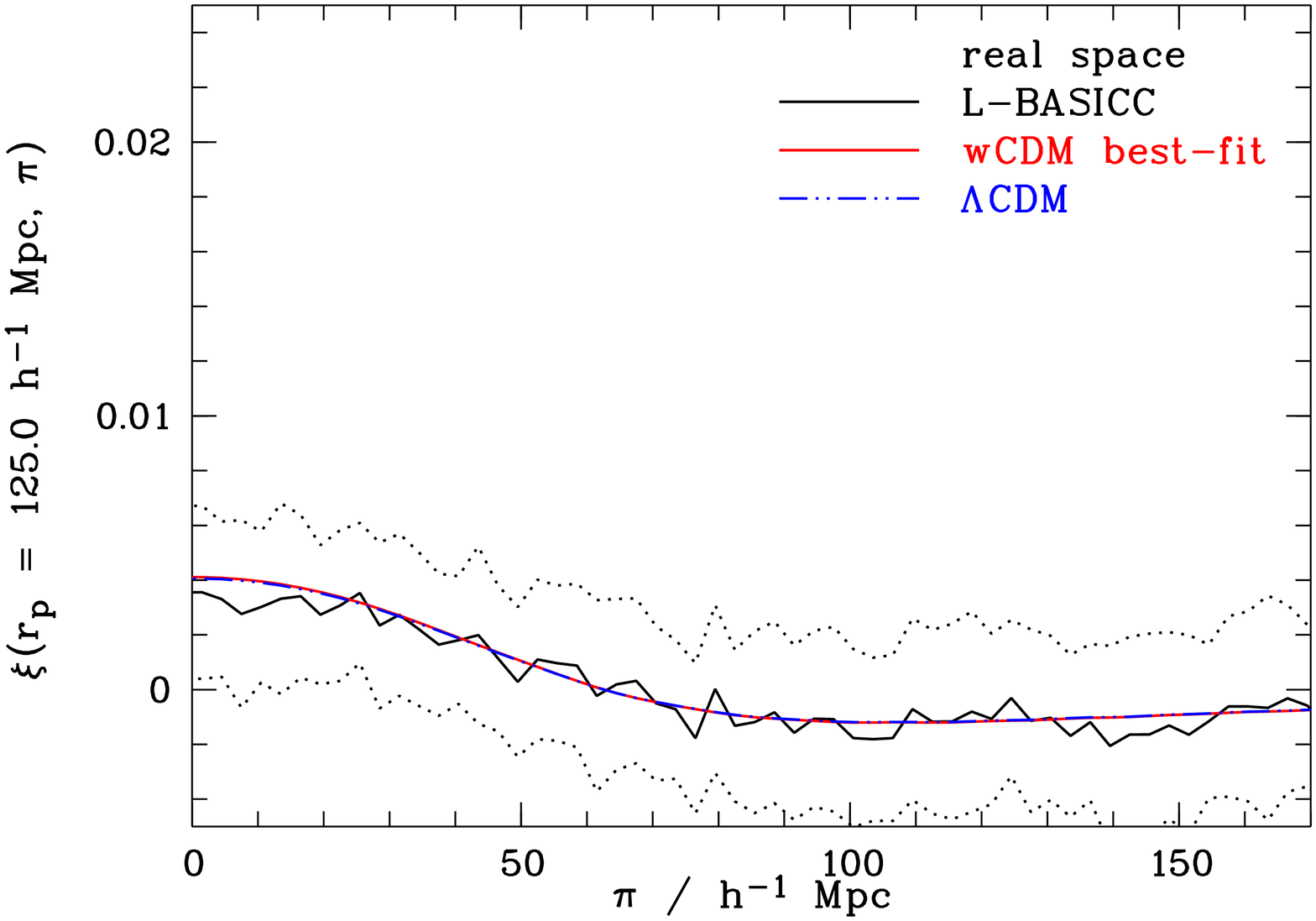,angle=270,clip=t,width=6.cm}
\psfig{figure=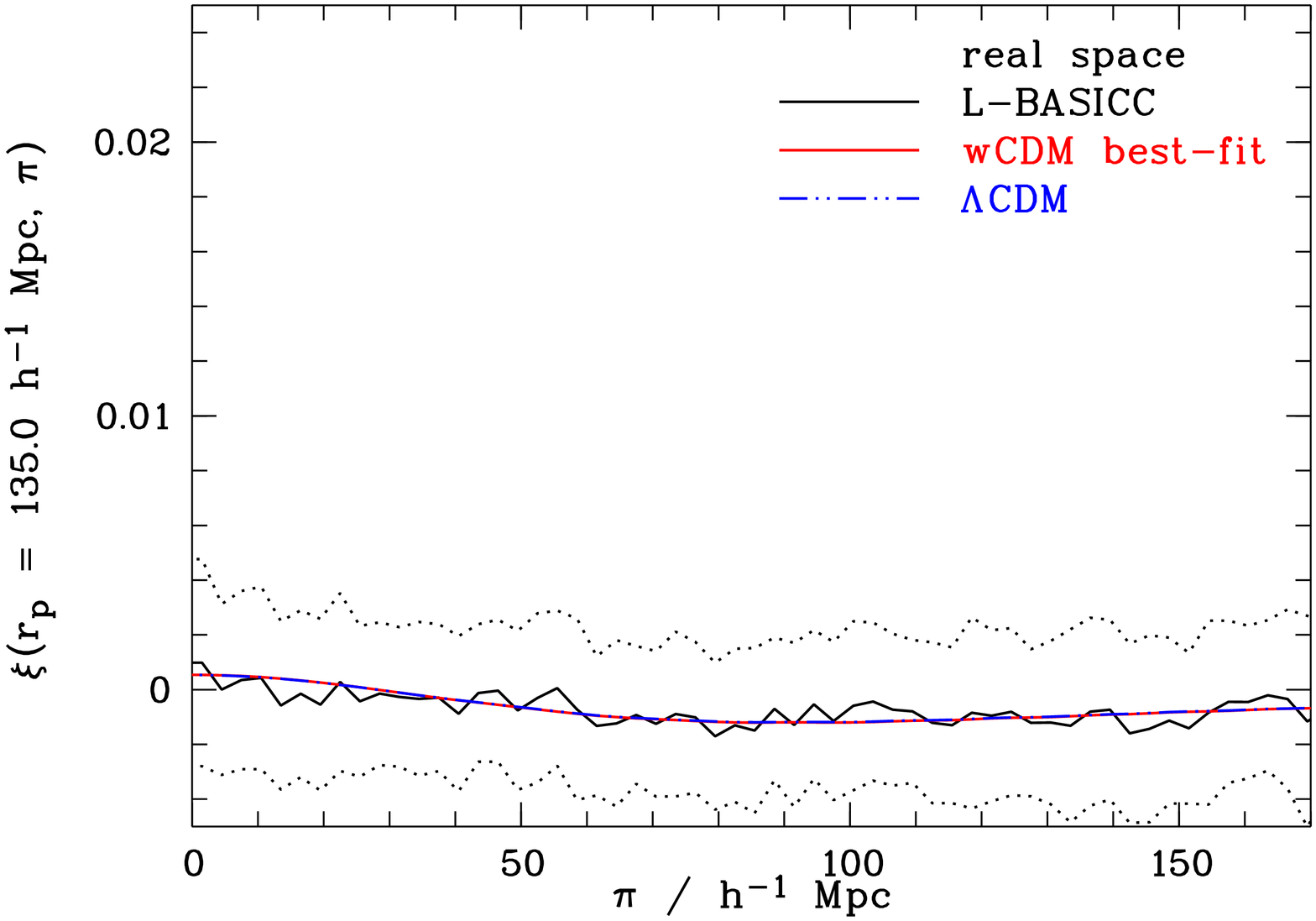,angle=270,clip=t,width=6.cm}
\psfig{figure=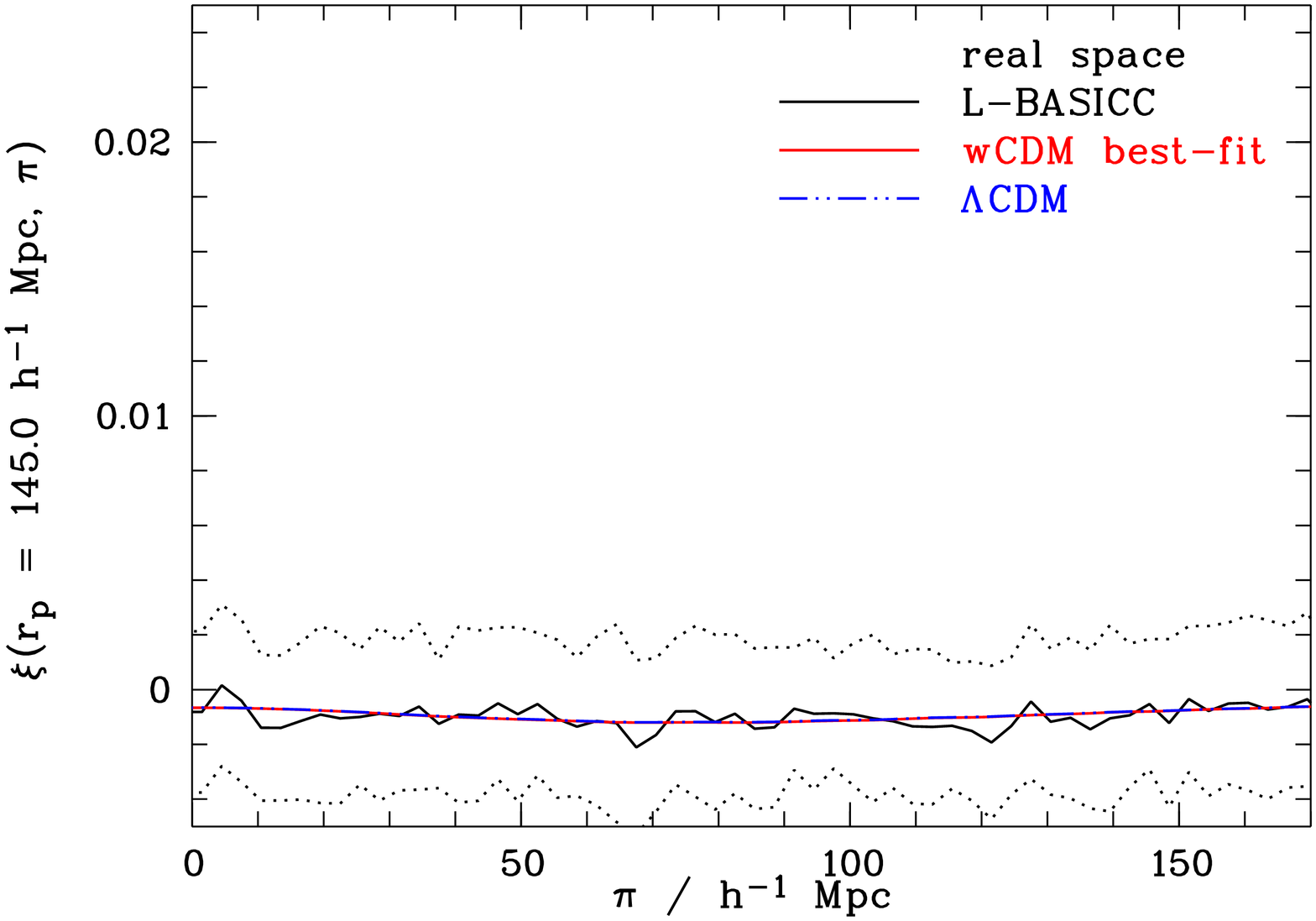,angle=270,clip=t,width=6.cm}}
\caption[ ]{Cuts through the real space correlation function $\xi_{rs}(r_p,\pi)$ of the L-BASICC II dark matter haloes along fixed $r_p$, black solid lines: mean, dotted lines: $1\sigma$-deviation calculated from the variance of the 50 boxes, red solid line: best-fitting wCDM model, blue dot-dot-dashed line: $\Lambda$CDM case.}\label{xirppicuts_rs}
\end{figure*}

In Figure \ref{xi_rp_pi_rs_zs} the contours of the best fitting model (black solid line) are plotted over the $\xi_{rs}(r_p,\pi)$ of the L-BASICC II dark matter haloes the values of $\xi_{rs}(r_p,\pi)$ are color coded in a logarithmic fashion; only the scales taken into account in the fit are shown.  Figure \ref{xirppicuts_rs} shows cuts through the $\xi_{rs}(r_p,\pi)$ plane along constant $r_p$ for both data and best-fitting model (including the $1\sigma$ uncertainty limits calculated from the variance of the correlation functions in the single L-BASICC II boxes). The deviations between the model and the L-BASICC II Simulations are small compared to the errors. The obtained value of $w_{{\mathrm DE}}$ is in good agreement with the fiducial value of the simulation.

\subsection{Redshift Space}
\label{redshift space measurement}

In redshift space the exact positions of the galaxies (and therefore the correlation function) are distorted due to the additional Doppler shift induced by their peculiar velocities, and thus mainly affects the line-of-sight components of $\xi(r_p,\pi)$. On large (BAO) scales coherent infall dominates, which in previous models of the anisotropic correlation function was assumed to be linear and modeled following the desciption in \citet{1987MNRAS.227....1K}. In this work the non-linear Kaiser effect \citep{2004PhRvD..70h3007S} is applied to the model of $\xi_{zs}(r_p, \pi)$. The validity of the approach of \citet{2004PhRvD..70h3007S} has been tested by \citet{Jennings11}, who found a good match to simulations. Since the size and angular dependence of the effect depends on the bias of the objects (the bias is also contained in the quadrupole and hexadecapole needed to evaluate the model $\xi_{zs}(r_p,\pi)$, see Section \ref{model}), in redshift space it is possible to infer the value of $b$ from the shape of the correlation function alone, in contrast to real space. The results of the MCMC analysis (again fitted in the range $64.0 \leq \sqrt{r_p^2 +\pi^2} \leq 165.0 h^{-1}$\,Mpc) are summarized in Table \ref{table_zs}, the comparison of the best fitting model (black solid line) with the logarithmically color coded L-BASICC II $\xi_{zs}(r_p,\pi)$ is shown in Figure \ref{xi_rp_pi_rs_zs}.

\begin{table}
\begin{center}
\begin{tabular}{c|c|c}
&$w_{{\mathrm DE}}$  & $b$\\
\hline
\hline
shape only&$-1.012 \pm 0.139$ & $2.518\pm 0.646$\\
\hline
shape and amplitude&$-1.020 \pm 0.147$ &$ 2.633\pm 0.222$\\
\end{tabular}
\end{center}
\caption{$w_{{\mathrm DE}}$ and $b$ from the model $\xi_{zs}(r_p, \pi)$: Mean and variance of the 50 L-BASICCS II boxes, fit in the range $64.0 \leq \sqrt{r_p^2 +\pi^2} \leq 165.0 h^{-1}$\,Mpc. The fit has been carried out for two cases, one where only the shape was input to the fit and the amplitude was marginalized over, and one where both shape and amplitude have been taken into account.}
\label{table_zs}
\end{table}

\begin{figure*}
\centerline{\psfig{figure=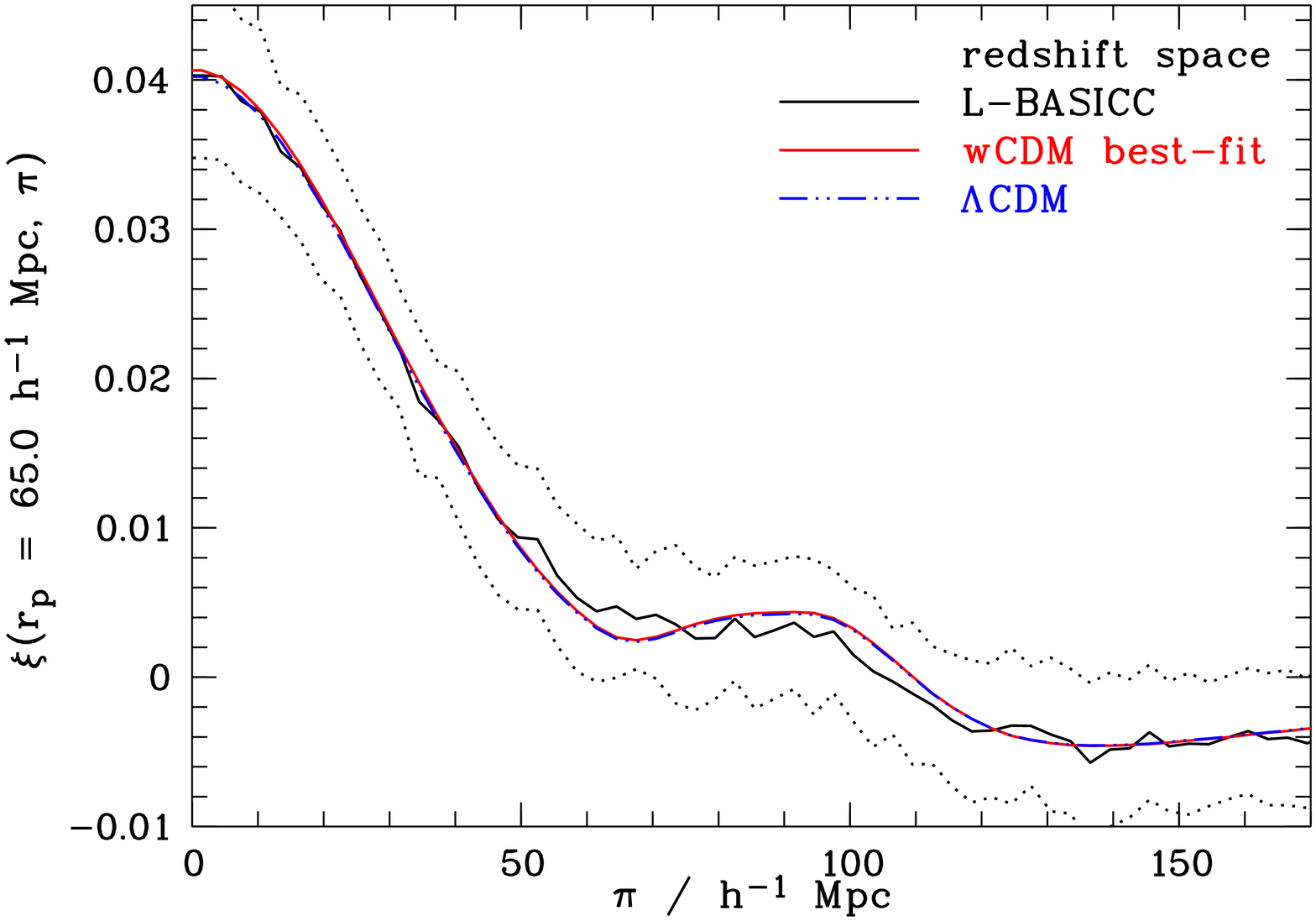,angle=270,clip=t,width=6.cm}
\psfig{figure=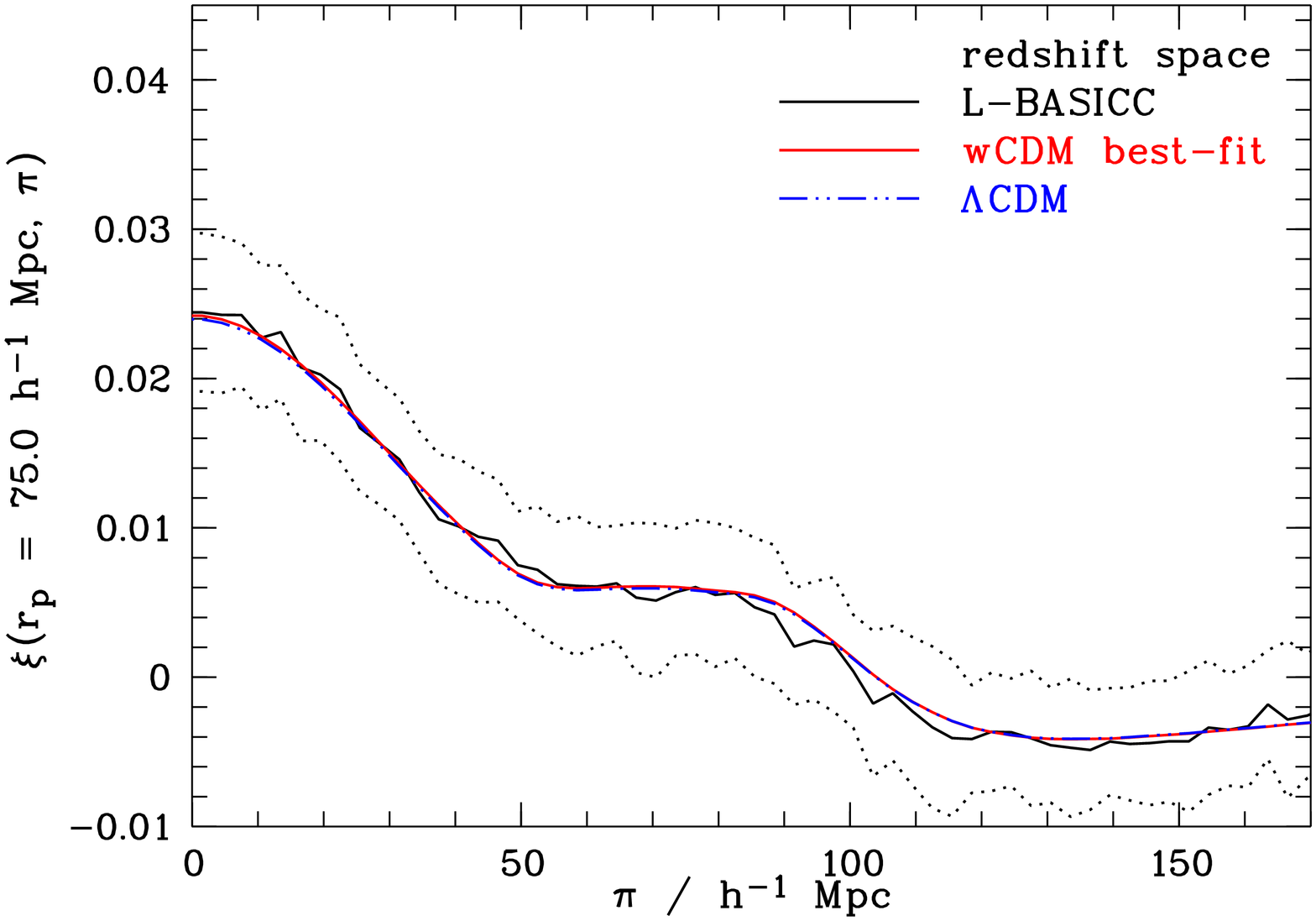,angle=270,clip=t,width=6.cm}
\psfig{figure=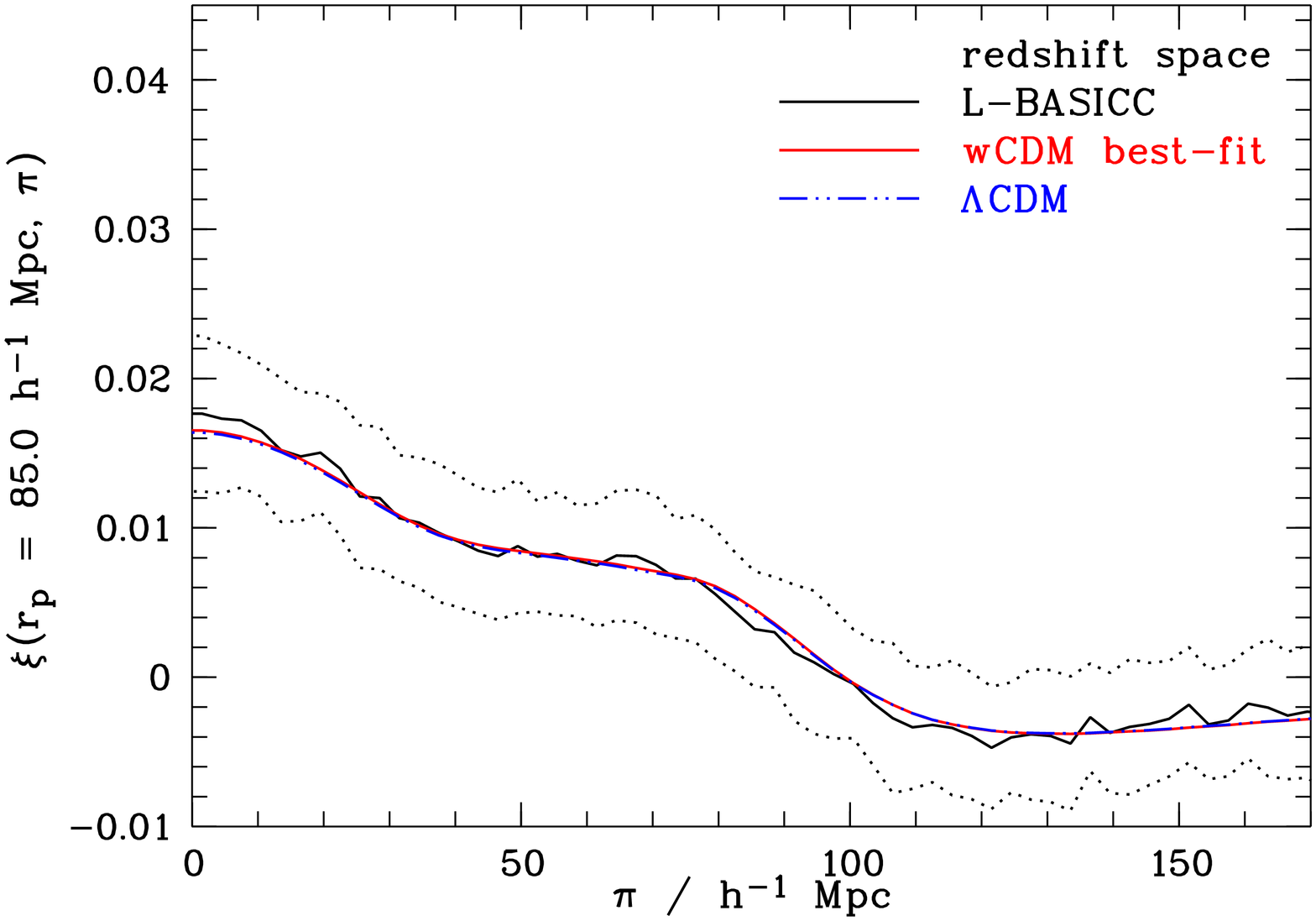,angle=270,clip=t,width=6.cm}}
\centerline{\psfig{figure=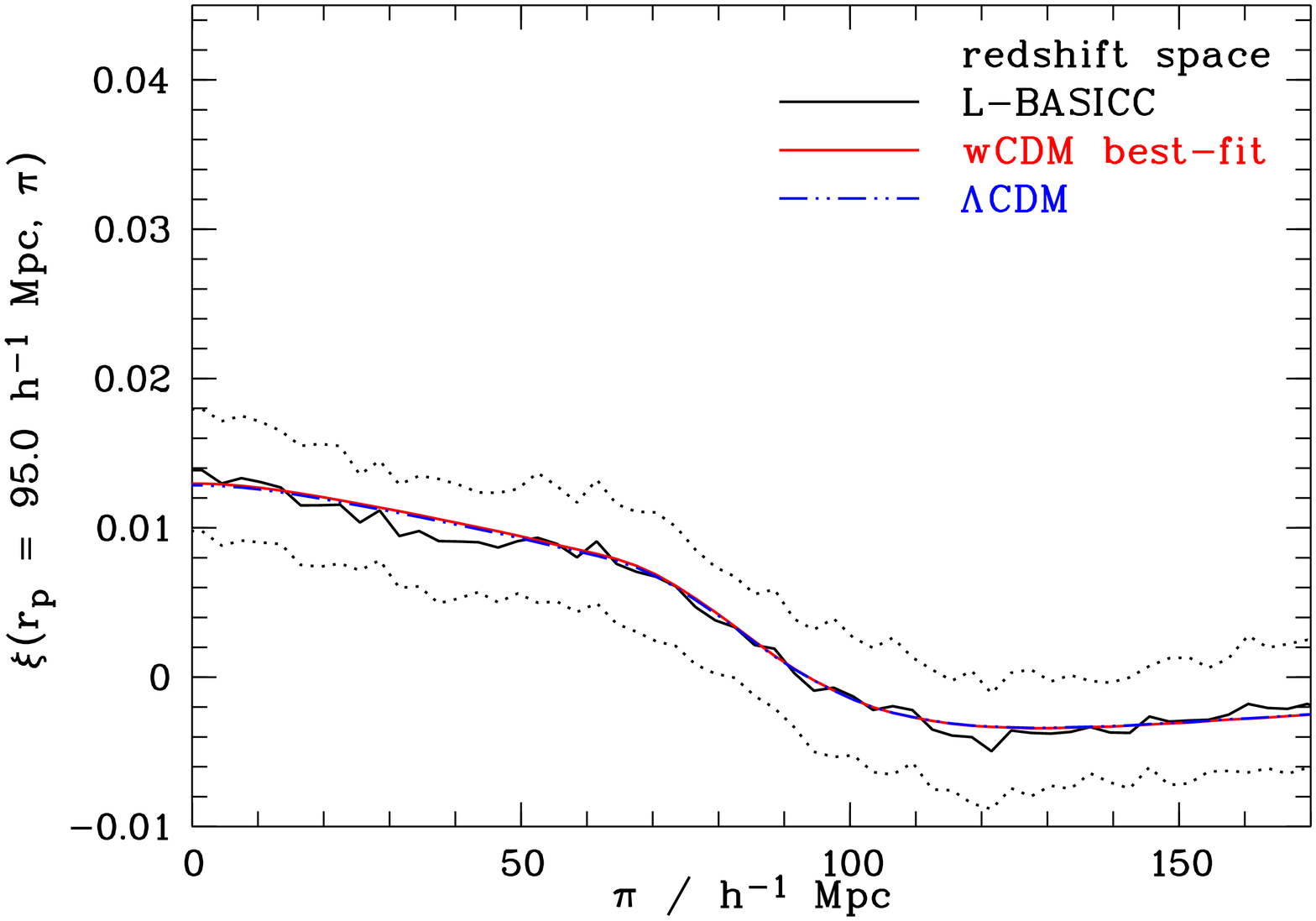,angle=270,clip=t,width=6.cm}
\psfig{figure=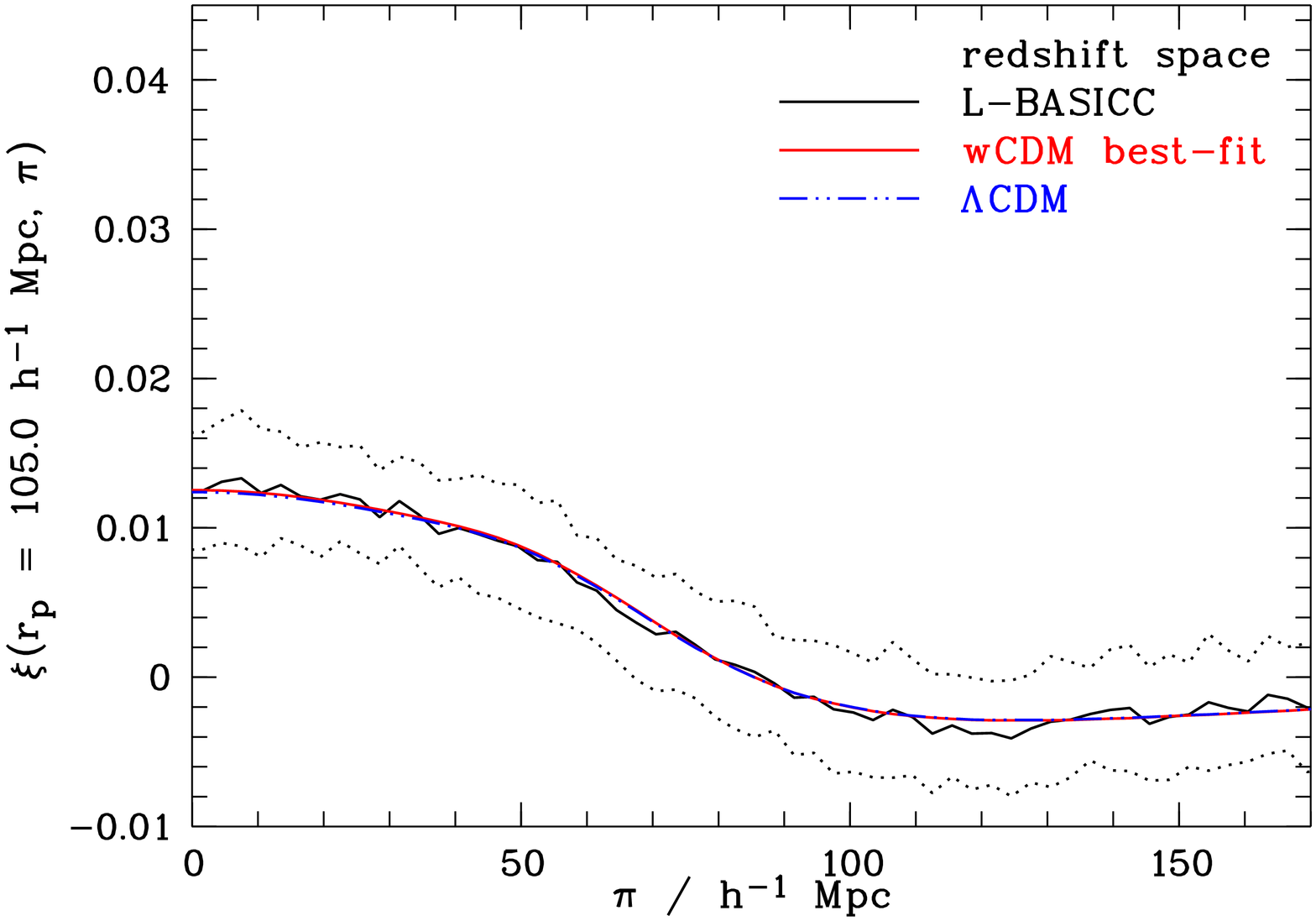,angle=270,clip=t,width=6.cm}
\psfig{figure=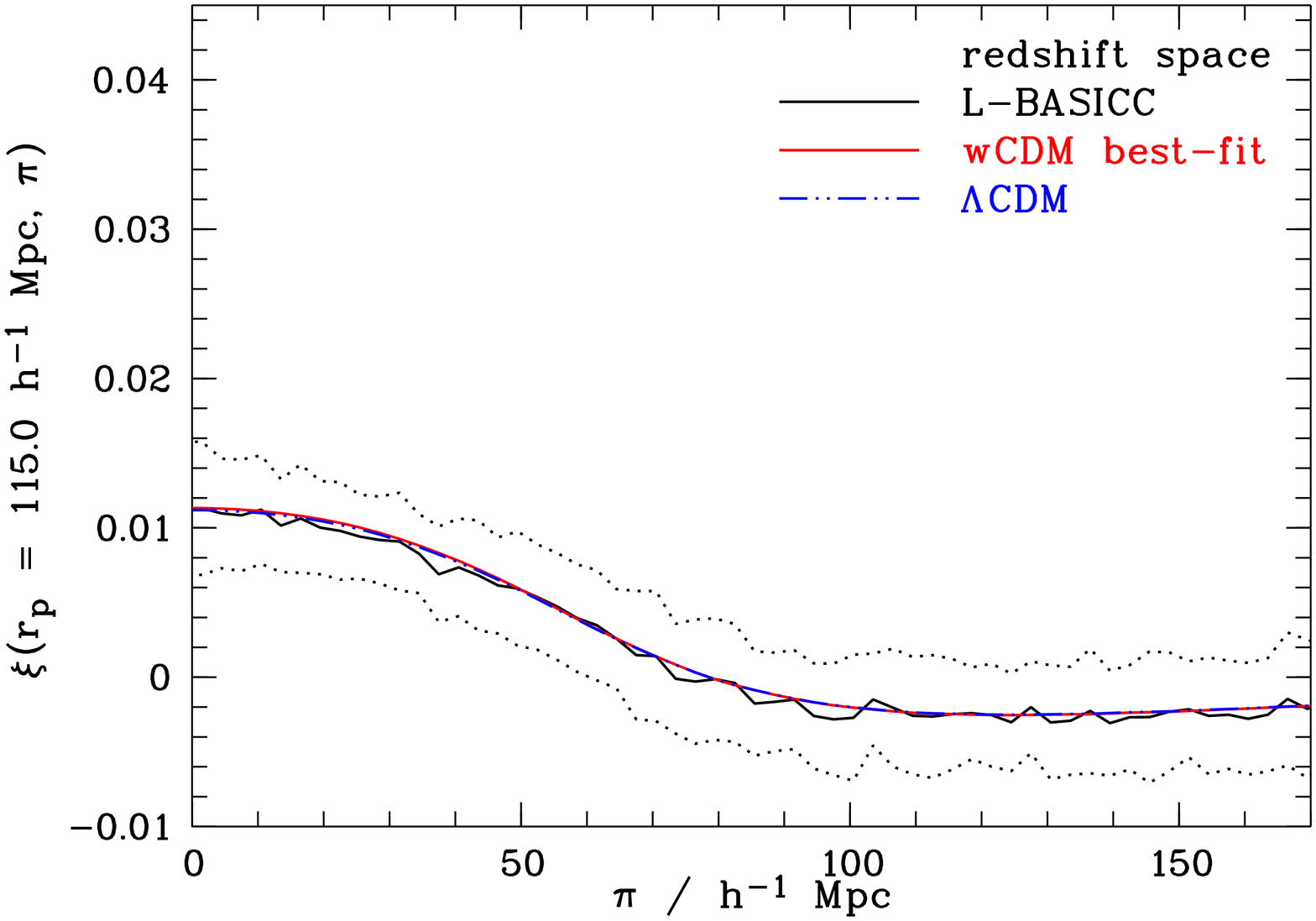,angle=270,clip=t,width=6.cm}}
\centerline{\psfig{figure=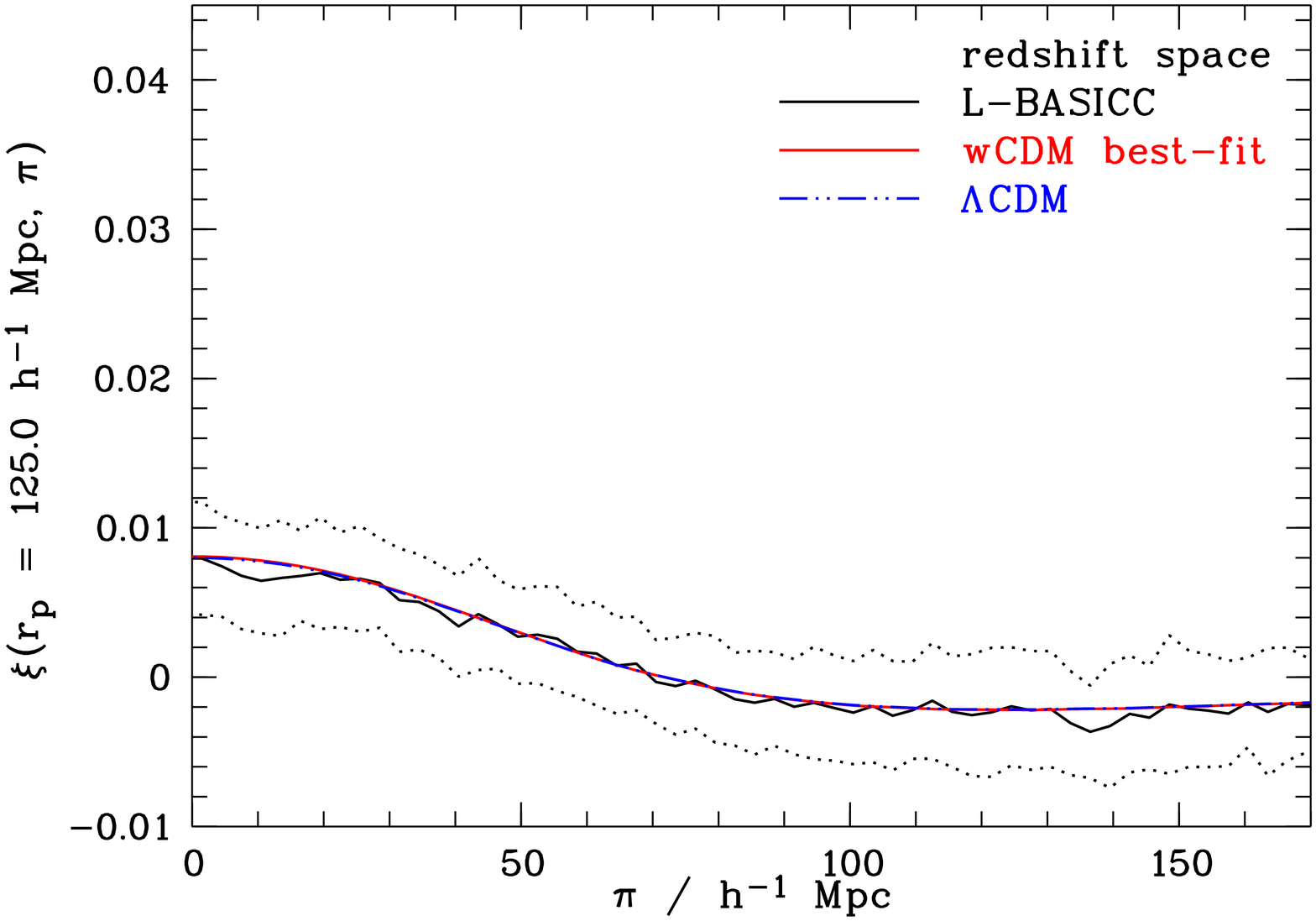,angle=270,clip=t,width=6.cm}
\psfig{figure=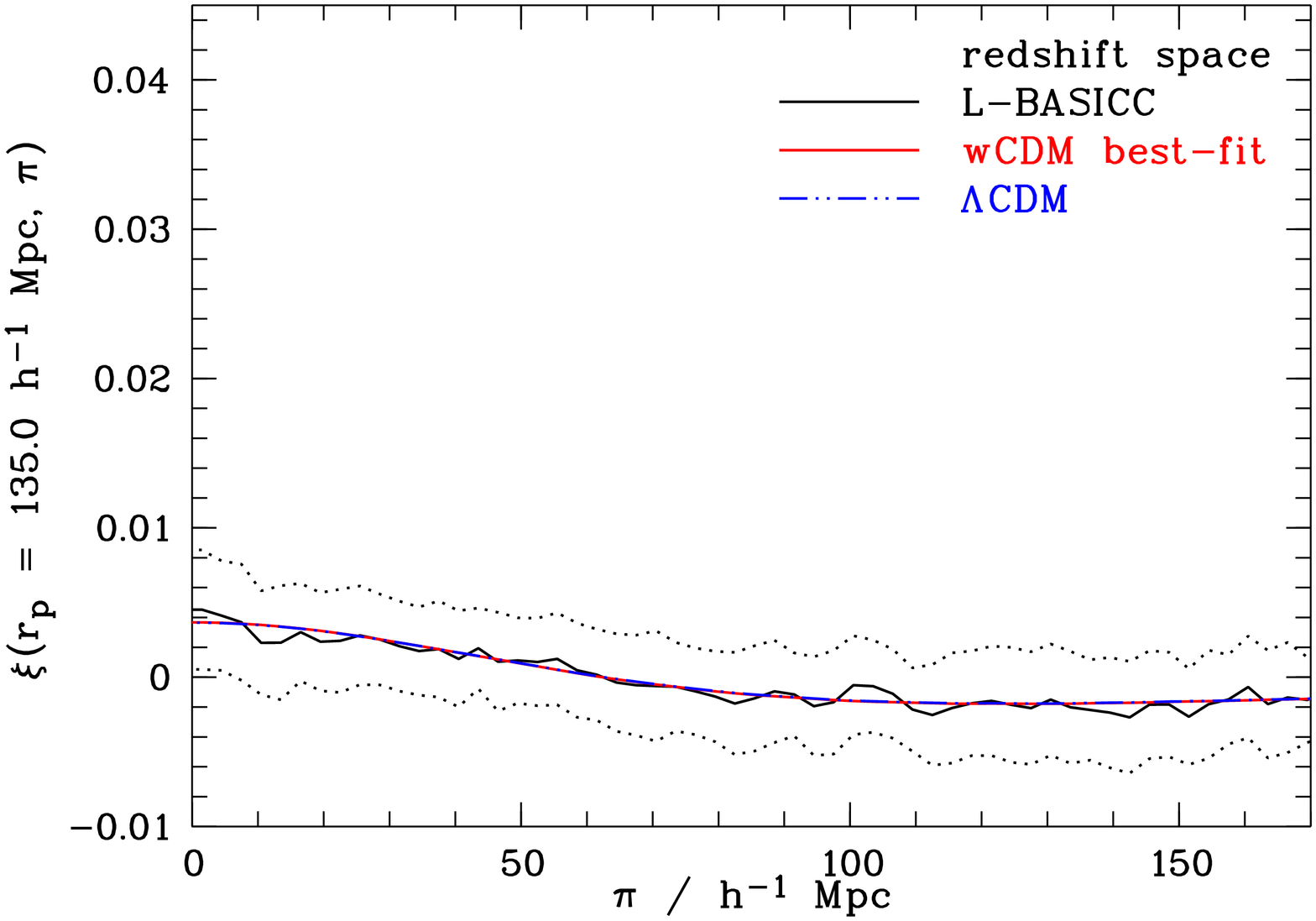,angle=270,clip=t,width=6.cm}
\psfig{figure=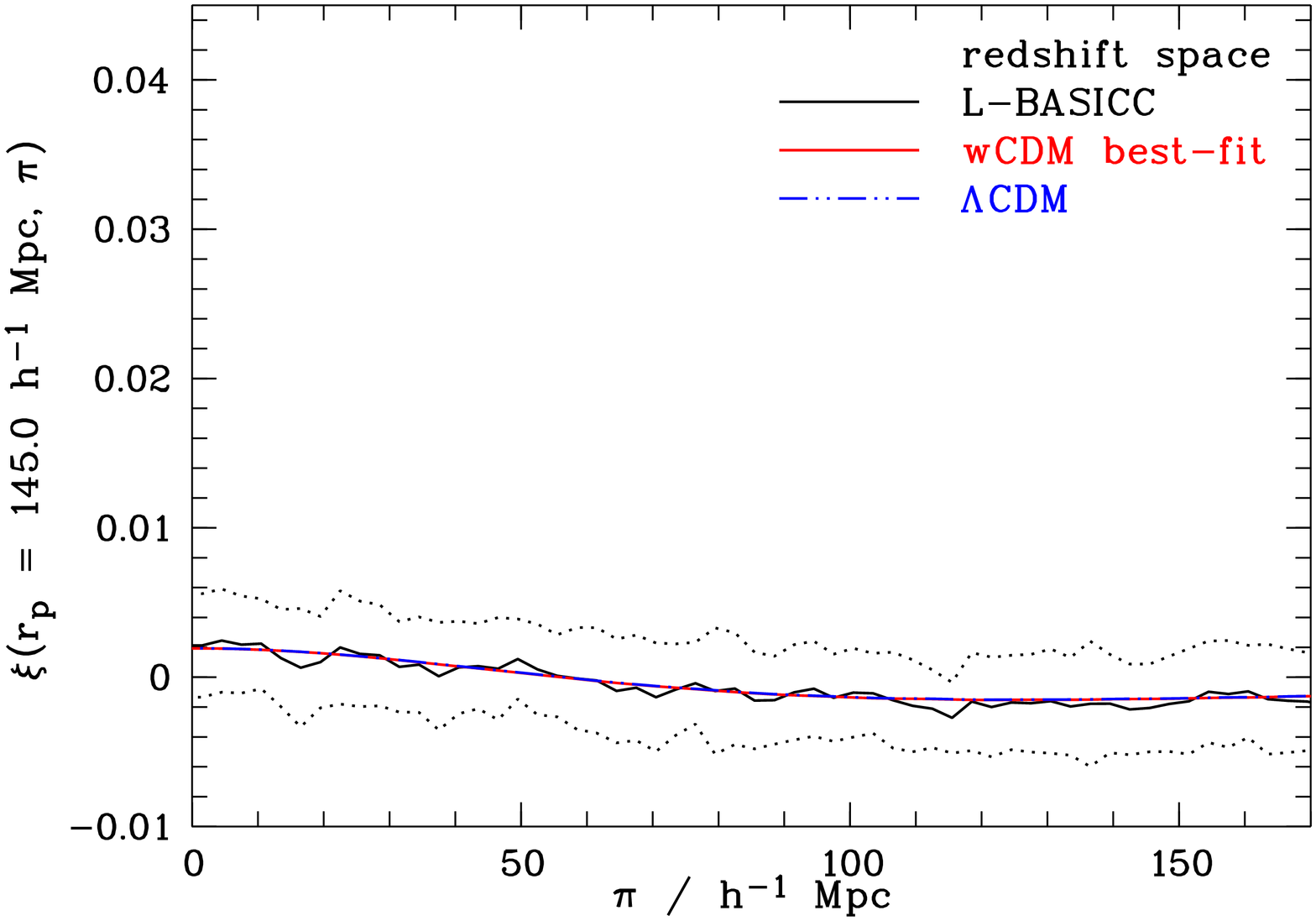,angle=270,clip=t,width=6.cm}}
\caption[ ]{Cuts through the redshift space correlation function $\xi_{zs}(r_p,\pi)$ of the L-BASICC II dark matter haloes along fixed $r_p$, black solid lines: mean, dotted lines: $1\sigma$-deviation calculated from the variance of the 50 boxes, red solid line: best-fitting wCDM model, blue dot-dot-dashed line: $\Lambda$CDM case.}\label{xirppicuts_zs}
\end{figure*}

The correlation function of the L-BASICCS II halo catalogue can be well described by the model. The model contours match the L-BASICC II $\xi_{zs}(r_p, \pi)$ almost perfectly (see Figure \ref{xirppicuts_zs}, and the resulting values of $w_{{\mathrm DE}}$ and $b$ are in good agreement with the real space estimates, too. As expected the errors are larger in redshift space than in real space. Also the error on $b$ is larger if the information contained in the amplitude is ignored, the determined values of  $w_{{\mathrm DE}}$ and $b$ are however consistent and do not depend on whether the amplitude is taken into account or not.

\subsection{Redshift Error Space}
\label{redshift error measurement}

One of the advantages of using the anisotropic correlation function $\xi(r_p,\pi)$ (or its Fourier space equivalent $P(k_p, k_{\parallel})$) to infer cosmological parameters is that in the presence of redshift space distortions the clustering measurement perpendicular to the line-of-sight remains almost unaffected, while distortions along the line-of-sight can be modeled and thus properly taken into account. This makes it a perfect tool to use in the case of photometric redshifts, the large errors of which lead to a rather dramatic distortion, as explained in Section \ref{model}. In order to investigate the effect of photometric redshift errors on the estimate of $w_{{\mathrm DE}}$ and $b$, we simulated the influence of a gaussian redshift error distribution with a rms of $\sigma_z= 0.015$, $0.03$, $0.06$, and $0.12$, respectively, on the measurement.

\begin{figure*}
\centerline{\psfig{figure=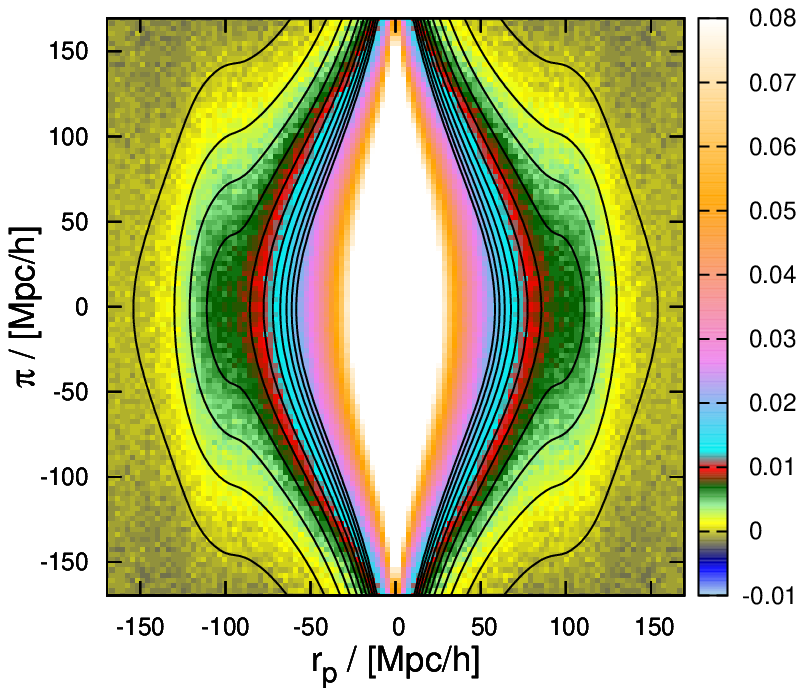,angle=0,clip=t,width=12.cm}
\psfig{figure=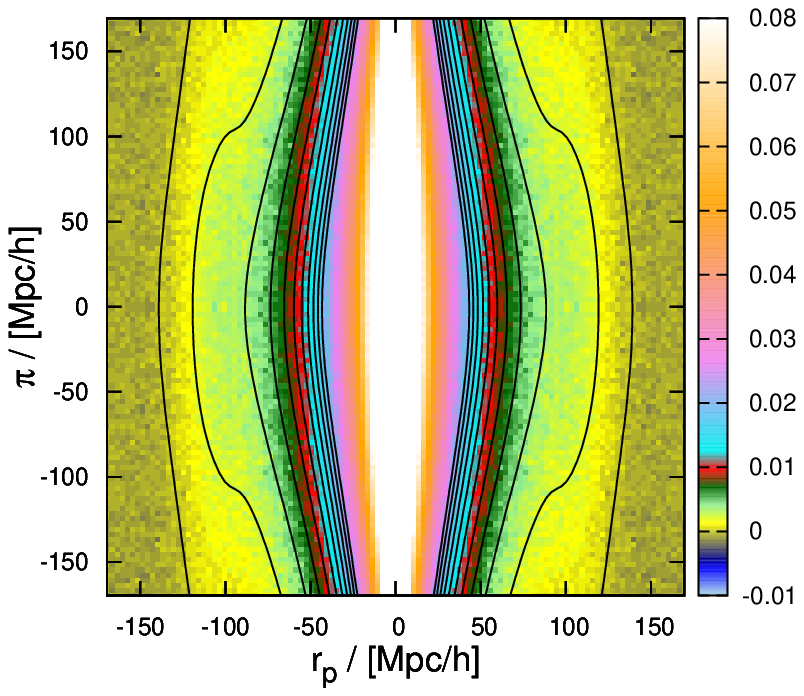,angle=0,clip=t,width=12.cm}}
\centerline{\psfig{figure=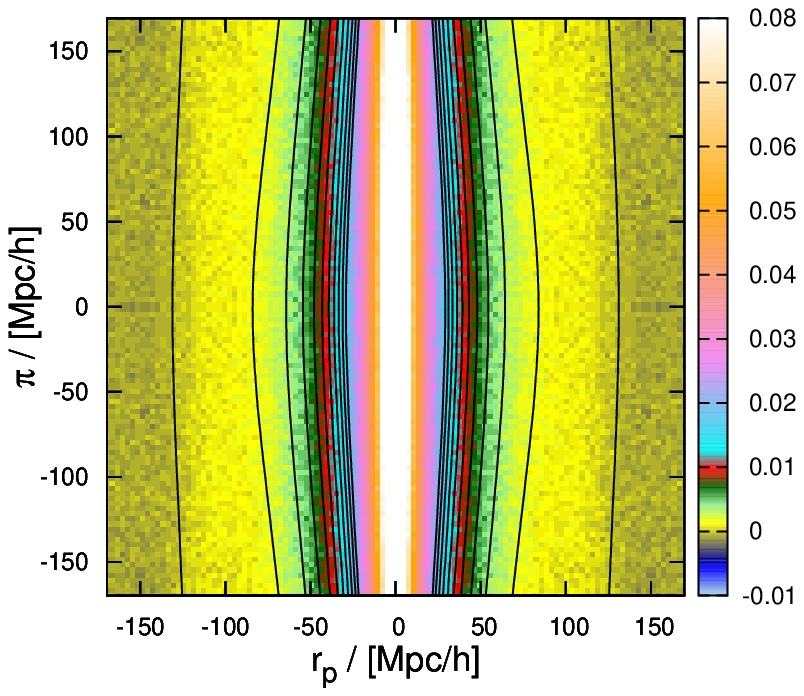,angle=0,clip=t,width=12.cm}
\psfig{figure=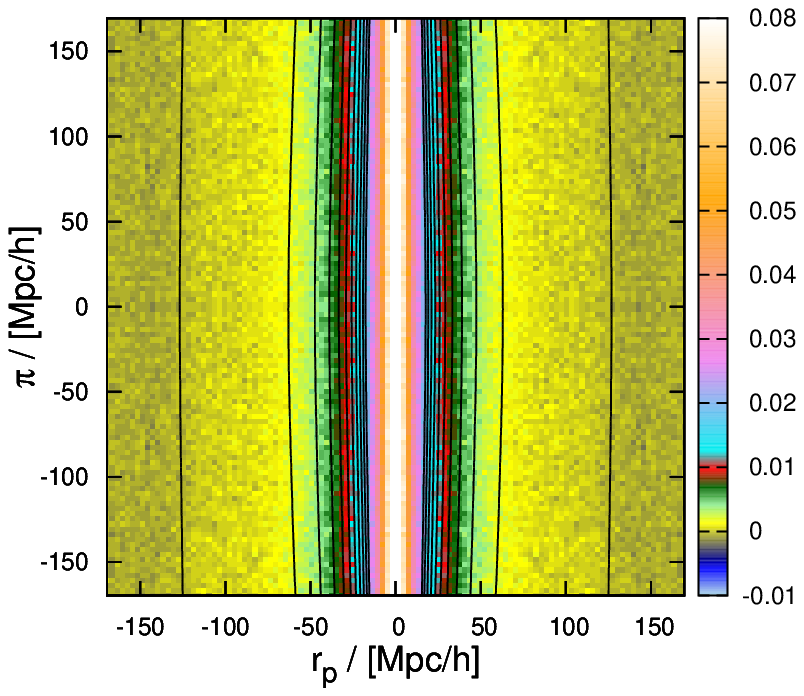,angle=0,clip=t,width=12.cm}}
\caption[ ]{L-BASICC II $\xi_{zerr}(r_p, \pi)$ and contours from the model $\xi_{zerr}(r_p, \pi)$ for $\sigma_z= 0.015$ (top left), $\sigma_z= 0.03$ ( top right) $\sigma_z = 0.06$ (bottom left) and $\sigma_z = 0.12$ (bottom right).}\label{compare_xi_zerr}
\end{figure*}

As described in Section \ref{analysis} the spherical shell in which the fit was carried out has now been replaced by a cylinder (indicated by the rectangular box in the $r_p,\pi$-plane in Figure (\ref{redshifterrorpicture})) in order to compare (to zeroth order) the same information. We found that in the last case a large part of the clustering signal is smeared out to distances much larger than the $300 h^{-1}$\,Mpc we are calulating our model for, the BAO ring disappears, and the noise increases such that an accurate estimate of $w_{{\mathrm DE}}$ and $b$ becomes impossible. While still fitting $\xi(r_p,\pi)$ only up to $165\,h^{-1}$\,Mpc, extending the model to distances $\pi=2000\,h^{-1}$\,Mpc before convolving it with the pairwise redshift error distribution allows us to recover some of the clustering signal for redshift errors (at least for $\sigma_z\leq 0.06$), but its information content is limited due to the low signal-to-noise of the data on these scales. The values of $w_{{\mathrm DE}}$  and $b$ we find for $\sigma_z=0.015$, $0.03$, $0.06$, and $0.12$ are summarized in Table \ref{table_zerr}; contours of the corresponding models of  $\xi_{zs}(r_p,\pi)$ are shown in comparison to the logarithmically color coded measurement from the data in Figure \ref{compare_xi_zerr}.

\begin{table}
\begin{center}
\begin{tabular}{c|c|c|c}
$\sigma_z$&&$w_{{\mathrm DE}}$ & $b$ \\
\hline
\hline
\multirow{2}{*}{0.015}&only shape&$-0.965\pm0.298$ & $3.660\pm2.694$ \\ 
&shape and amplitude&$-0.980 \pm0.296 $ & $2.704\pm0.333$\\
\hline
\hline
\multirow{2}{*}{0.030}&only shape &$-0.883\pm0.313 $ &$6.309 \pm3.603 $\\
&shape and amplitude&$-0.966\pm0.363 $ & $2.622\pm0.415 $\\

\hline
\hline
\multirow{2}{*}{0.060}&only shape &$-1.081\pm0.344 $ &$ 5.336\pm3.931 $\\
&shape and amplitude&$-1.036\pm0.402 $ & $2.609\pm0.512 $\\
\hline
\hline
\multirow{2}{*}{0.120}&only shape &$-1.316\pm0.348 $ &$4.904 \pm 4.537$\\
&shape and amplitude&$-1.199\pm0.424 $ & $2.295\pm0.554 $\\
\end{tabular}
\end{center}
\caption{$w_{{\mathrm DE}}$ and $b$ from the fit of the model $\xi_{zerr}(r_p,\pi)$ to the 50 L-BASICCS II boxes for $\sigma_z= 0.015$, $0.03$, $0.06$, and $0.12$ (from top to bottom). Again the fit has been carried out for two cases, one where only the shape was input to the fit and the amplitude was marginalized over, and one where both shape and amplitude have been taken into account.}
\label{table_zerr}
\end{table}

\begin{figure*}
\centerline{\psfig{figure=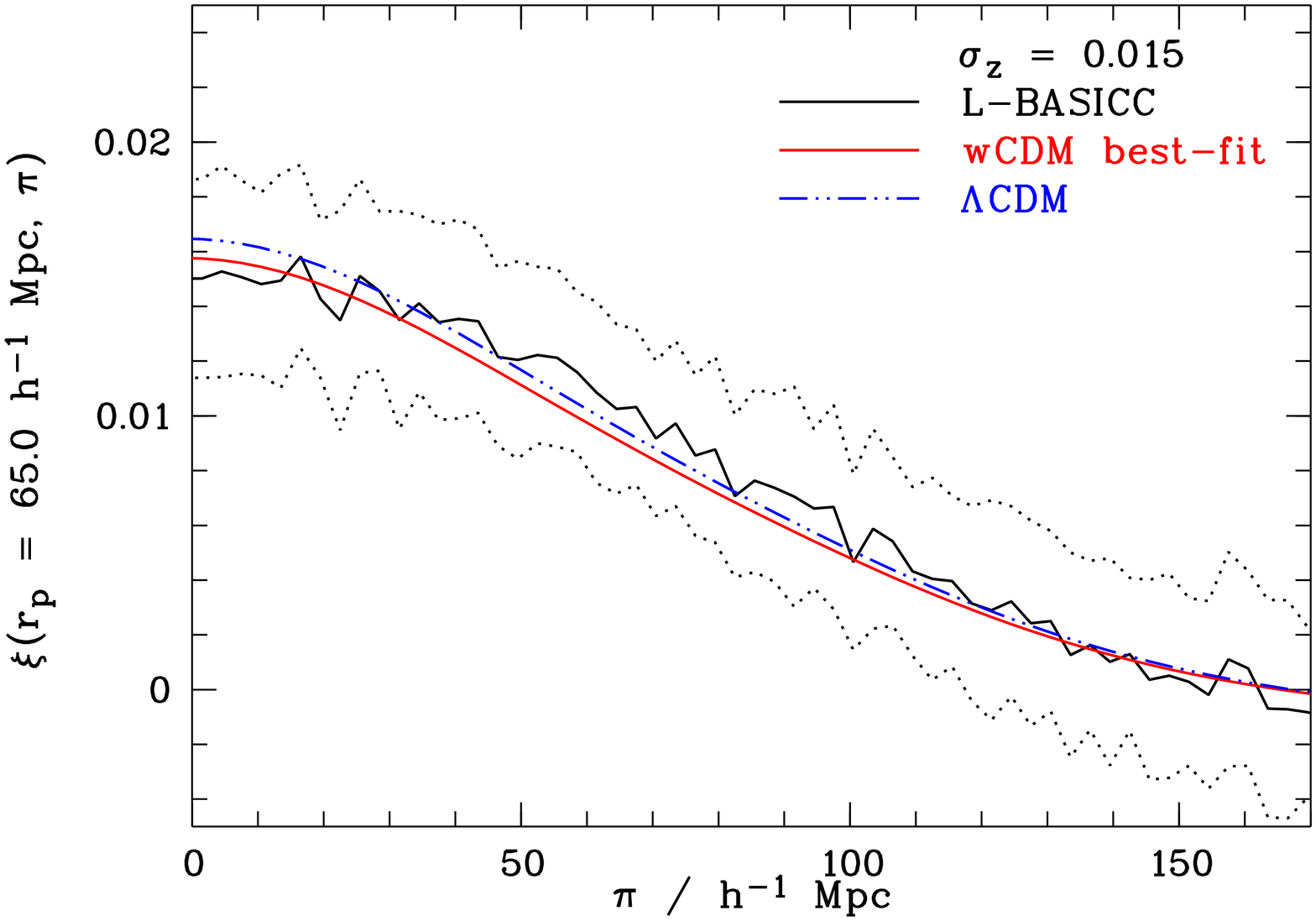,angle=270,clip=t,width=6.cm}
\psfig{figure=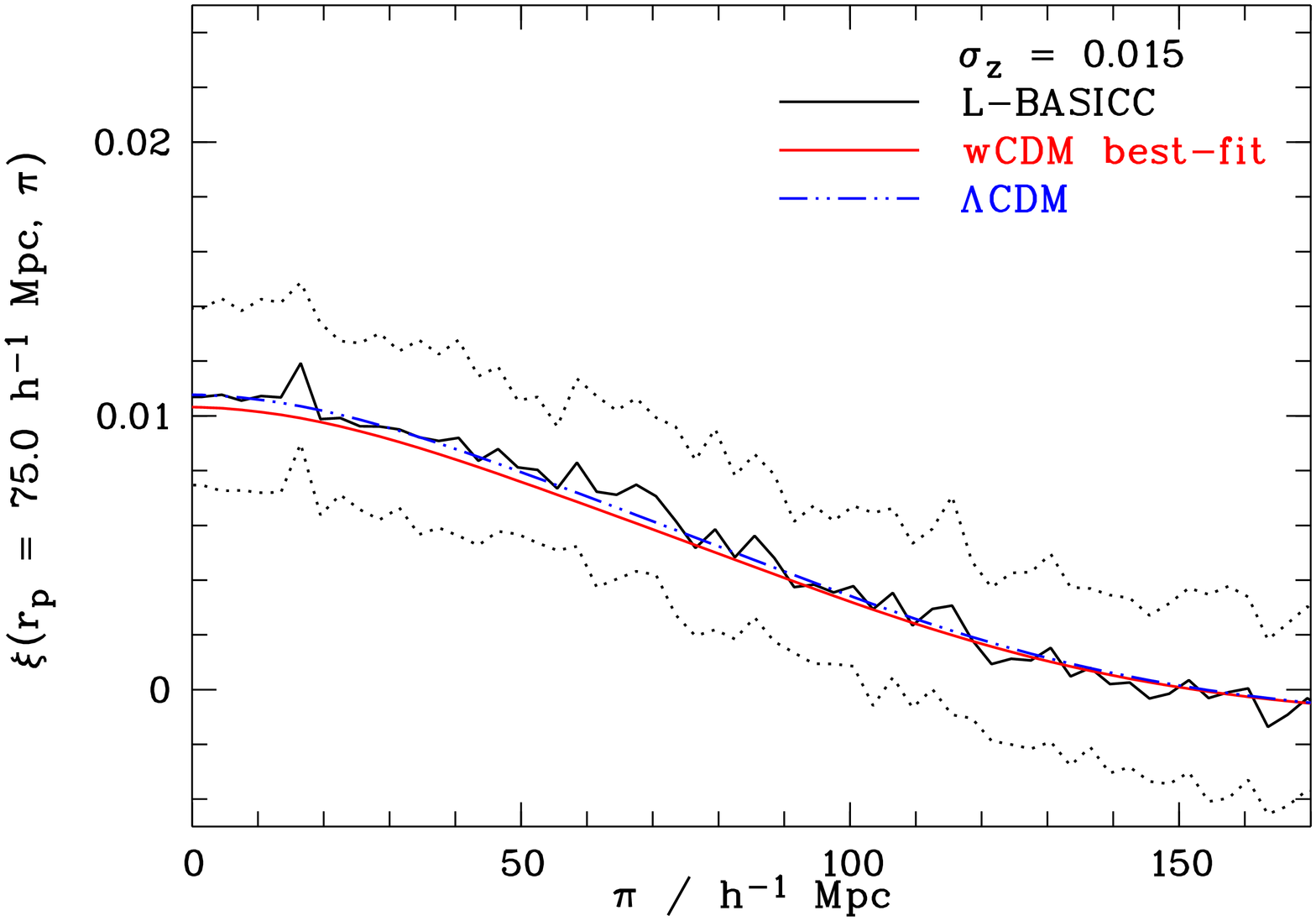,angle=270,clip=t,width=6.cm}
\psfig{figure=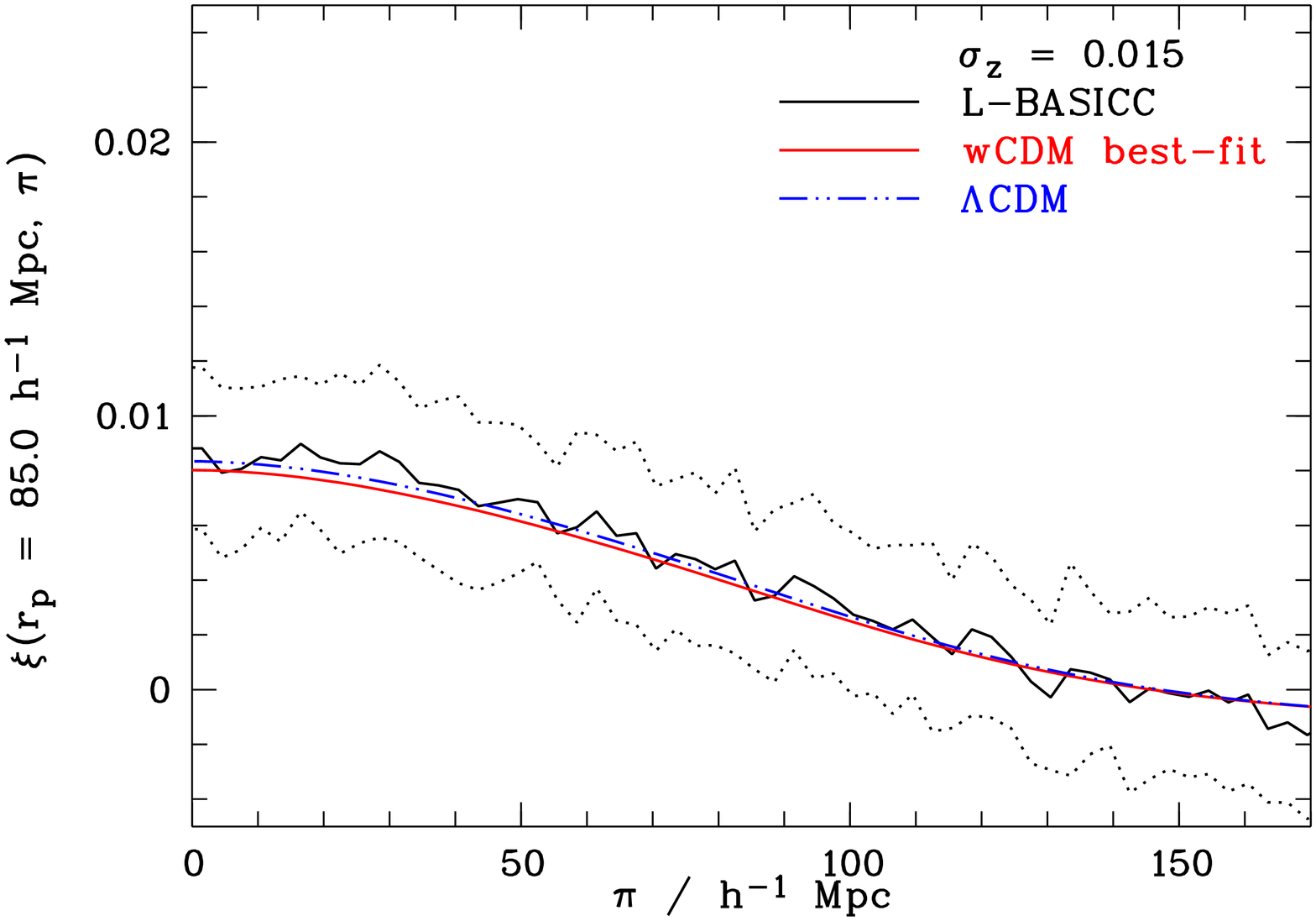,angle=270,clip=t,width=6.cm}}
\centerline{\psfig{figure=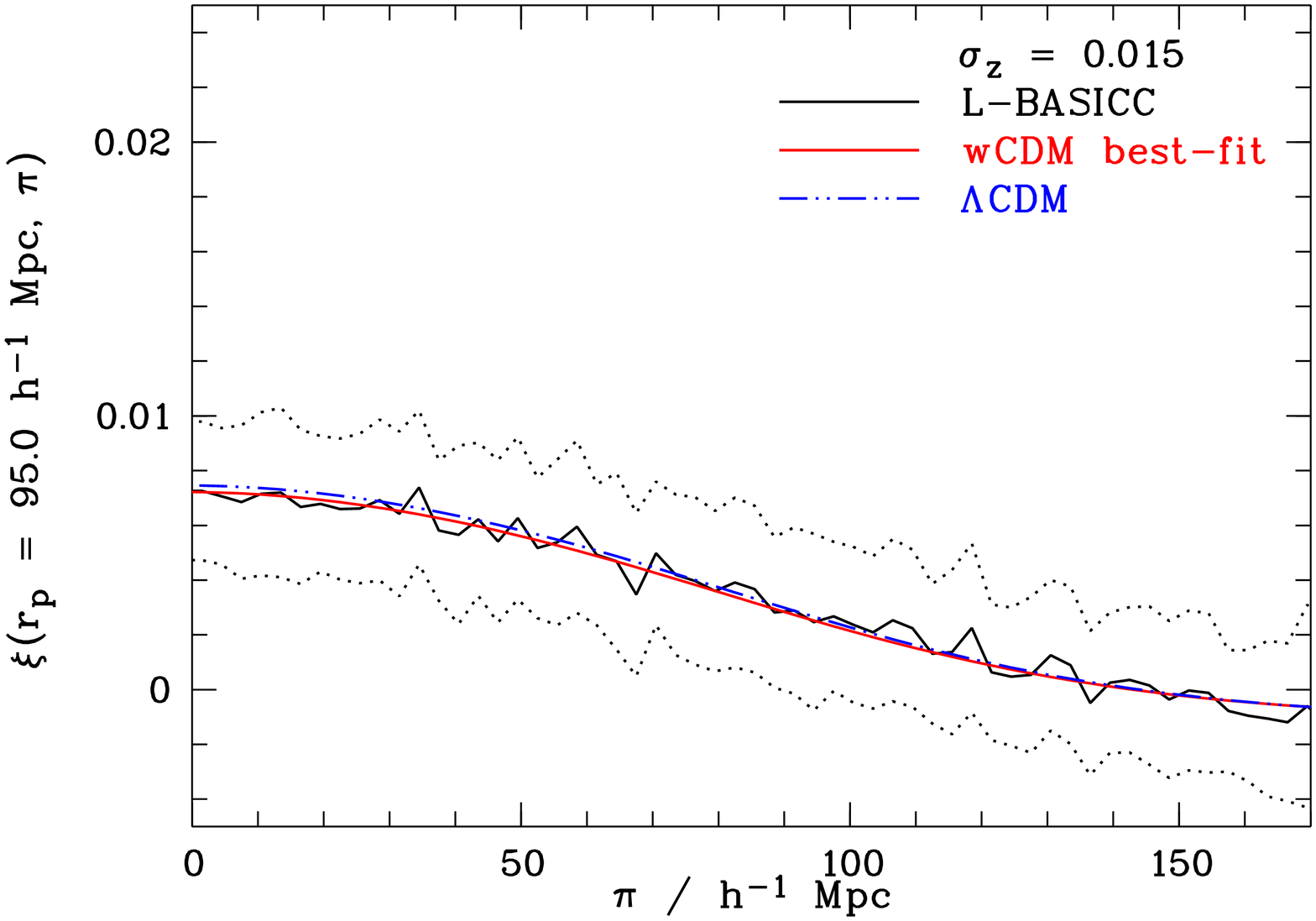,angle=270,clip=t,width=6.cm}
\psfig{figure=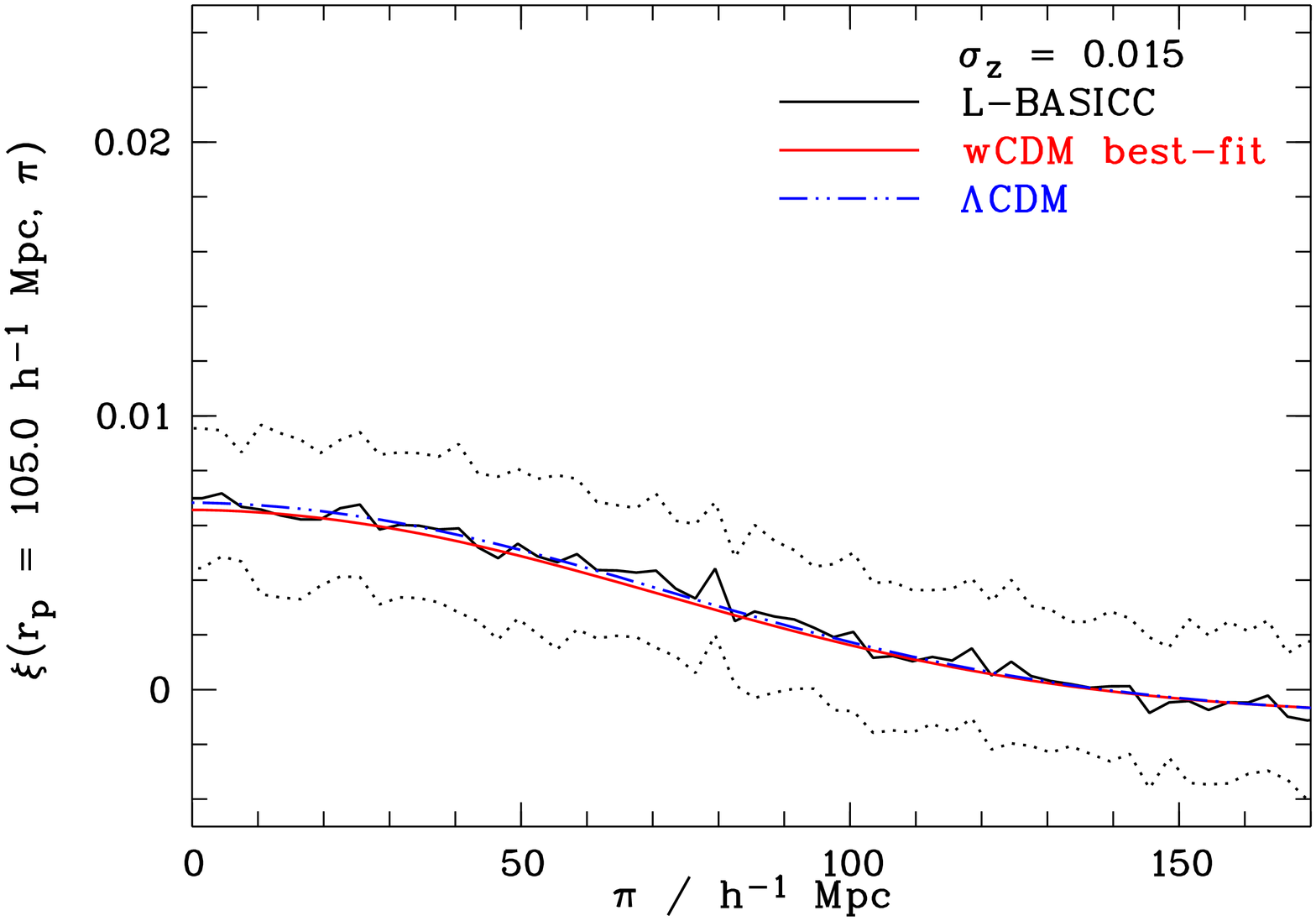,angle=270,clip=t,width=6.cm}
\psfig{figure=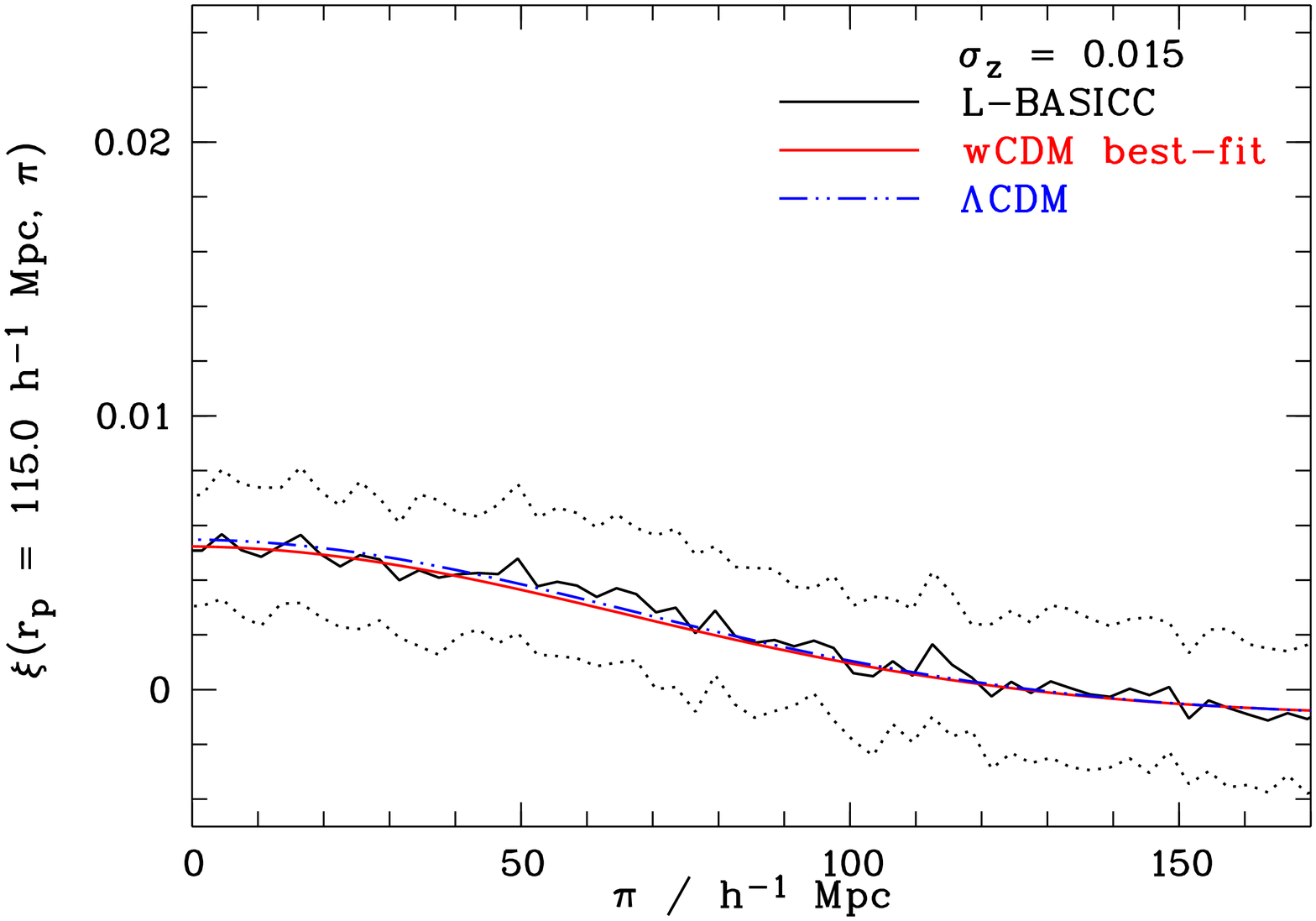,angle=270,clip=t,width=6.cm}}
\centerline{\psfig{figure=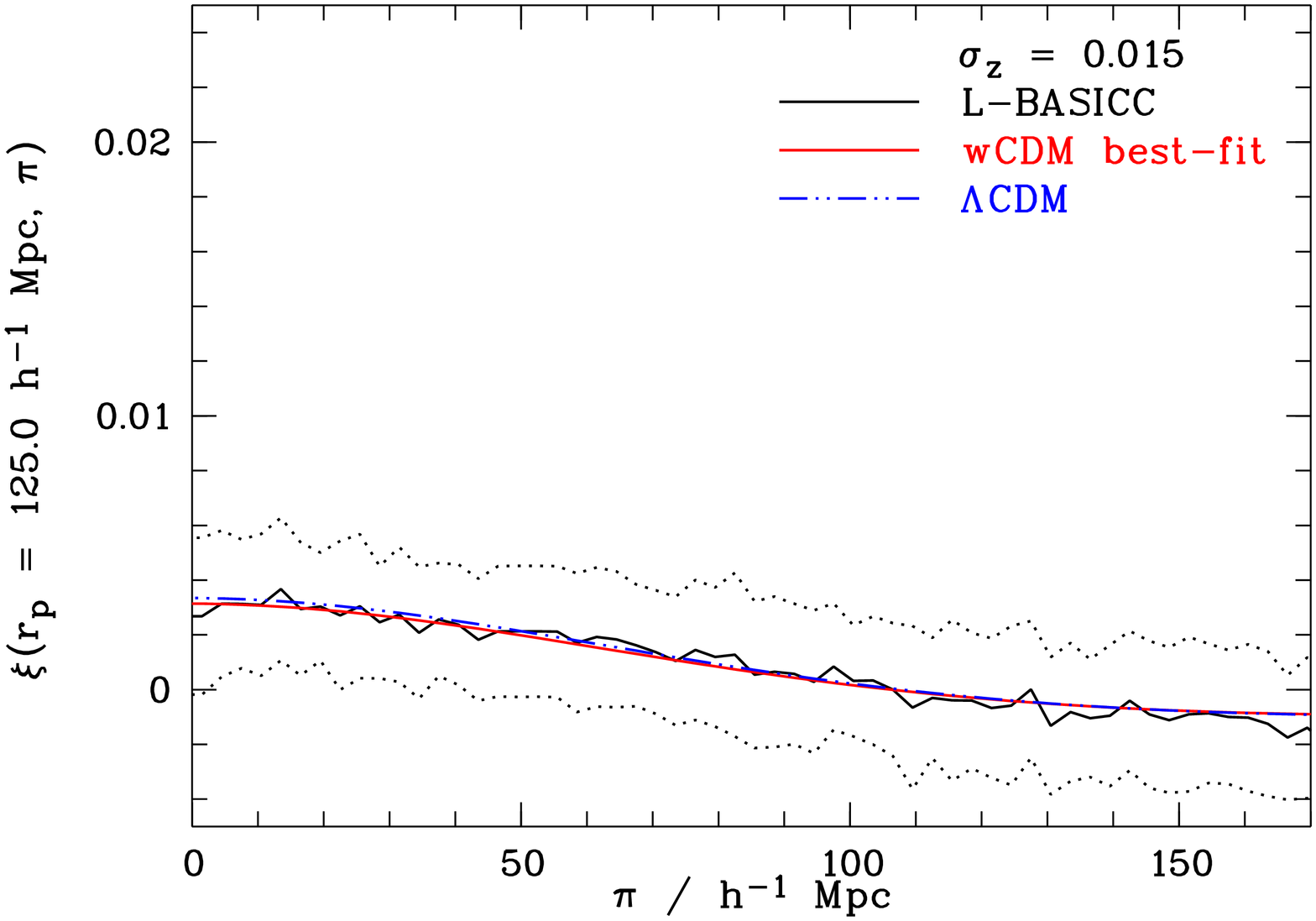,angle=270,clip=t,width=6.cm}
\psfig{figure=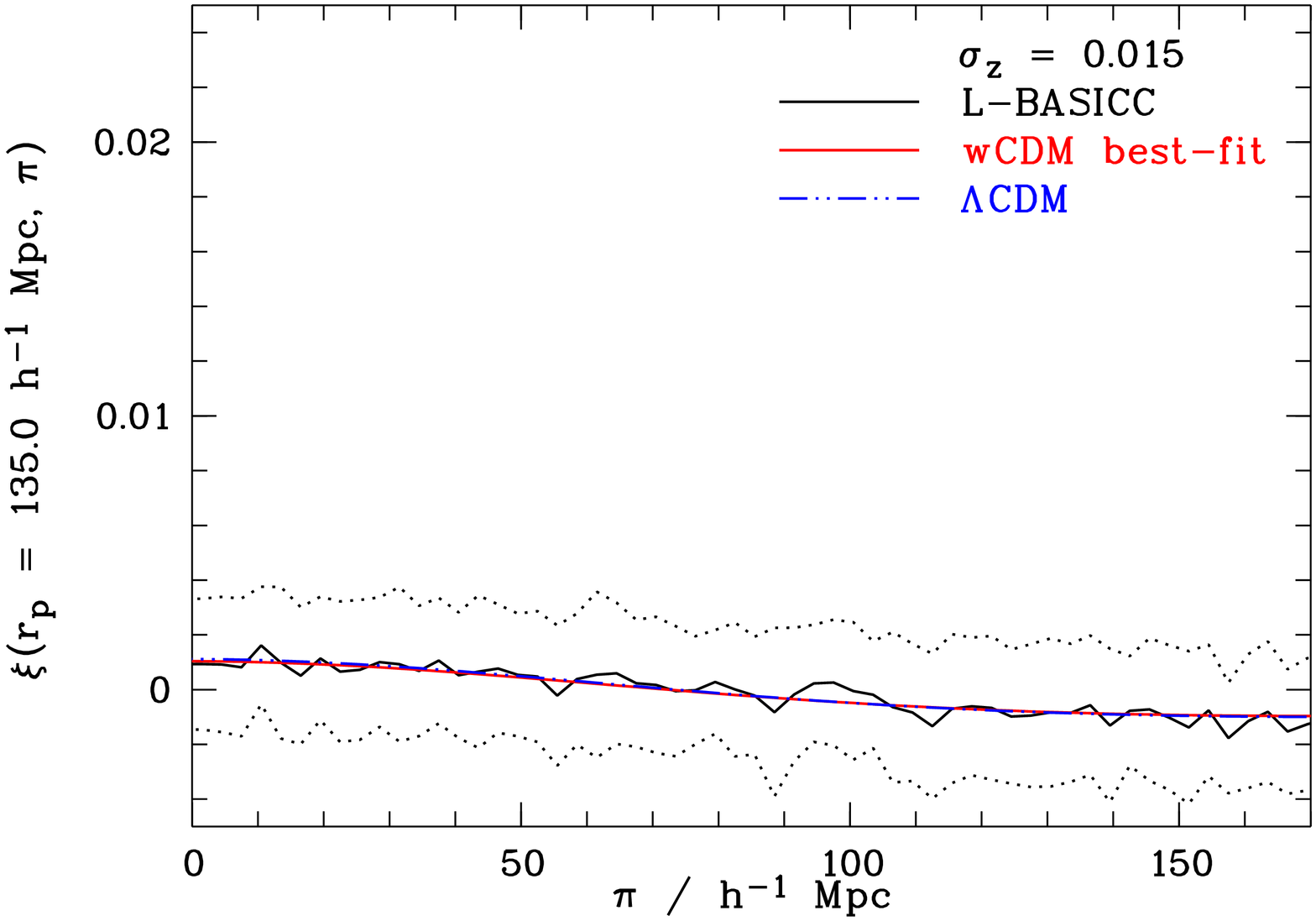,angle=270,clip=t,width=6.cm}
\psfig{figure=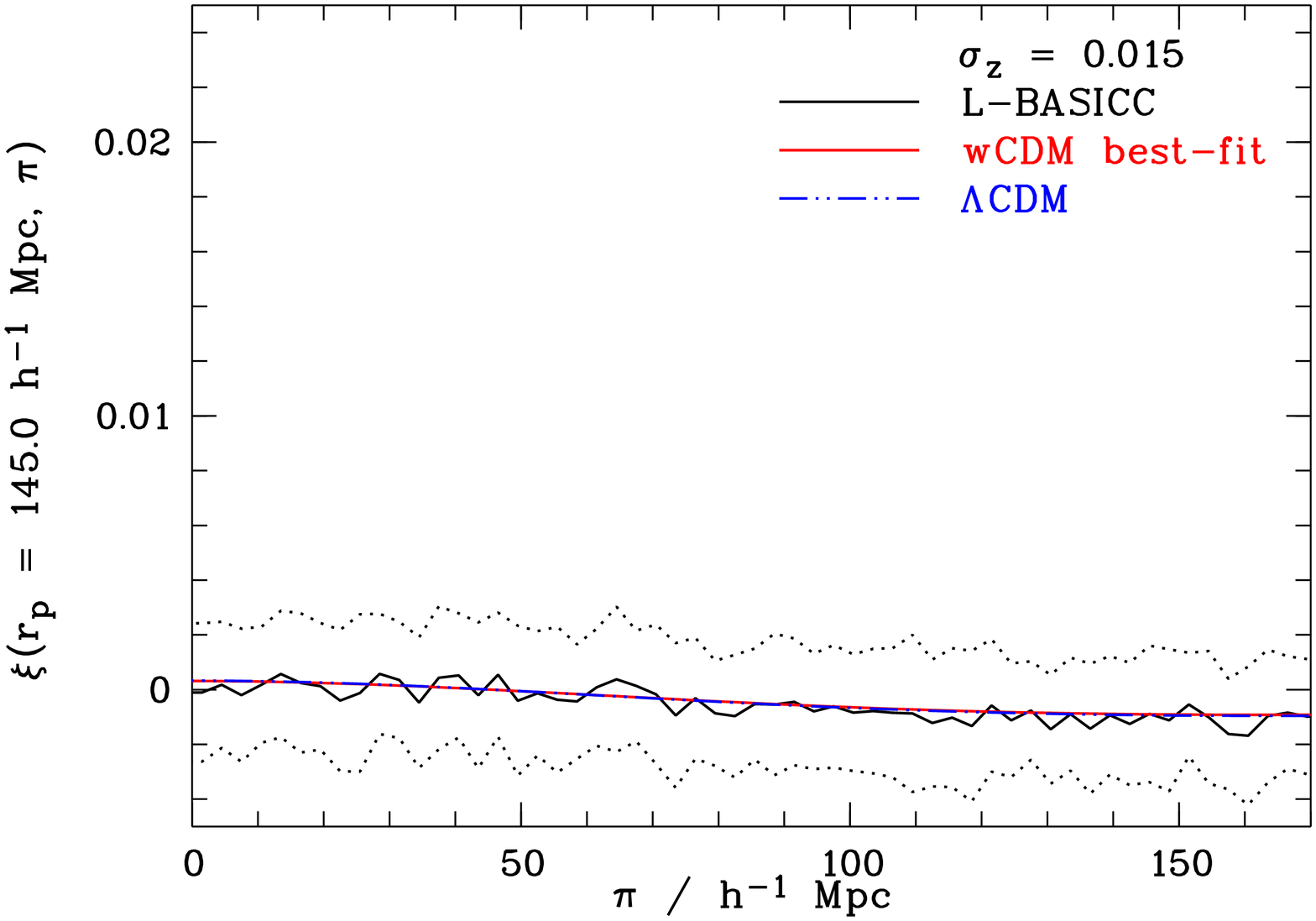,angle=270,clip=t,width=6.cm}}
\caption[ ]{Cuts through the correlation function $\xi_{zerr}(r_p,\pi)$ of the L-BASICC II dark matter haloes along fixed $r_p$, for redshift errors of $\sigma_z= 0.015$, black solid lines: mean, dotted lines: $1\sigma$-deviation calculated from the variance of the 50 boxes, red solid line: best-fitting wCDM model, blue dot-dot-dashed line: $\Lambda$CDM case.}\label{xirppicuts_zerr015}
\end{figure*}

\begin{figure*}
\centerline{\psfig{figure=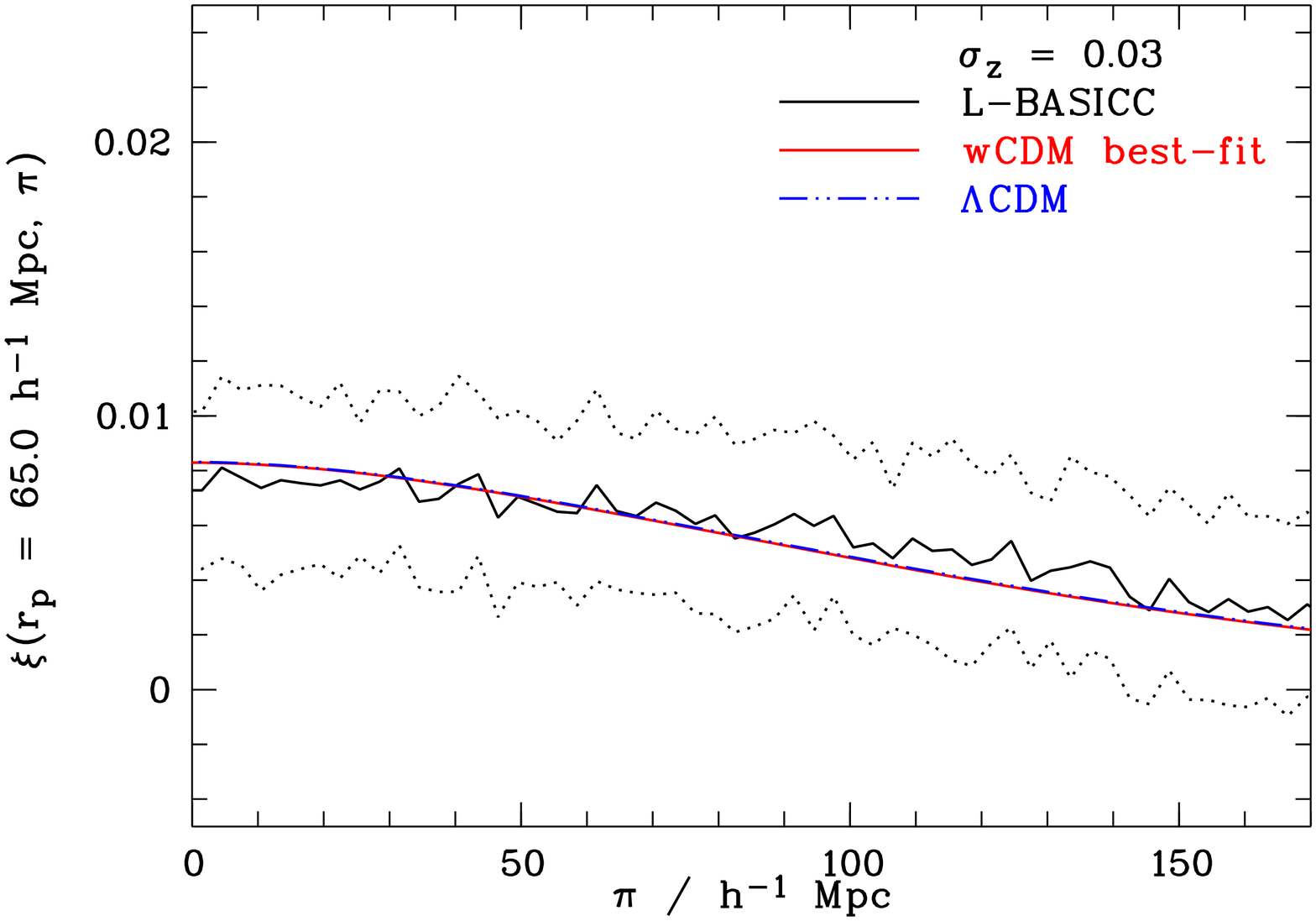,angle=270,clip=t,width=6.cm}
\psfig{figure=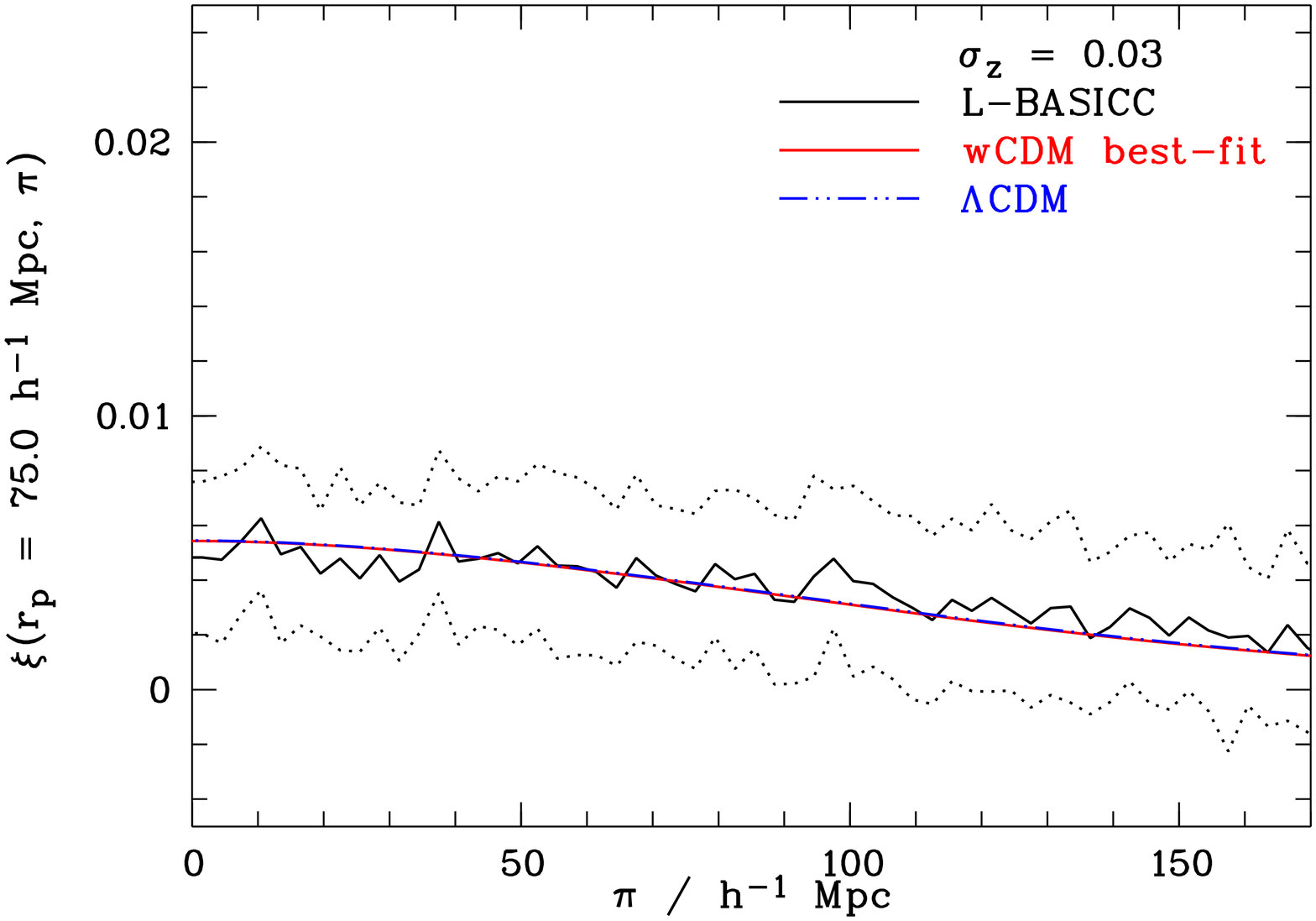,angle=270,clip=t,width=6.cm}
\psfig{figure=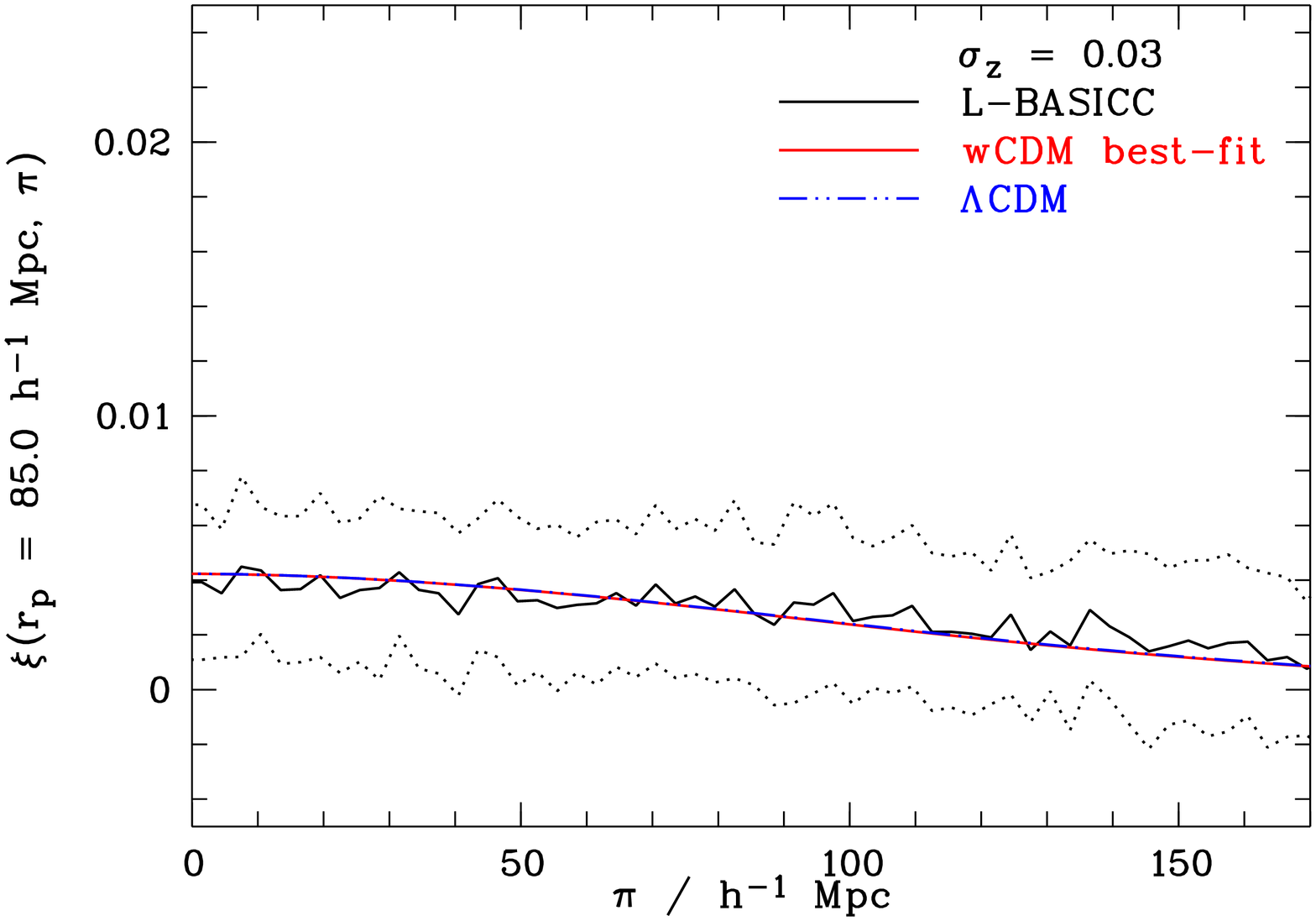,angle=270,clip=t,width=6.cm}}
\centerline{\psfig{figure=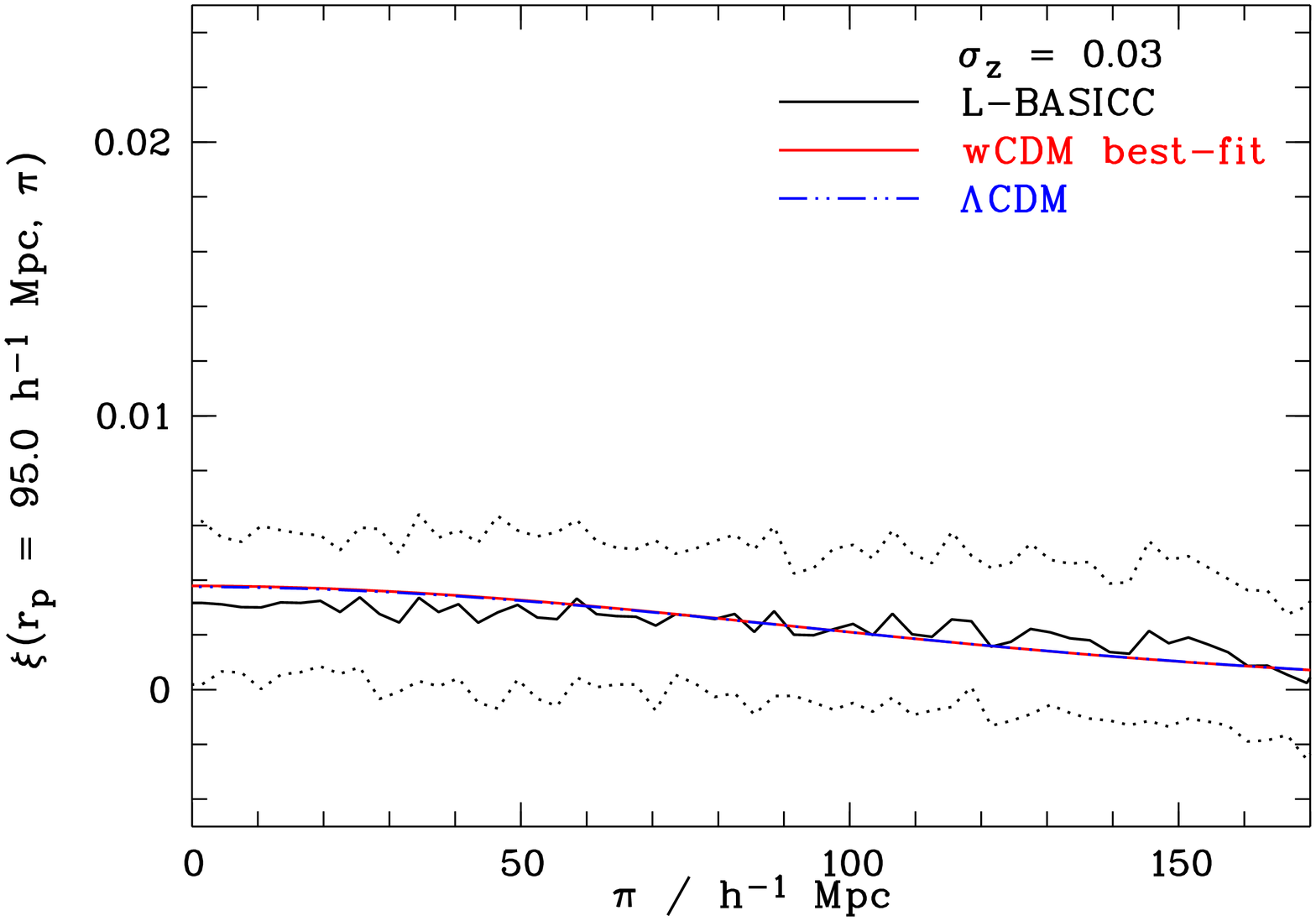,angle=270,clip=t,width=6.cm}
\psfig{figure=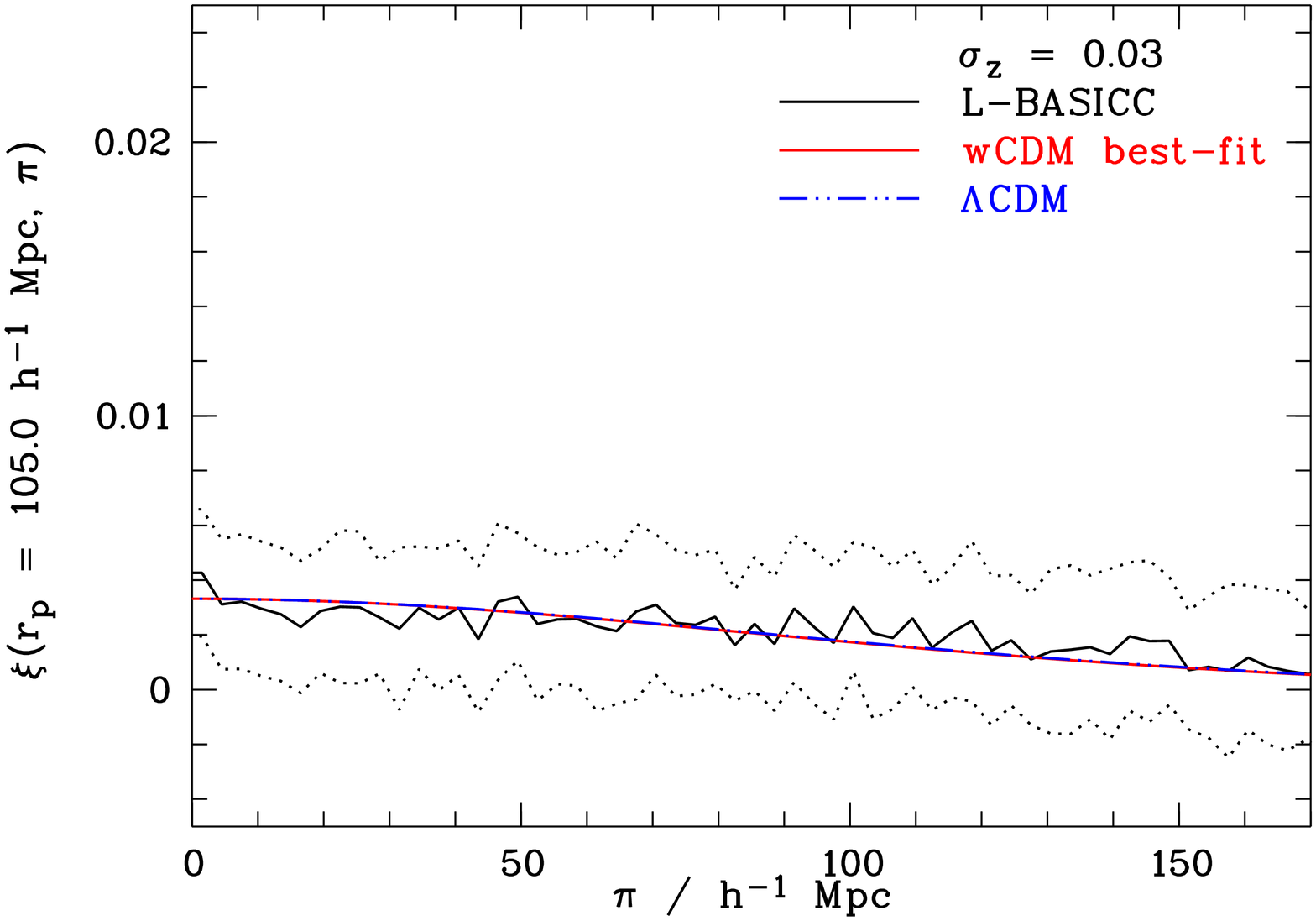,angle=270,clip=t,width=6.cm}
\psfig{figure=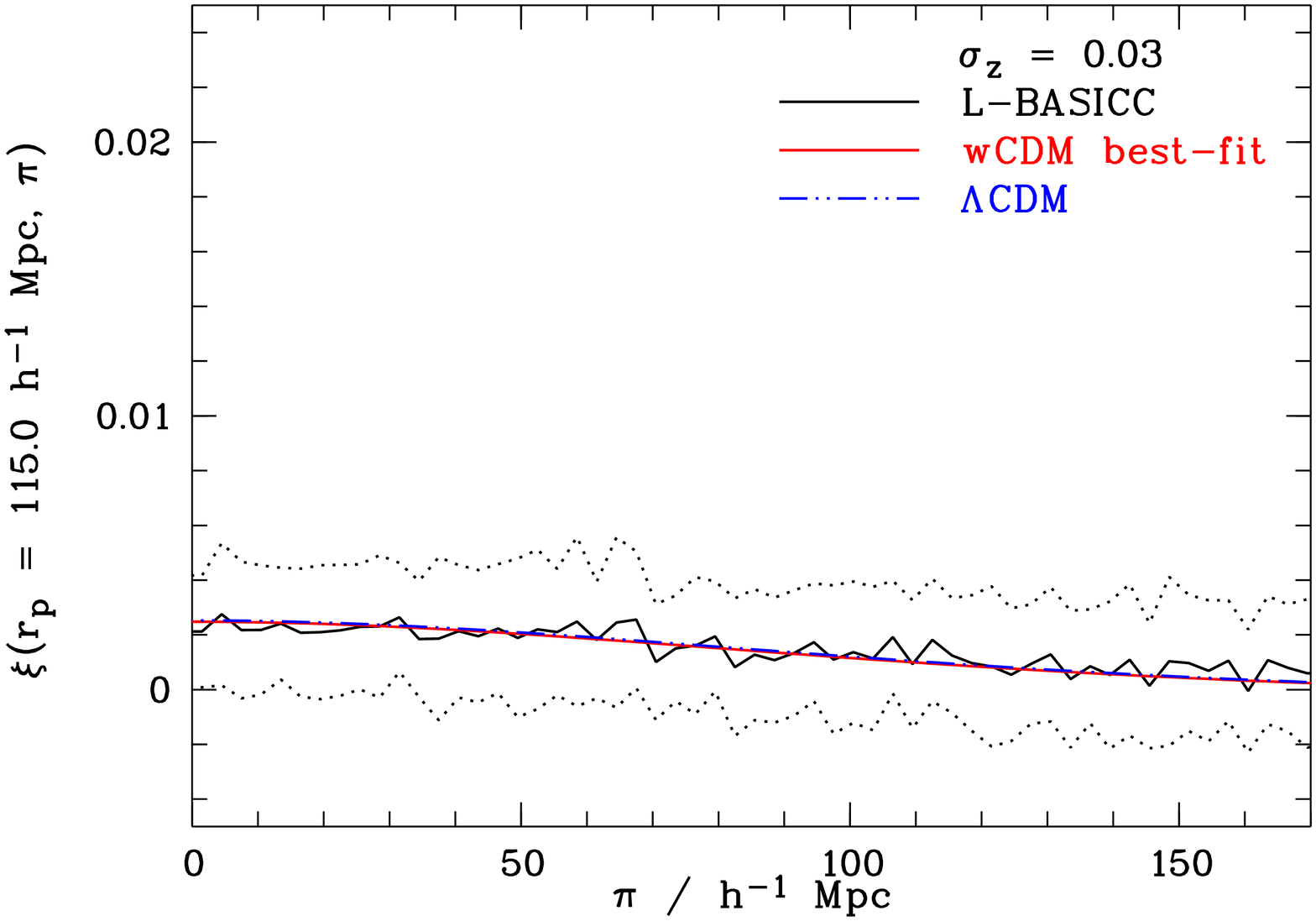,angle=270,clip=t,width=6.cm}}
\centerline{\psfig{figure=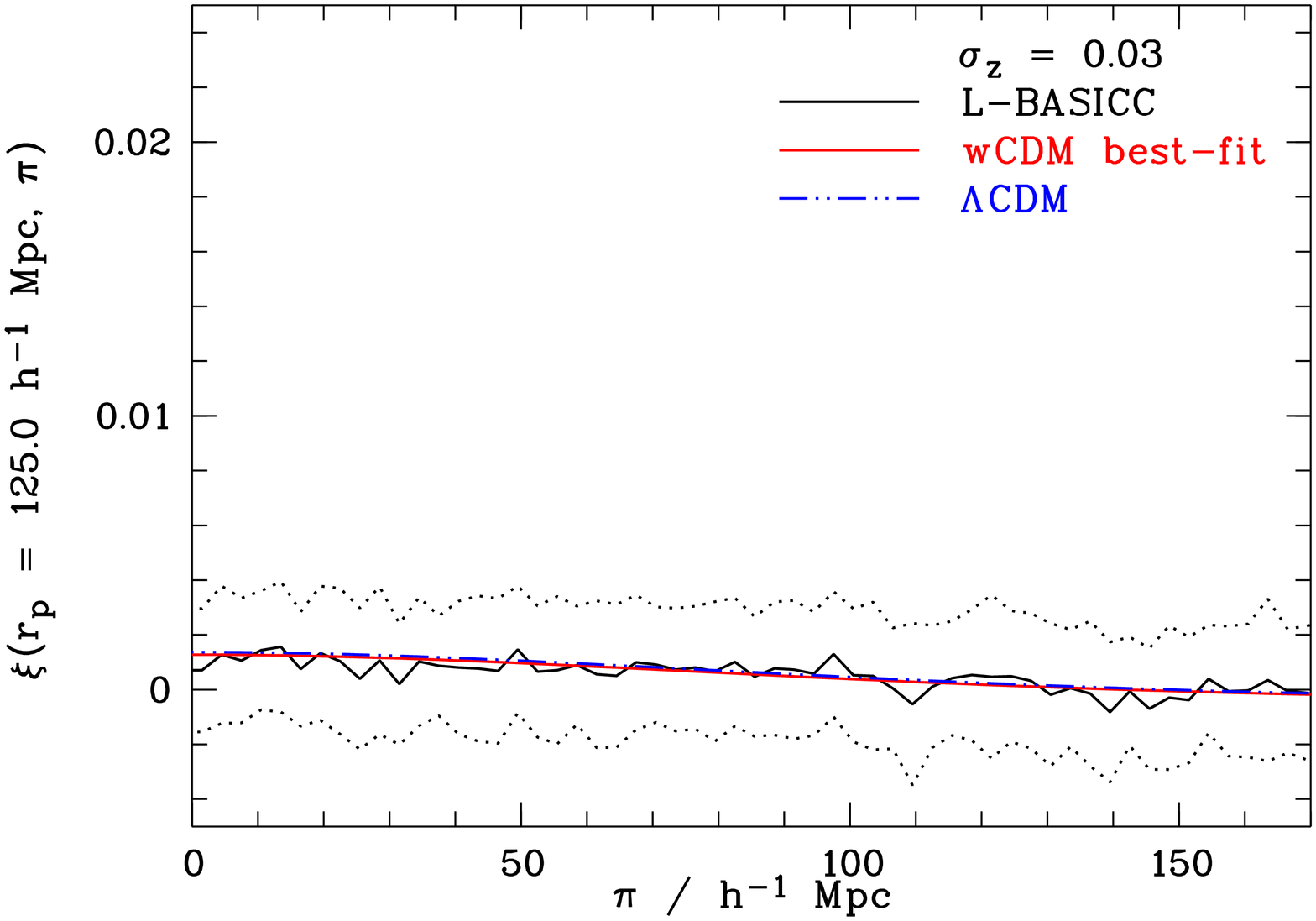,angle=270,clip=t,width=6.cm}
\psfig{figure=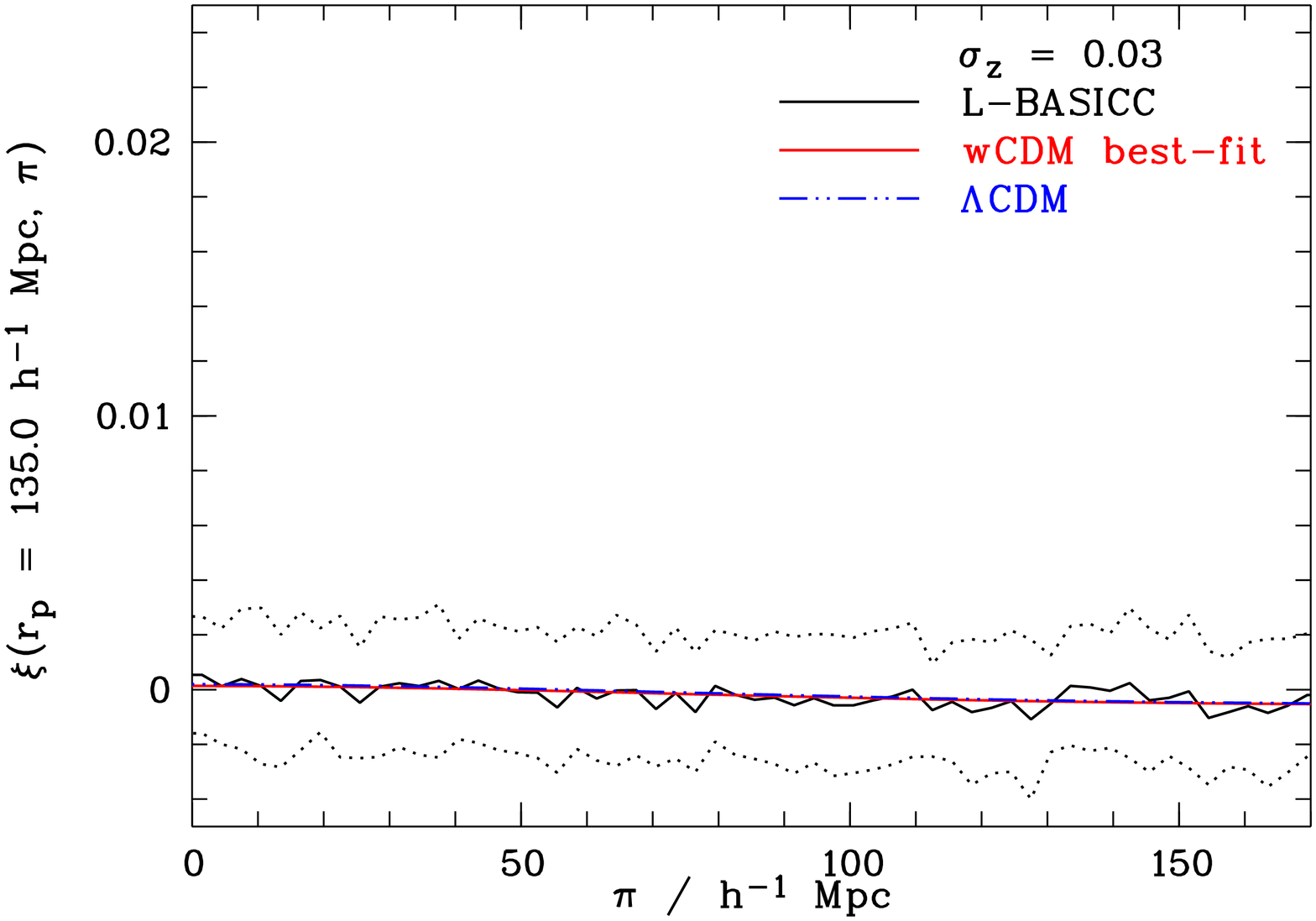,angle=270,clip=t,width=6.cm}
\psfig{figure=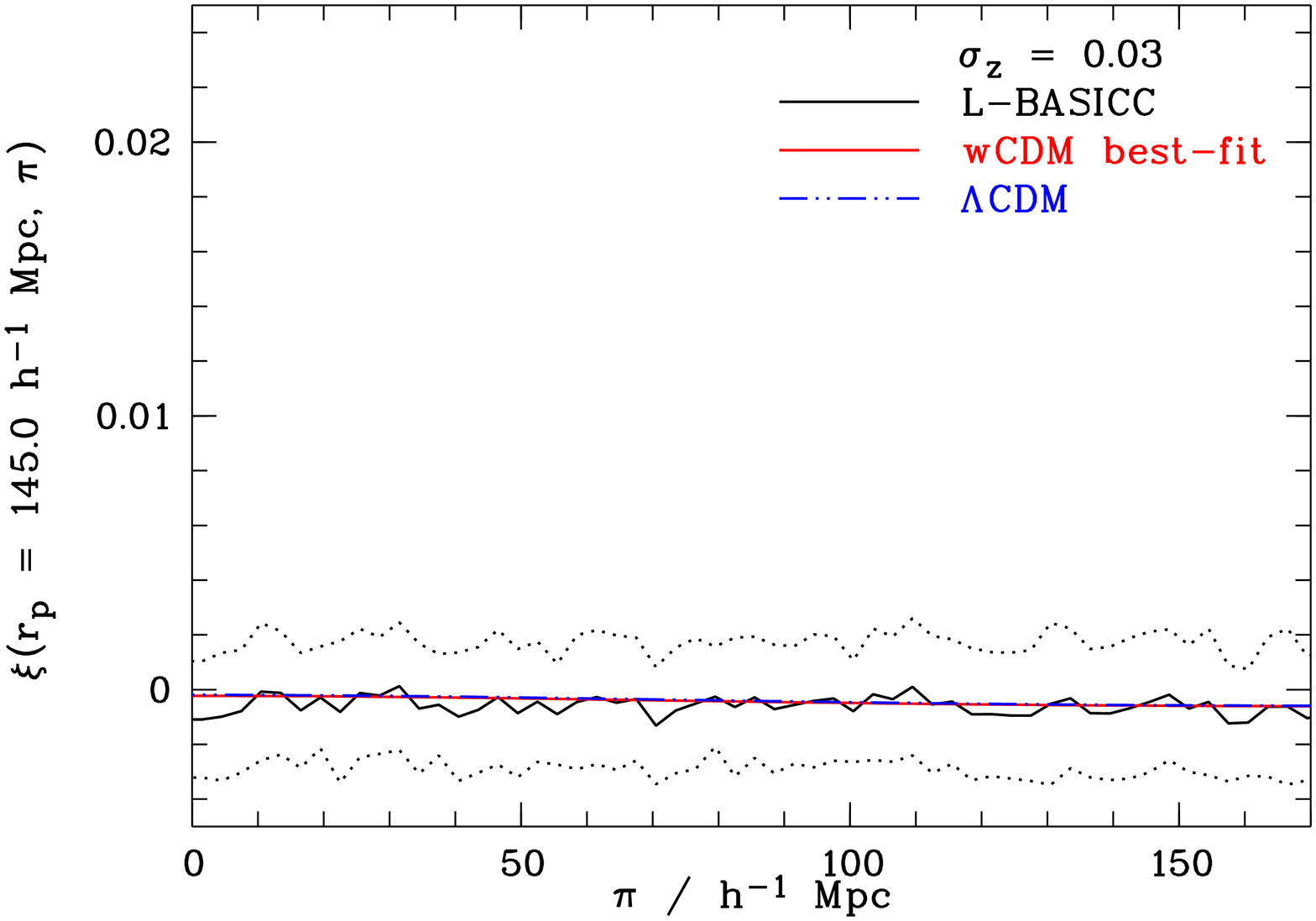,angle=270,clip=t,width=6.cm}}
\caption[ ]{As in Figure \ref{xirppicuts_zerr015}, but for $\sigma_z= 0.03$.}\label{xirppicuts_zerr03}
\end{figure*}

\begin{figure*}
\centerline{\psfig{figure=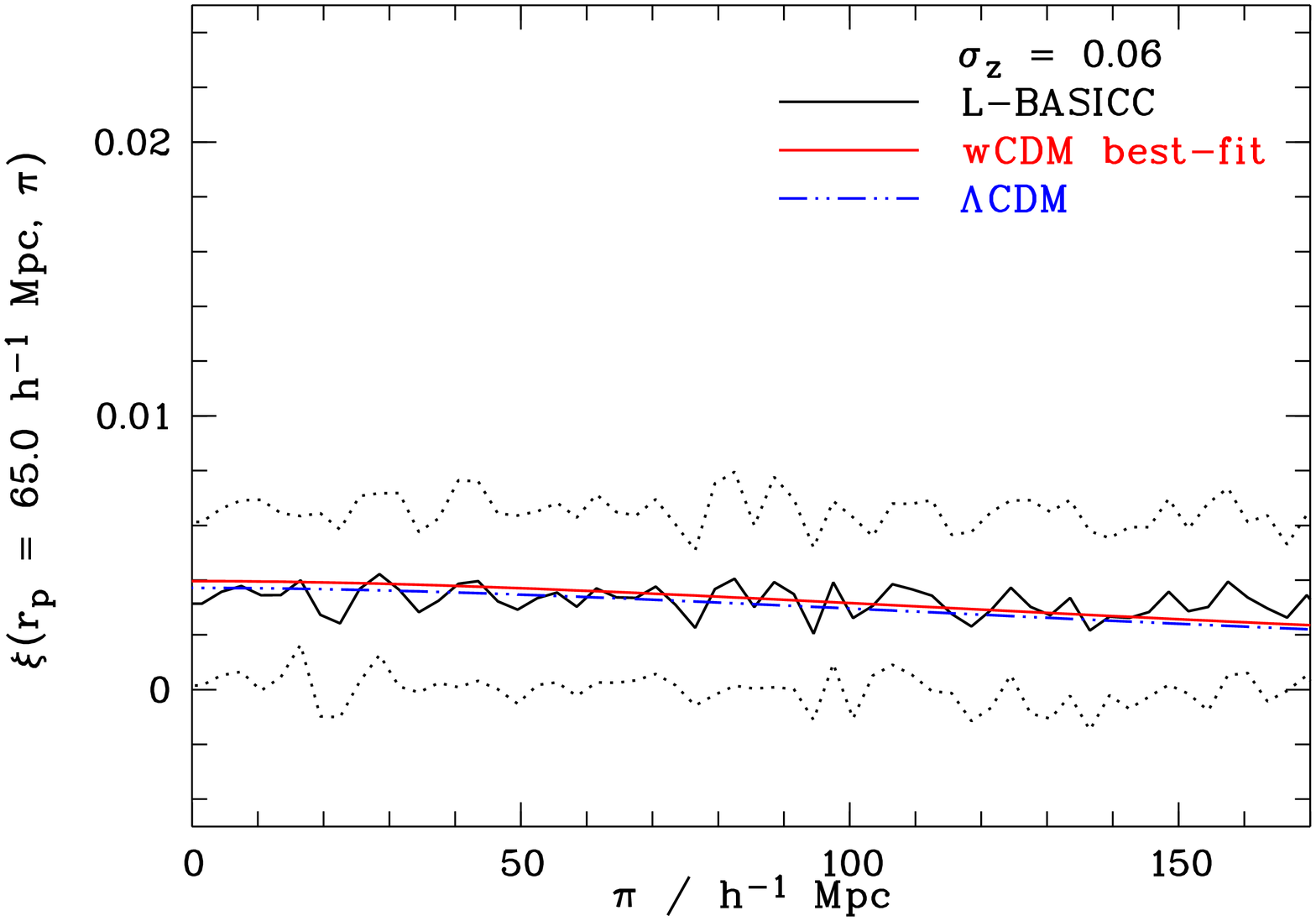,angle=270,clip=t,width=6.cm}
\psfig{figure=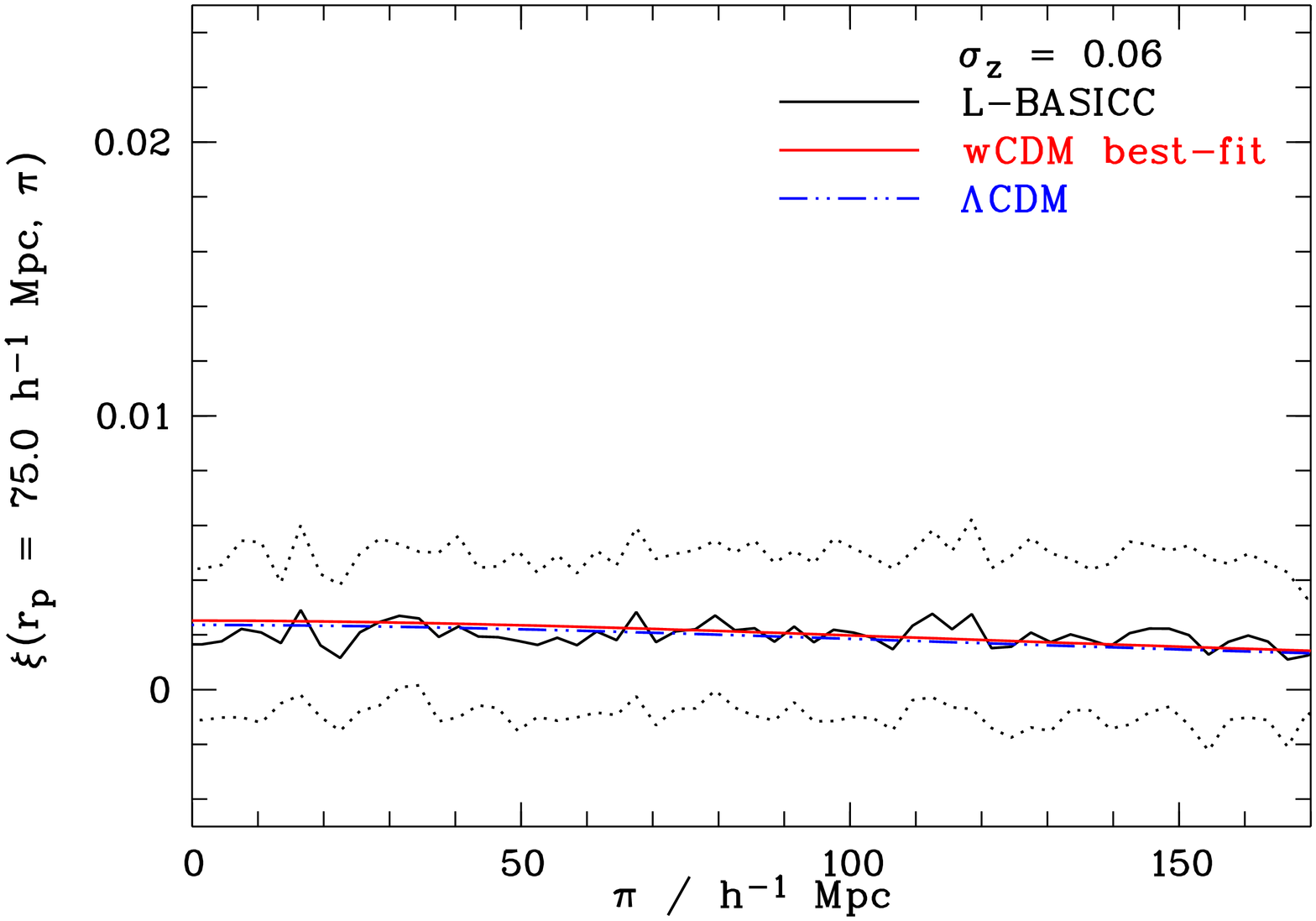,angle=270,clip=t,width=6.cm}
\psfig{figure=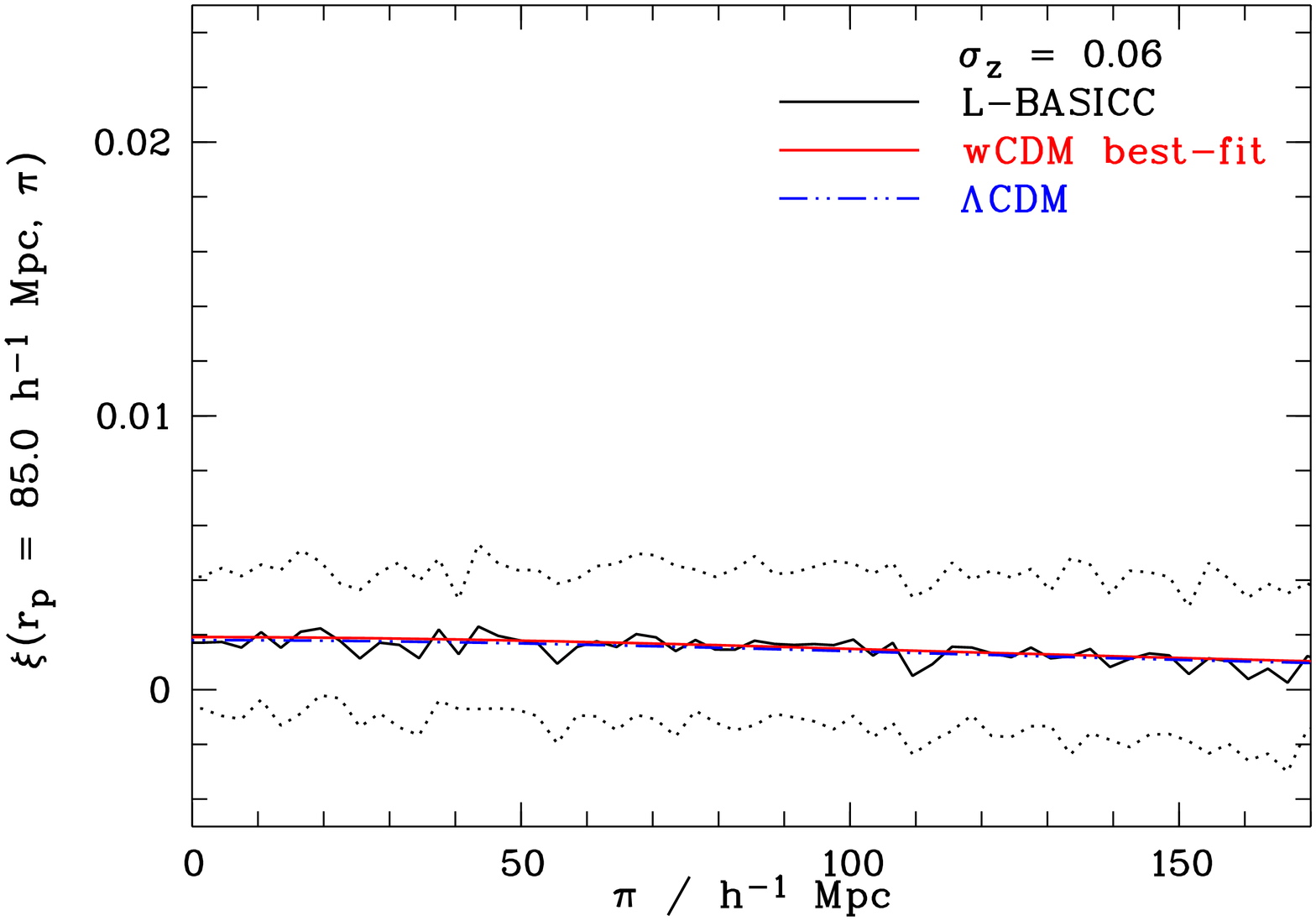,angle=270,clip=t,width=6.cm}}
\centerline{\psfig{figure=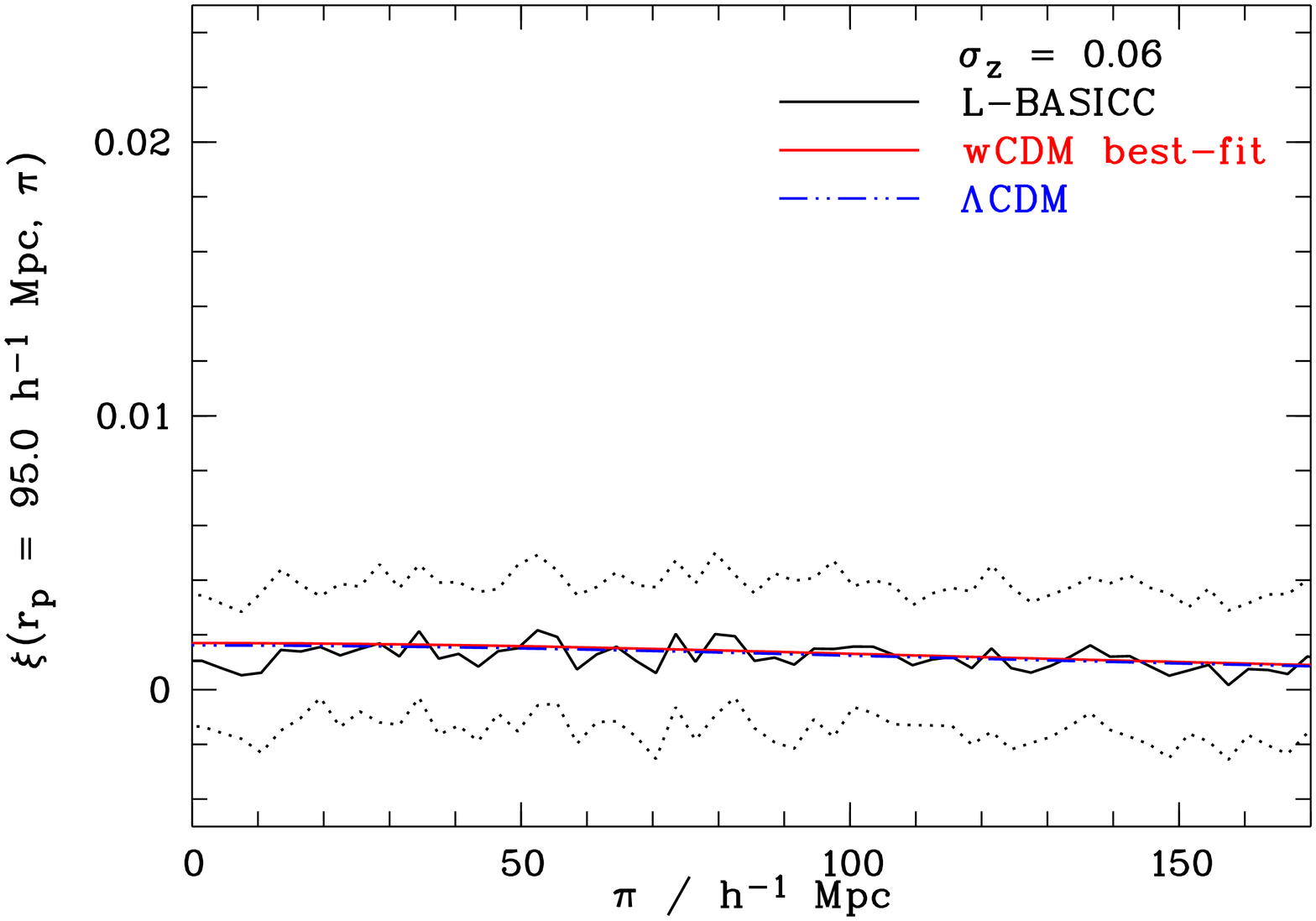,angle=270,clip=t,width=6.cm}
\psfig{figure=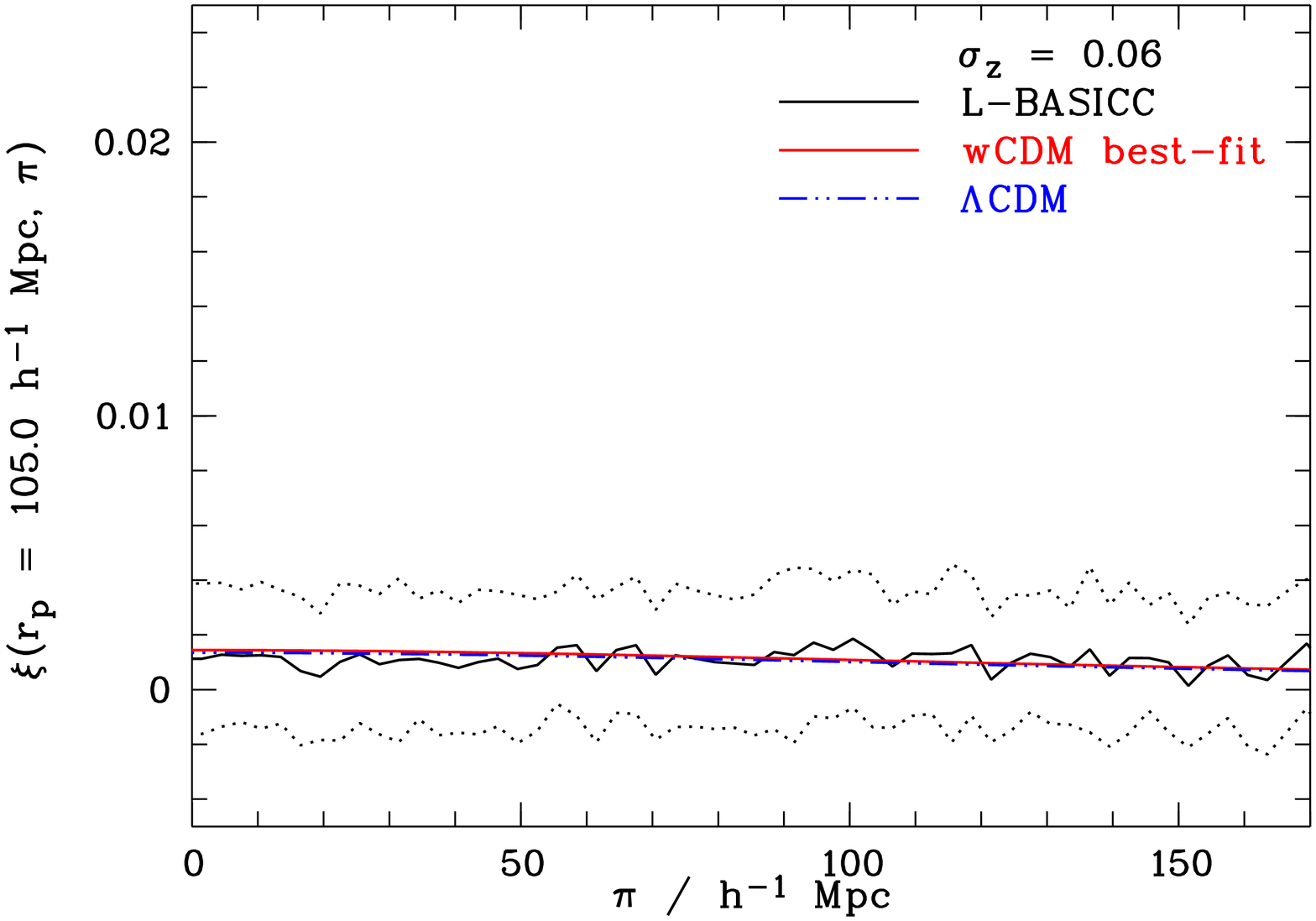,angle=270,clip=t,width=6.cm}
\psfig{figure=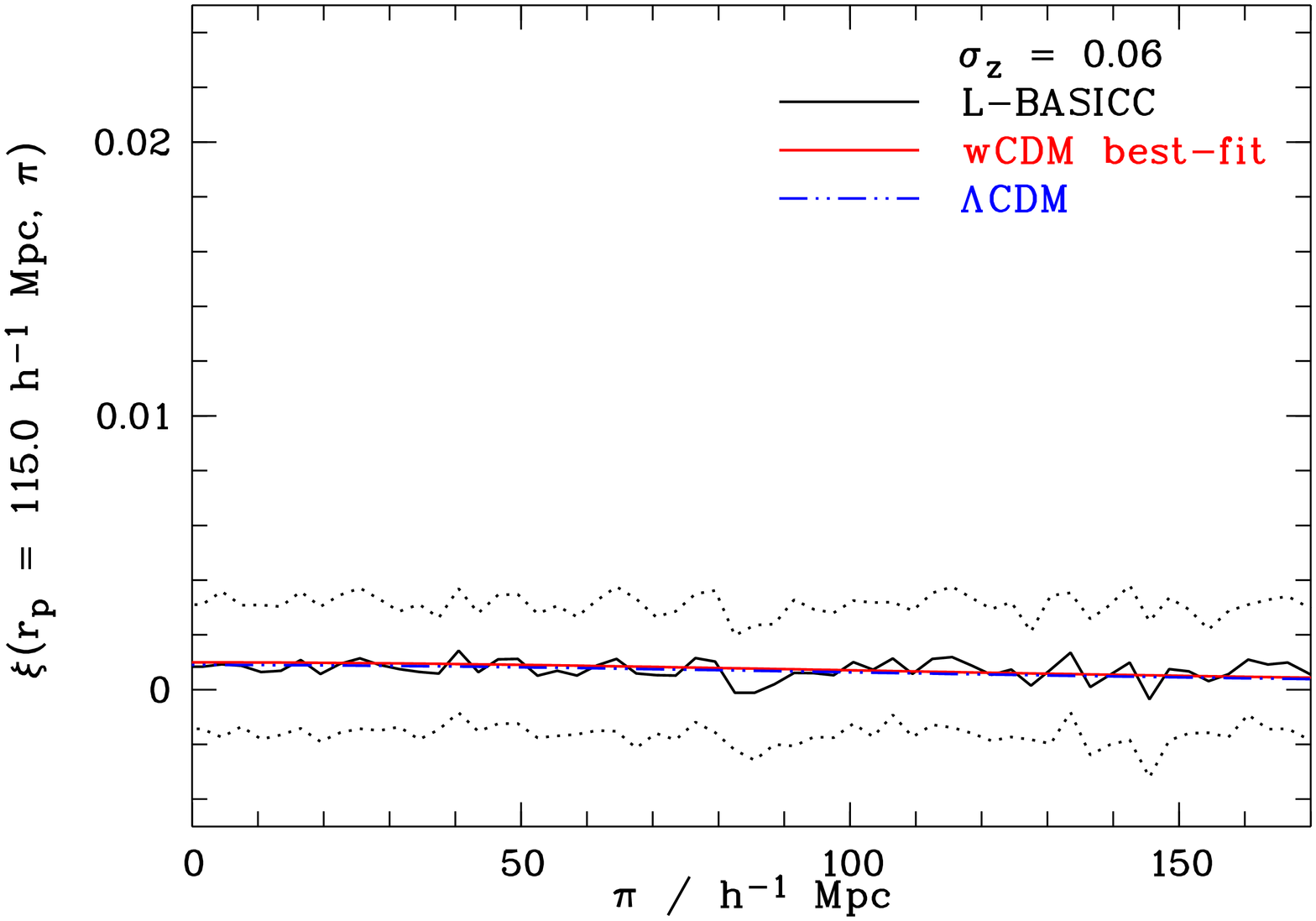,angle=270,clip=t,width=6.cm}}
\centerline{\psfig{figure=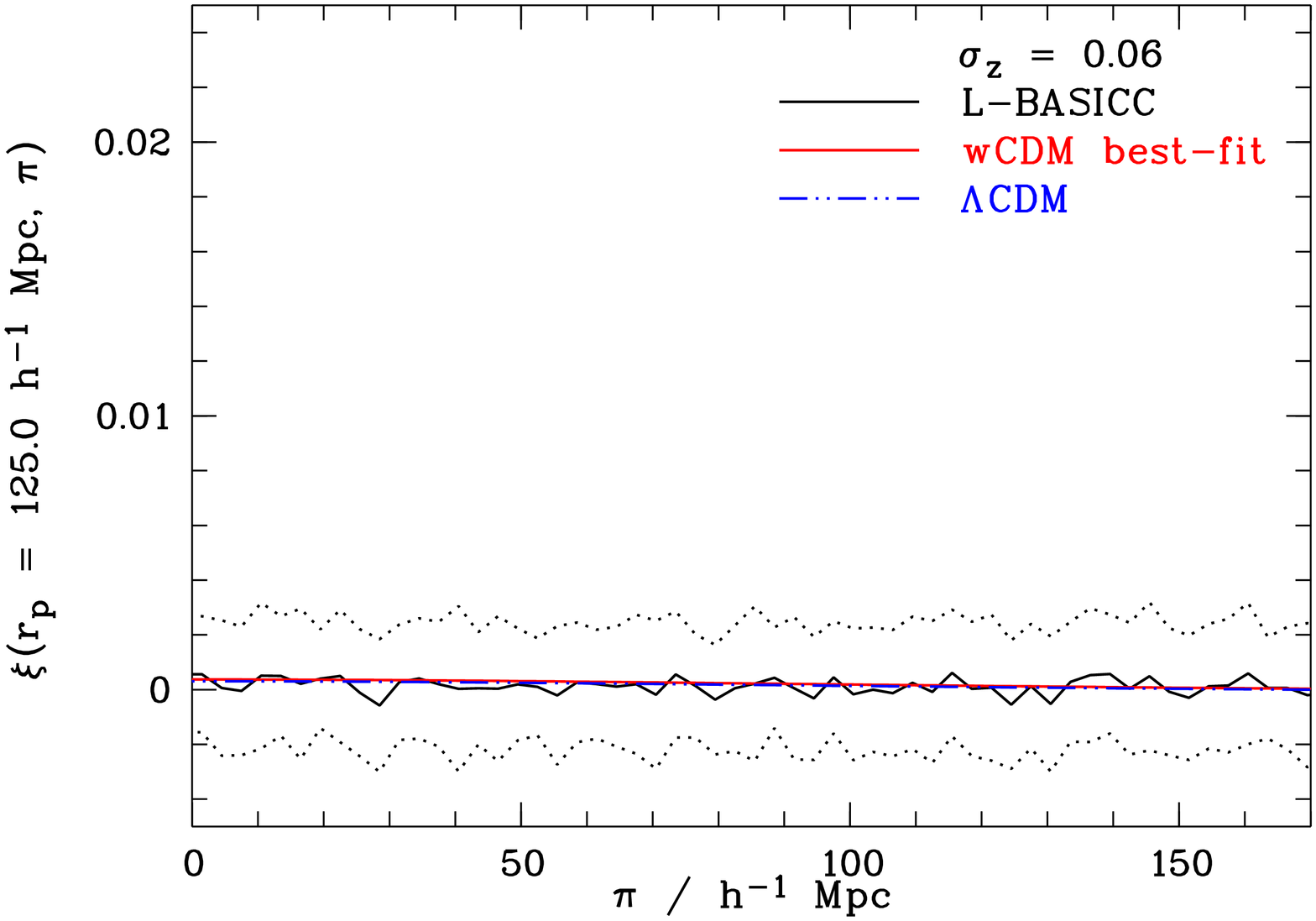,angle=270,clip=t,width=6.cm}
\psfig{figure=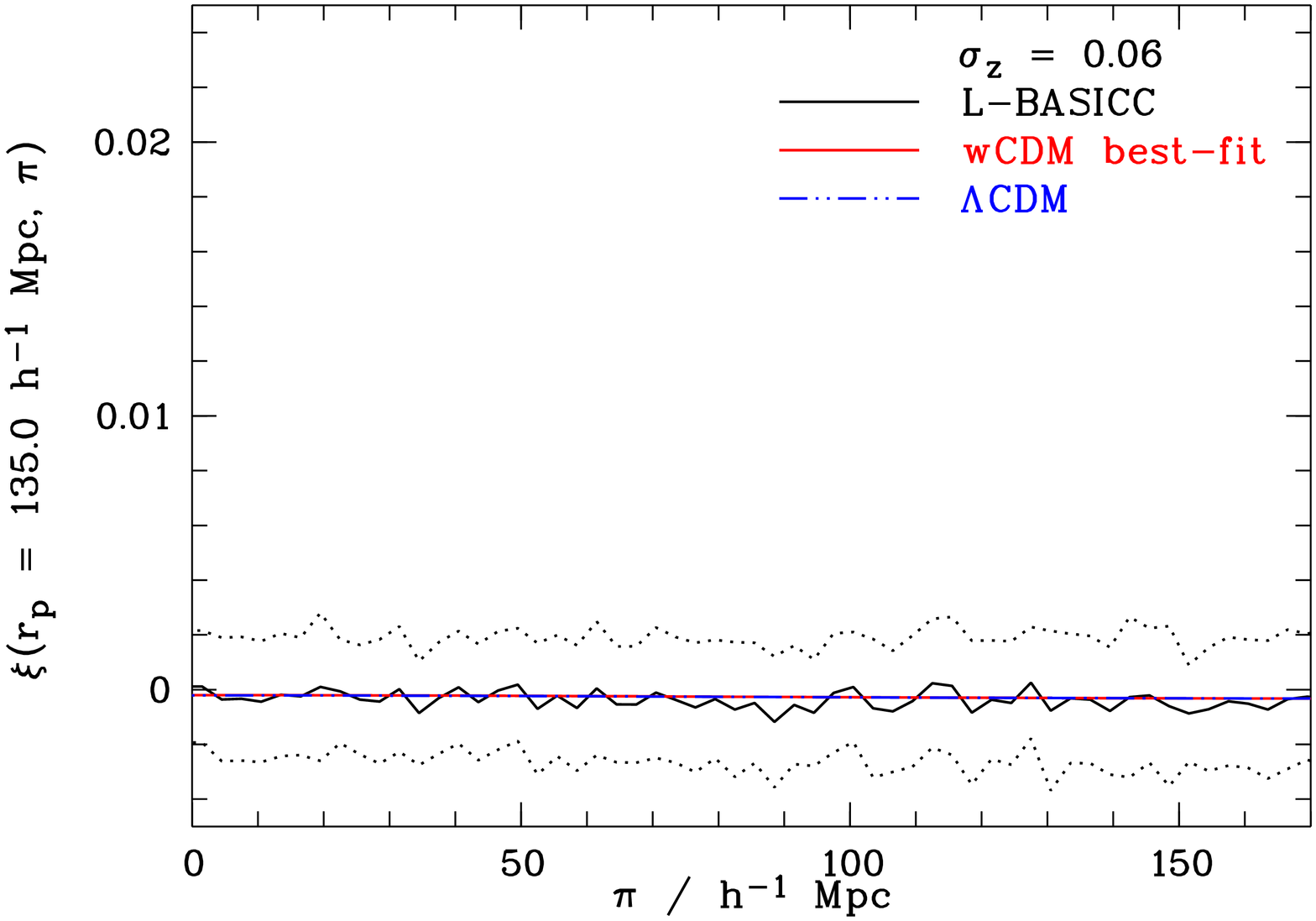,angle=270,clip=t,width=6.cm}
\psfig{figure=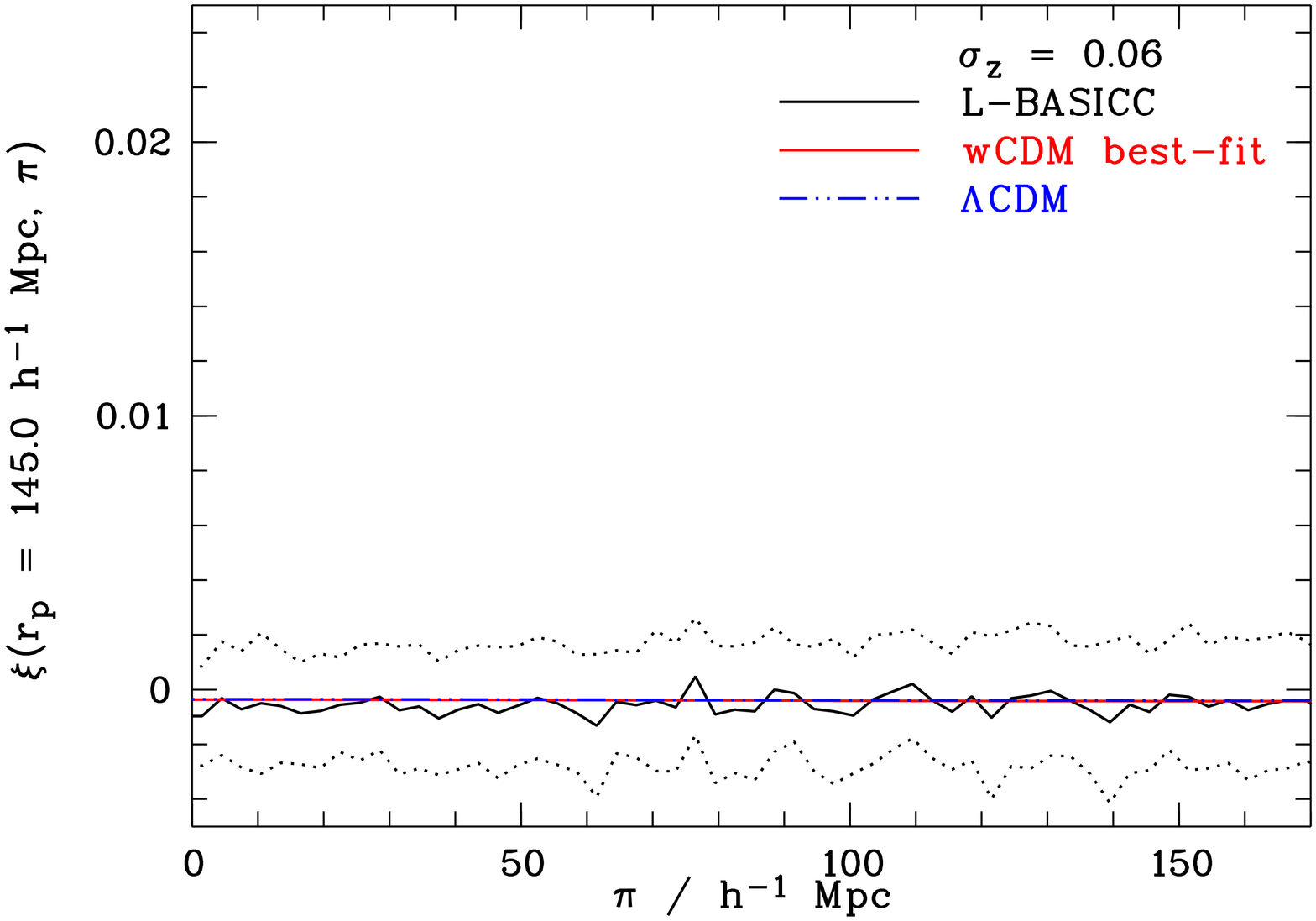,angle=270,clip=t,width=6.cm}}
\caption[ ]{As in Figure \ref{xirppicuts_zerr015}, but for $\sigma = 0.06$.}\label{xirppicuts_zerr06}
\end{figure*}

\begin{figure*}
\centerline{\psfig{figure=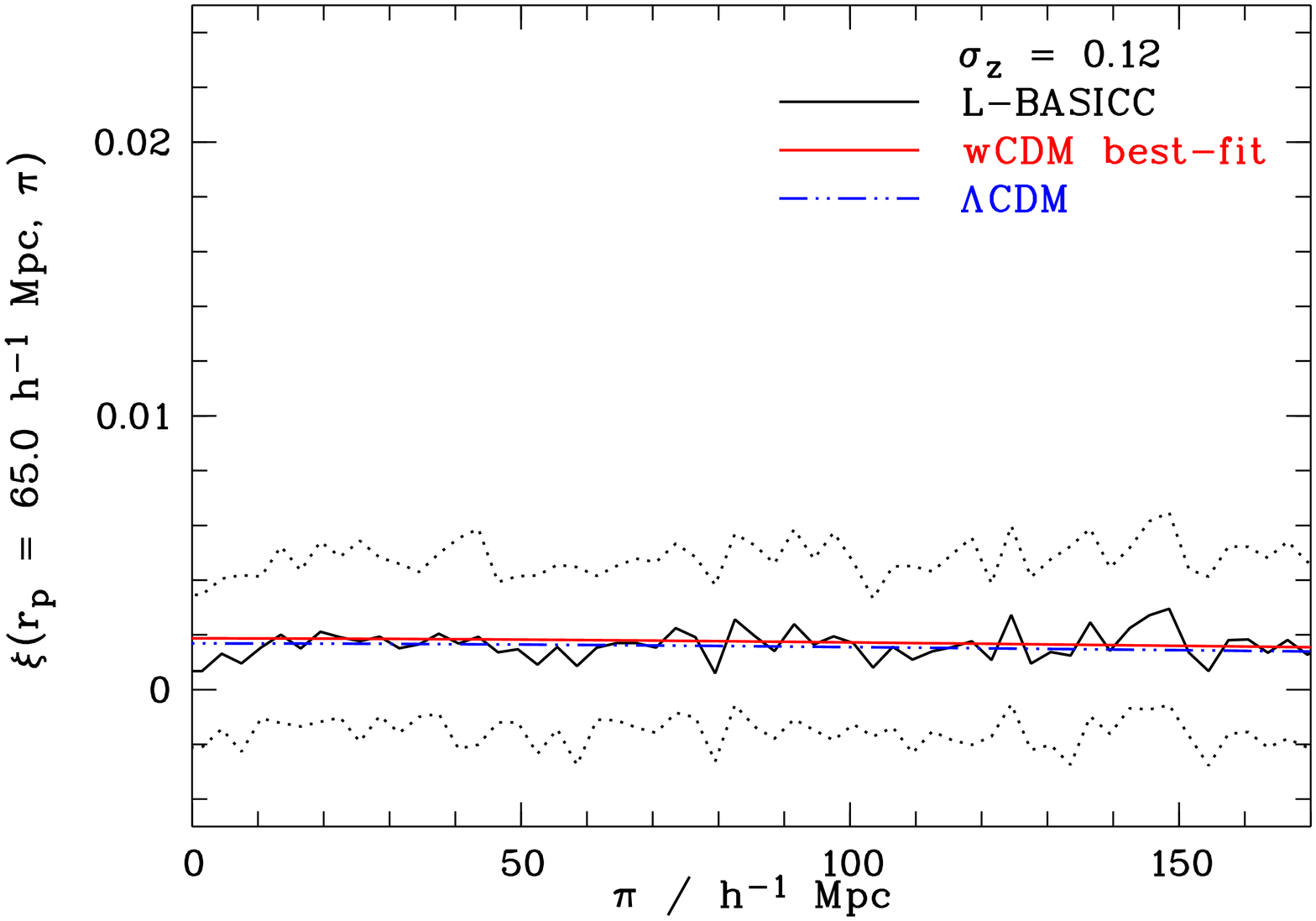,angle=270,clip=t,width=6.cm}
\psfig{figure=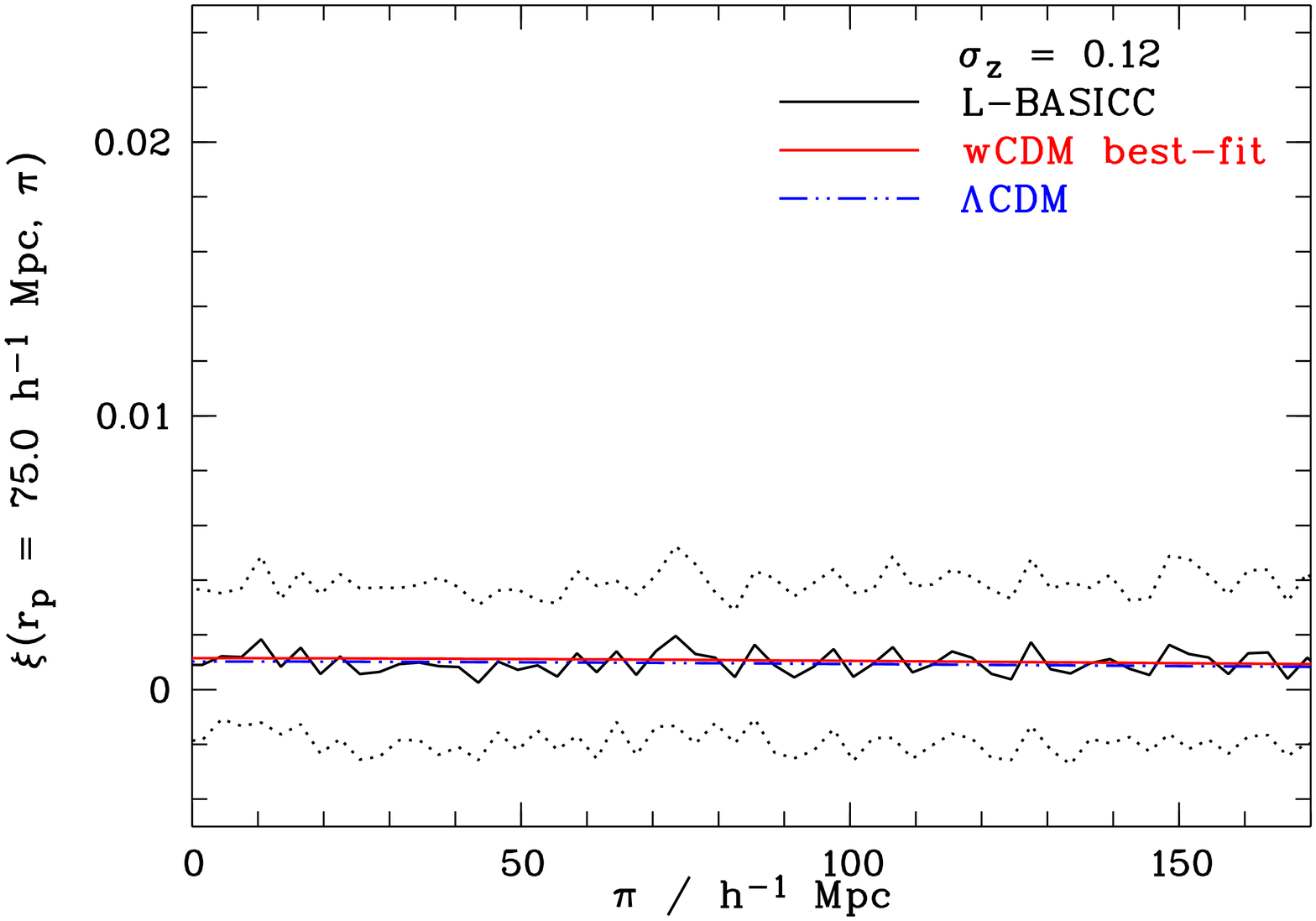,angle=270,clip=t,width=6.cm}
\psfig{figure=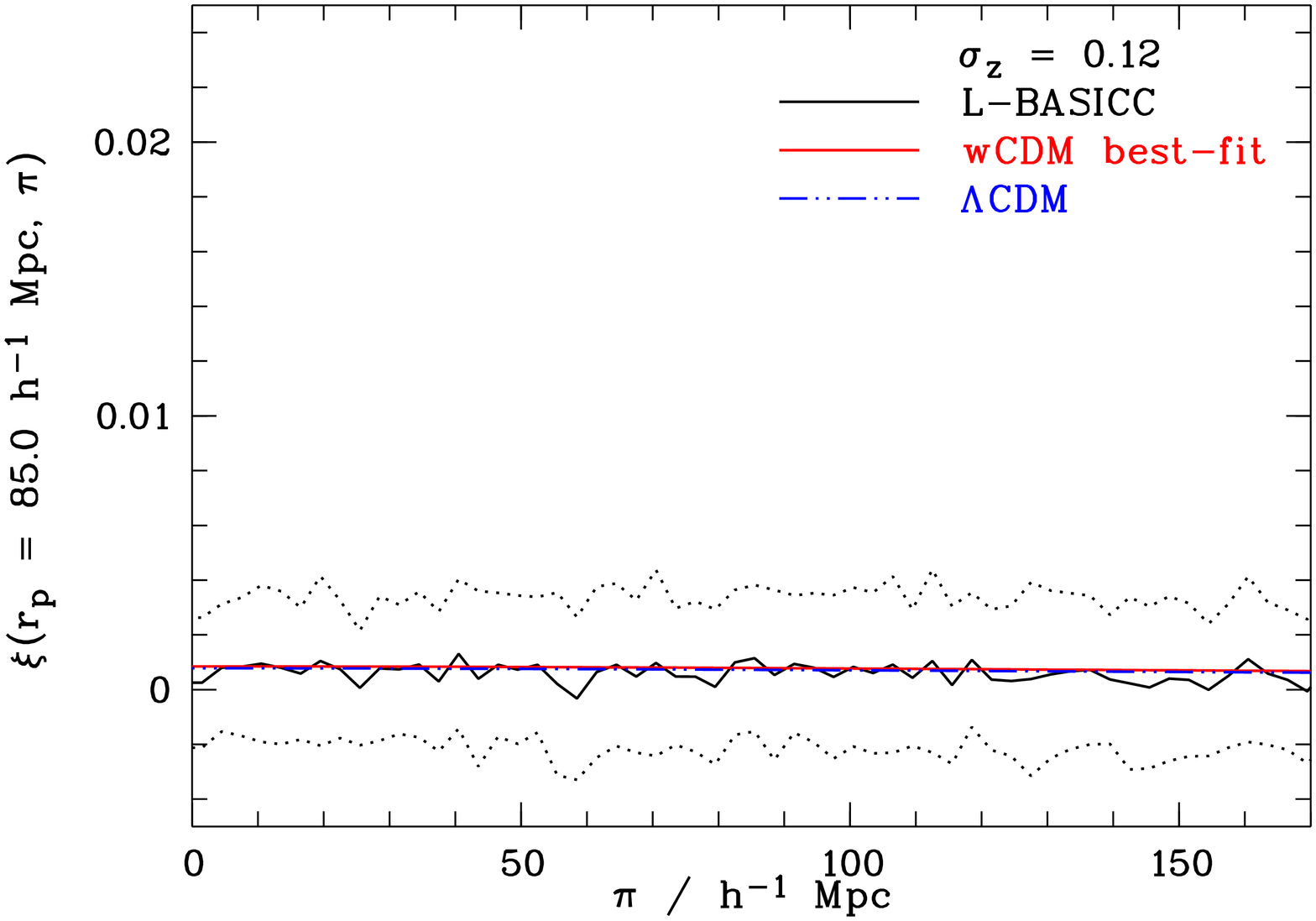,angle=270,clip=t,width=6.cm}}
\centerline{\psfig{figure=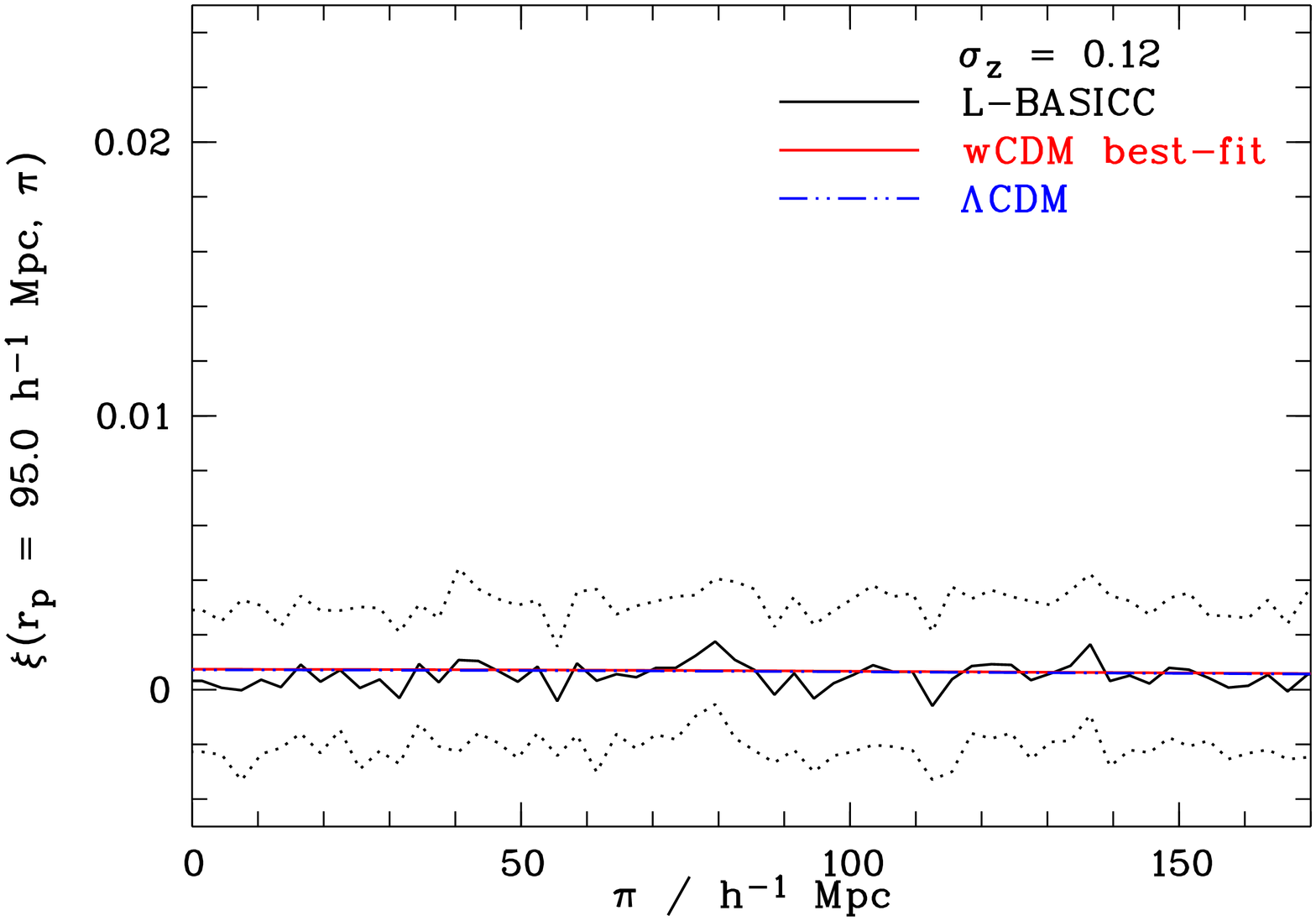,angle=270,clip=t,width=6.cm}
\psfig{figure=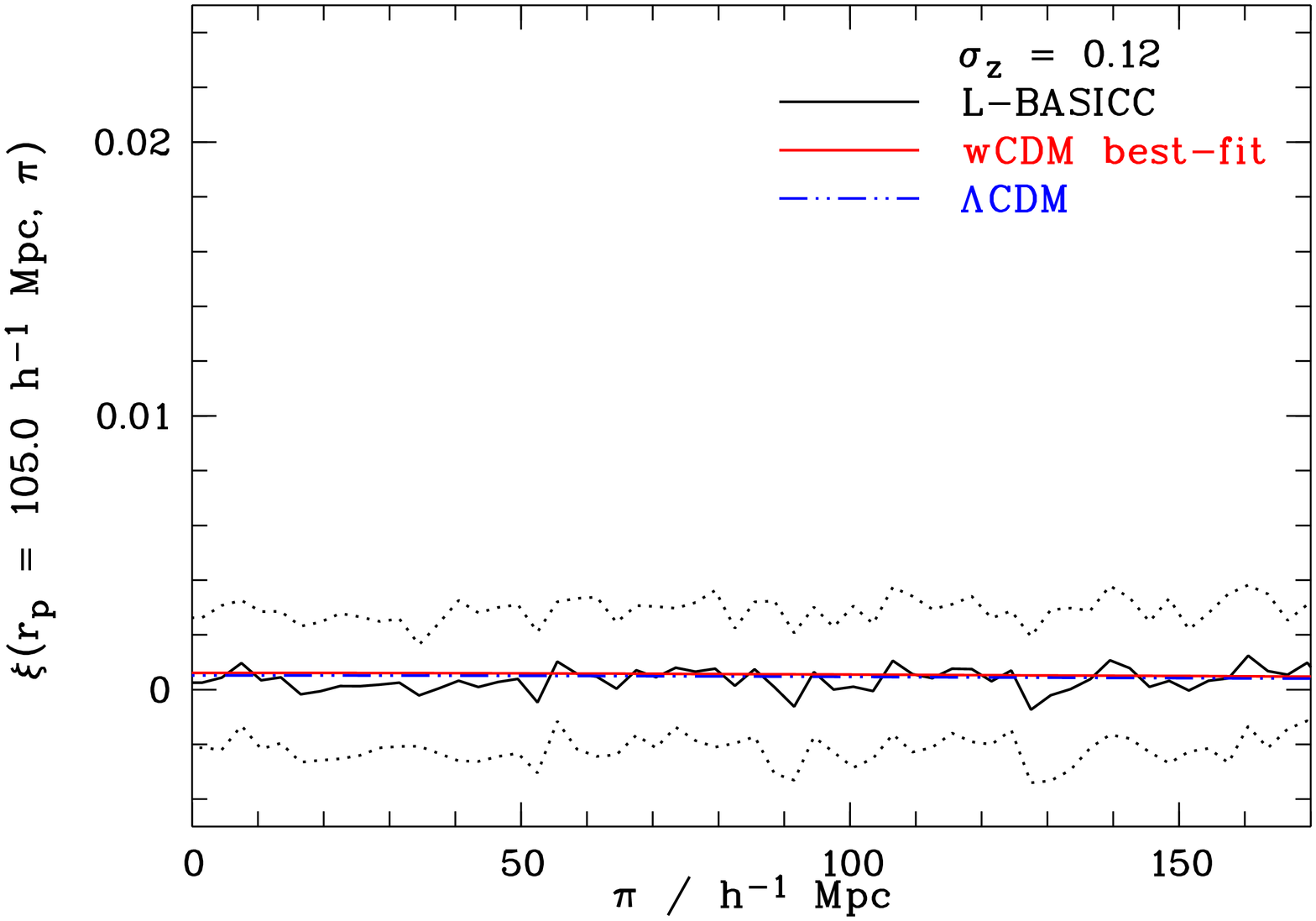,angle=270,clip=t,width=6.cm}
\psfig{figure=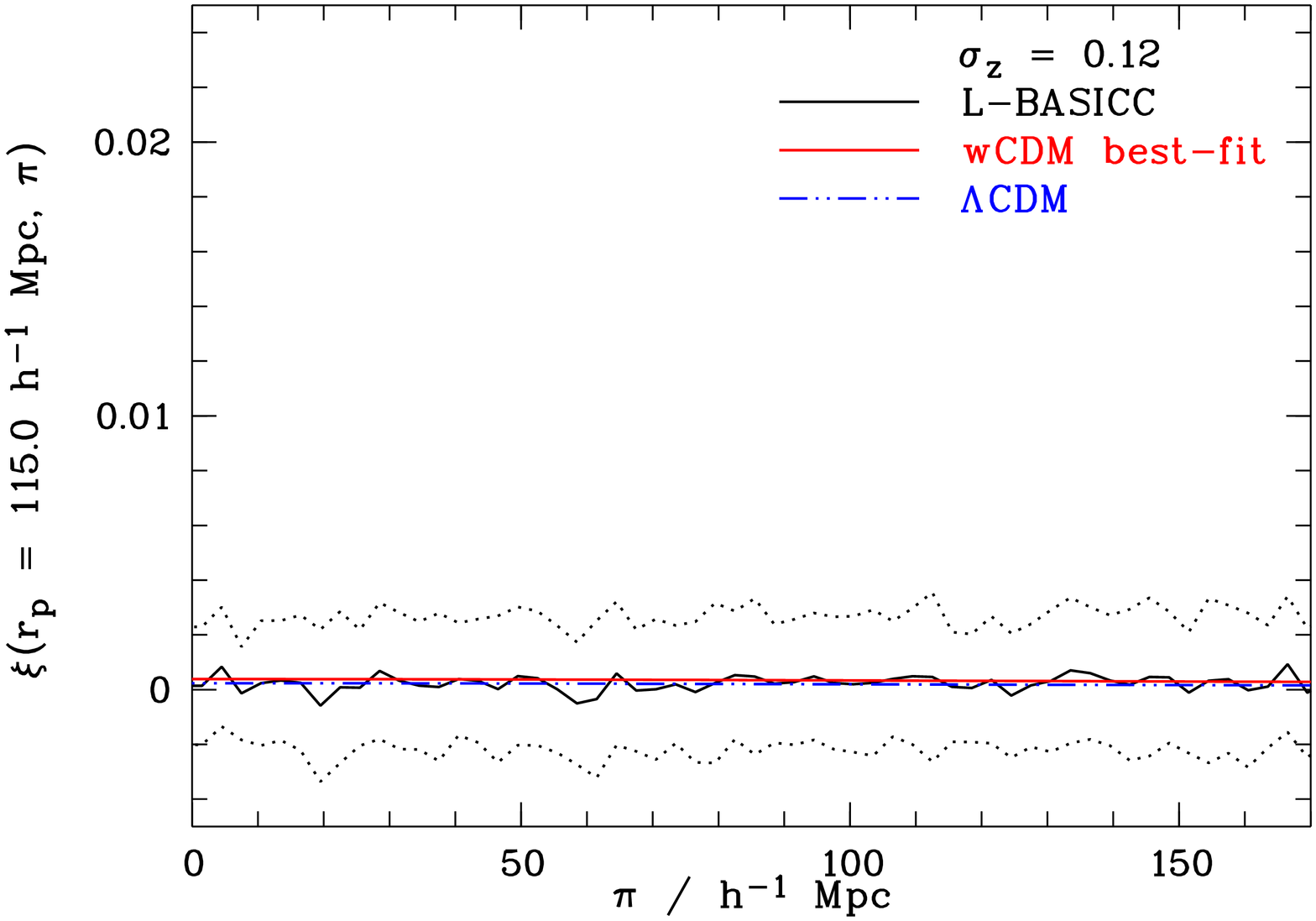,angle=270,clip=t,width=6.cm}}
\centerline{\psfig{figure=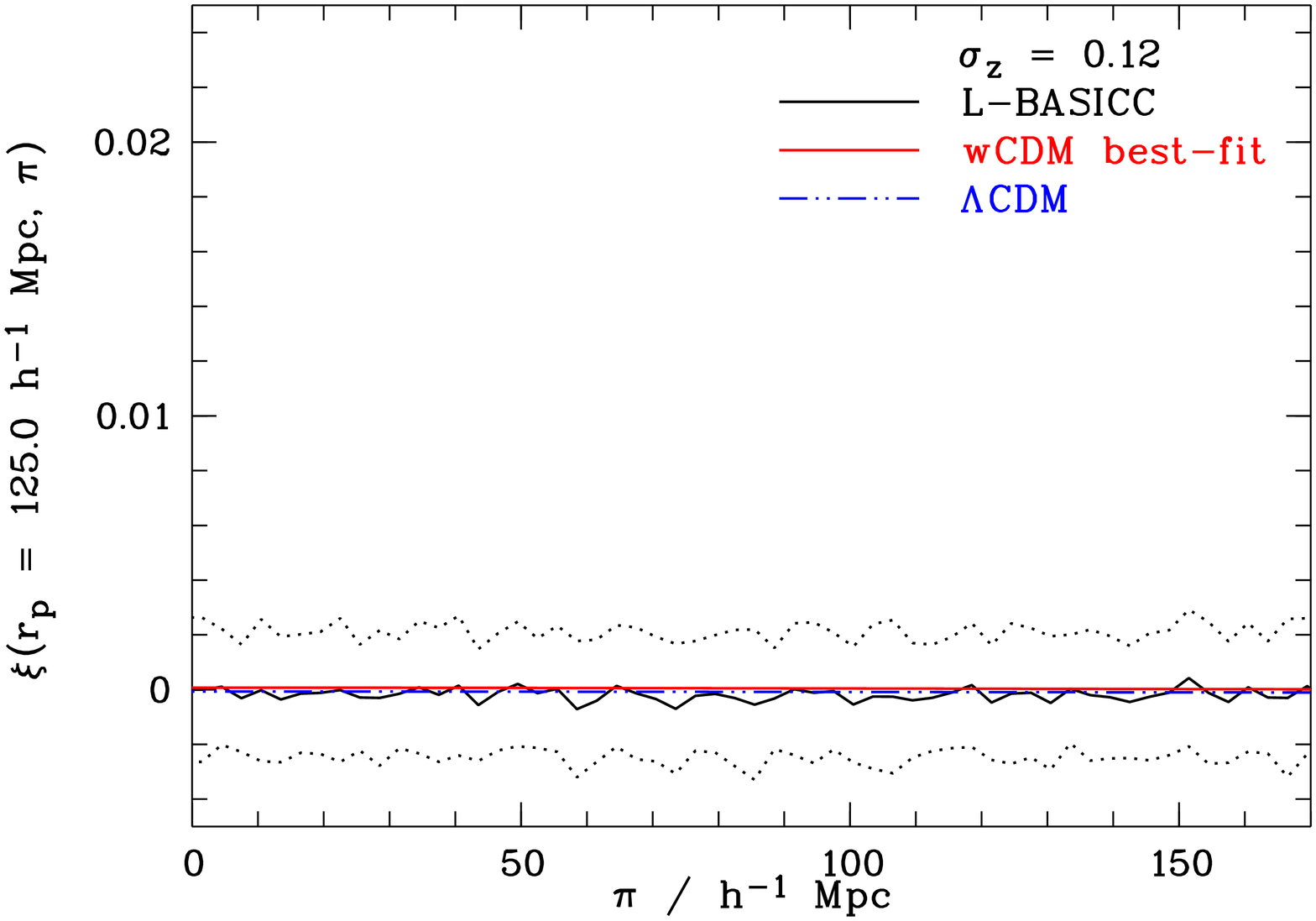,angle=270,clip=t,width=6.cm}
\psfig{figure=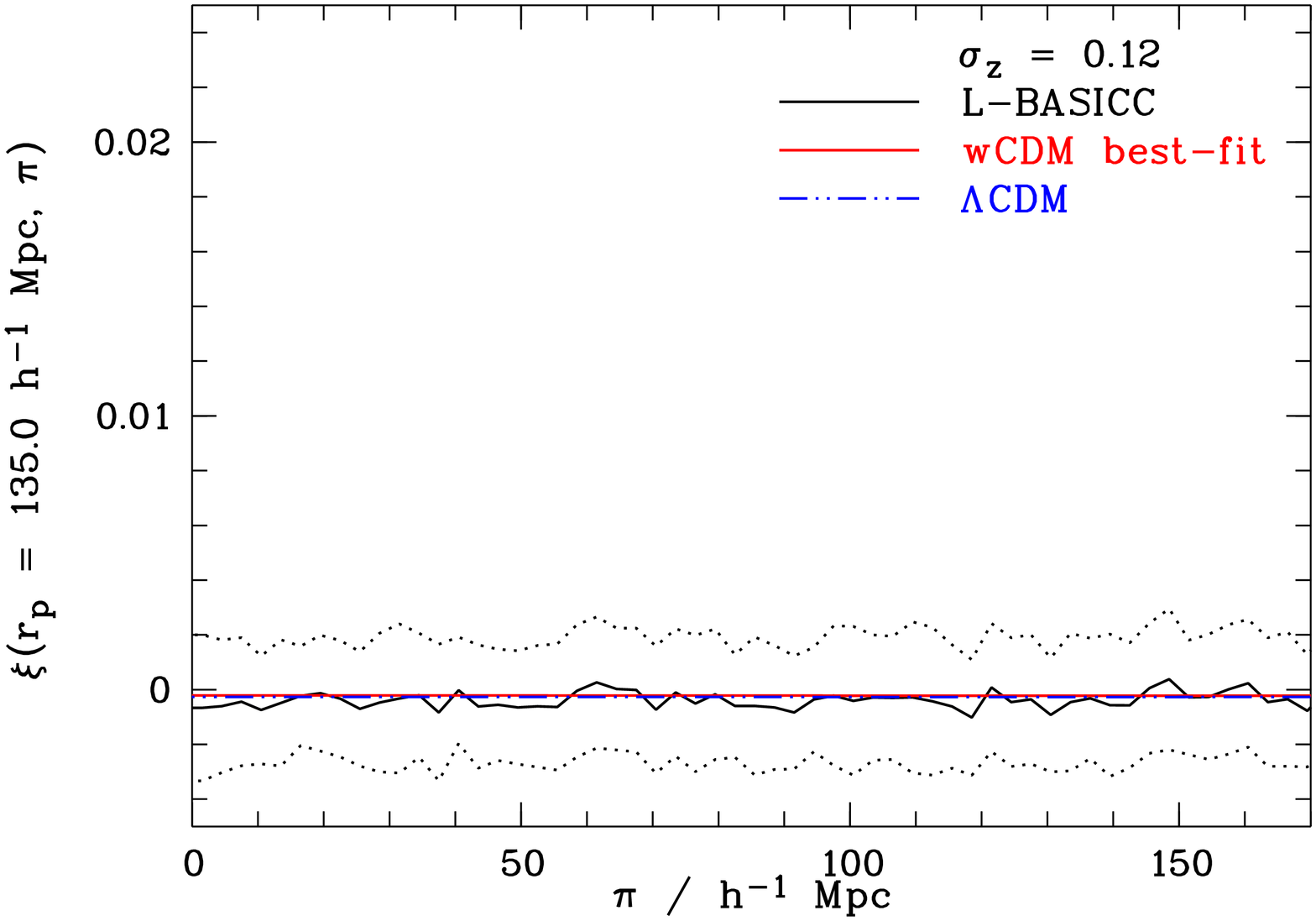,angle=270,clip=t,width=6.cm}
\psfig{figure=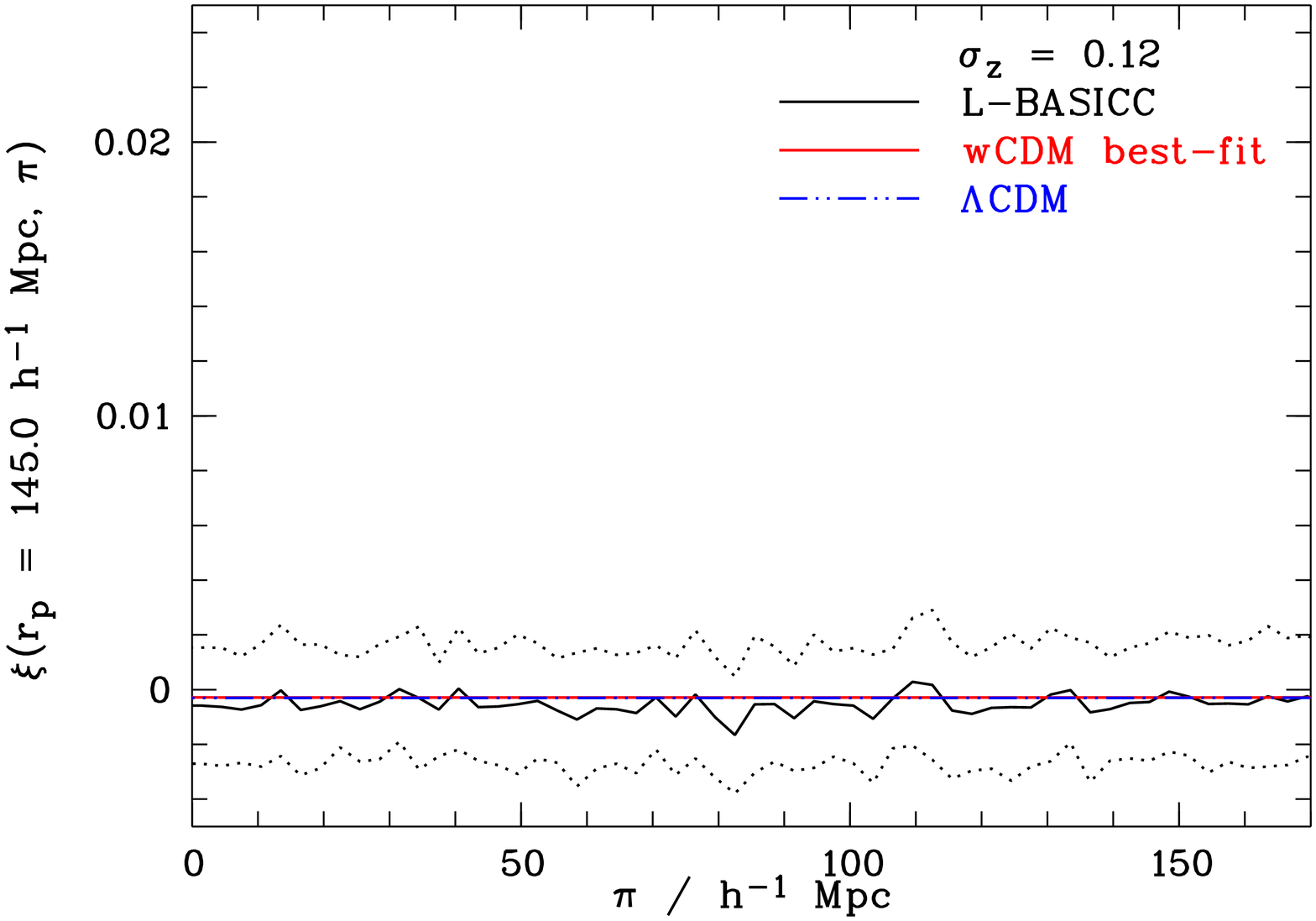,angle=270,clip=t,width=6.cm}}
\caption[ ]{As in Figure \ref{xirppicuts_zerr015}, but for $\sigma_z= 0.12$.}\label{xirppicuts_zerr12}
\end{figure*}

In Figure \ref{xirppicuts_zerr015}, Figure \ref{xirppicuts_zerr03}, Figure \ref{xirppicuts_zerr06}, and Figure \ref{xirppicuts_zerr12} we show again cuts along constant values of $r_p$ through  $\xi_{zerr}(r_p, \pi)$ for $\sigma_z= 0.015$, $0.03$, $0.06$, and $0.12$, respectively. The model fits the data well, although  as expected slightly worse than in real and redshift space (see Figures \ref{xirppicuts_rs} and \ref{xirppicuts_zs}).

\begin{figure*}
\centerline{\psfig{figure=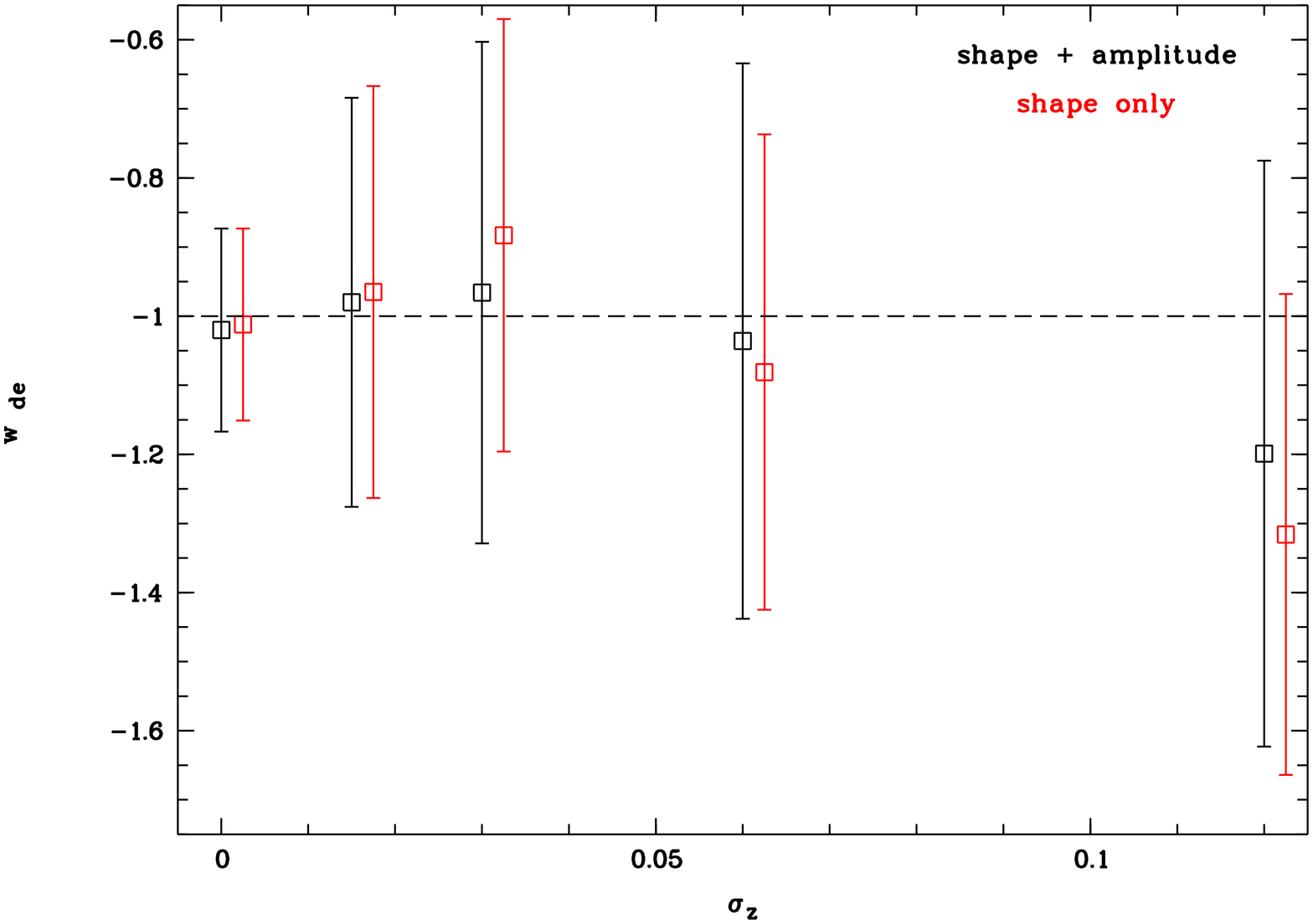,angle=270,clip=t,width=9.cm}
\psfig{figure=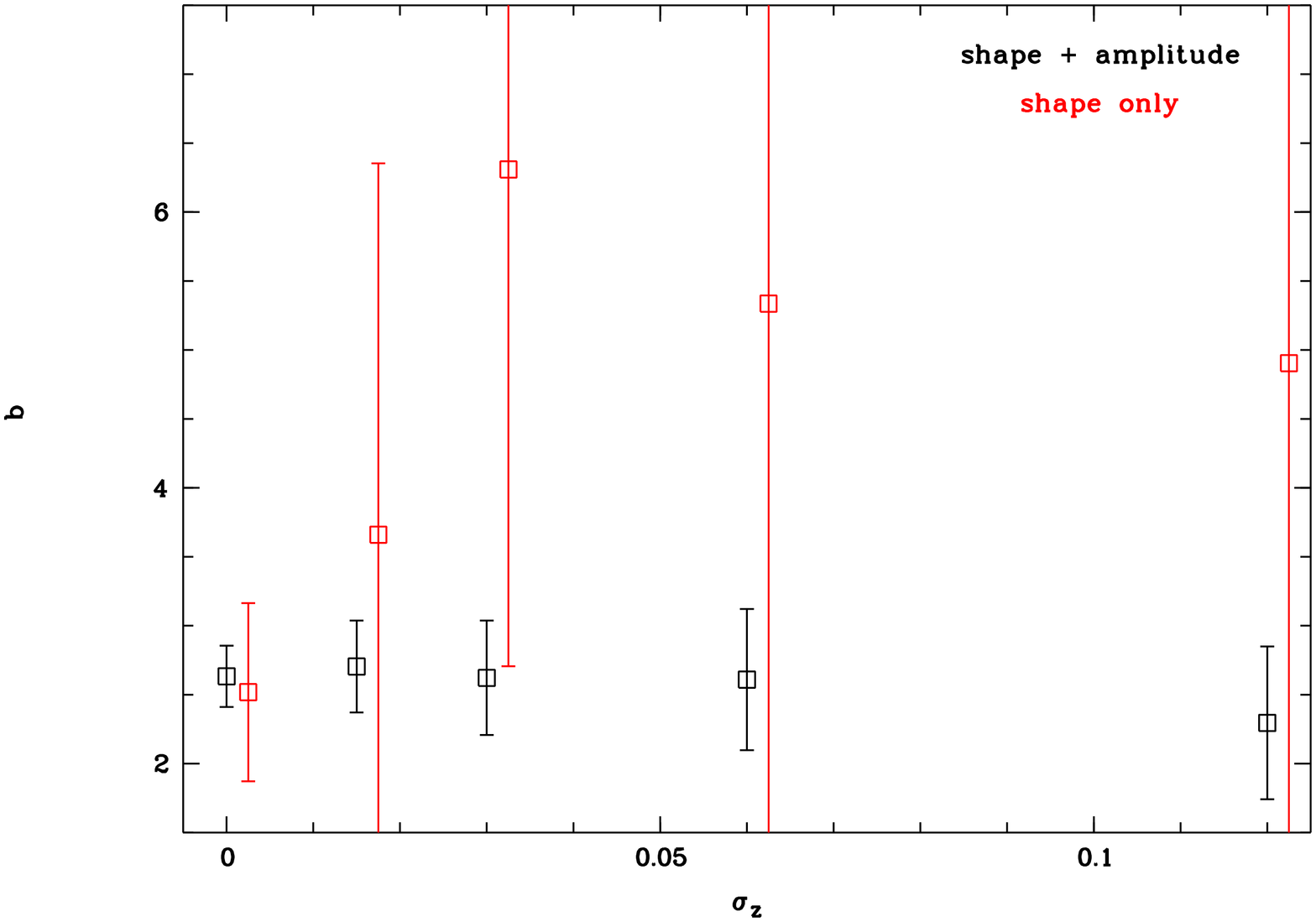,angle=270,clip=t,width=9.cm}}
\caption[ ]{Fitted values of the dark energy equation of state parameter $w_{{\mathrm DE}}$ (left) and the bias $b$  (right) against the width of the redshift errors applied to the L-BASICC II haloes and the model. In black the amplitude and the shape of $\xi(r_p, \pi)$ were taken into accout for the fit, whereas in red only the shape was considered.}\label{zerrwdeandb}
\end{figure*}
 
Figure \ref{zerrwdeandb} shows the values of $w_{{\mathrm DE}}$ and $b$ and their corresponding errors for the fit of the model to the 50 L-BASICCS II boxes including increasing widths of the redshift errors. As long as the errors are smaller than $\sigma_z\approx0.06$, the measurement is unbiased. The increase of the errors in redshift error space can be expected: Due to the convolution with the pairwise redshift error distribution not only the clustering signal is smeared out along the line-of-sight, but so is the noise and thus the errors on the cosmological parameters inferred from the measurement increase. Therefore, within a given bin in the $\xi_{zerr}(r_p,\pi)$ plane the variance is increased compared to redshift space, and the larger the redshift errors, the larger the increase. The loss of information contained in the multipoles due to the distortion by the convolution with the pairwise redshift error distribution function also means that the linear bias $b$ cannot be determined accurately by only using the shape of $\xi_{zerr}(r_p, \pi)$.

Large redshift errors also increase the probability for the MCMC not to converge within the allowed parameter space -- cosmic variance is still large even in boxes of the size of the L-BASICC II Simulations, and in some of them the BAO ring is almost invisible even in real space. In such cases redshift errors finally destroy all of the information that might have been there before, and the fit fails. The larger the redshift errors the more catastrophic failures are produced. Even if we use the information contained in the amplitude, the fraction of boxes where the correlation function can not be fit can be as high as $\sim40$\% for $\sigma_z=0.06$. For $\sigma_z=0.12$ the model tends to yield biased results, as can be seen from the ``best-fitting'' values. 

The exact size of the redshift error at which the values of $w_{{\mathrm DE}}$ and $b$ can not be measured accurately anymore and their errors become unacceptably large certainly depends on the exact shape of the redshift error distribution function (a more Lorentz-like distribution with broad wings will have a larger impact on $\xi(r_p,\pi)$ than a Gaussian with a comparable width of the core), and it most certainly also depends on the volume and/or number density of the survey: since the reason the fit fails is mainly that the BAO feature vanishes in the increasing noise, the larger the signal-to-noise on large scales, the larger the redshift errors may be at which the disappearance of the BAO ring occurs. There may also be the possibility to improve the signal-to-noise ratio for photometric data by using the full probability distribution function of the redshifts in combination with a set of spectroscopic redshifts in the same area and redshift range, a method that has been shown to be able to improve the clustering signal strength in a manner equivalent to increasing the survey size by a factor 4-5 by \citet{Myers09}.

\subsection{Is the projected correlation function $w(r_p)$ an alternative?}\label{wrpsection}
\begin{figure*}
\centerline{\psfig{figure=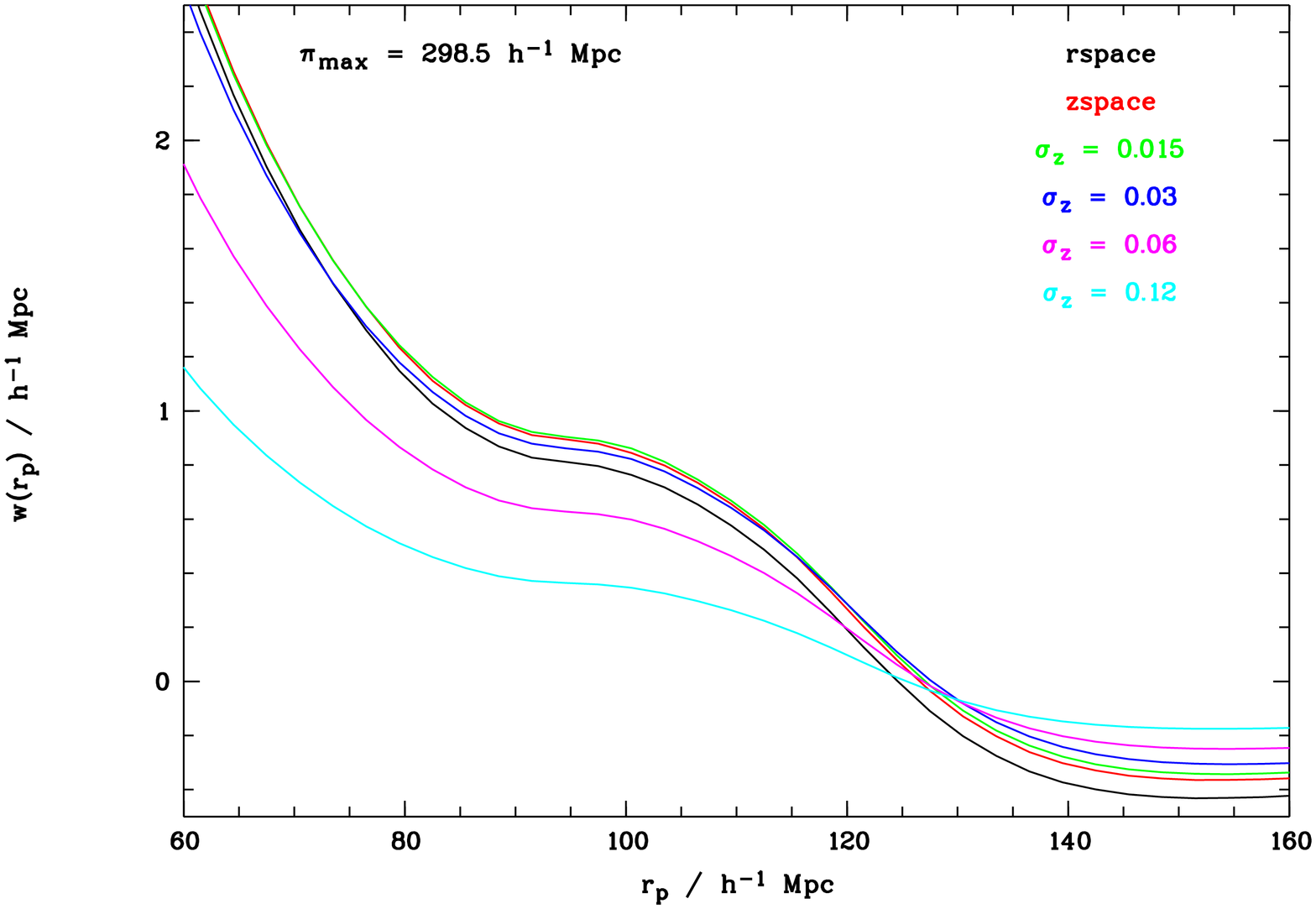,angle=270,clip=t,width=9.cm}
\psfig{figure=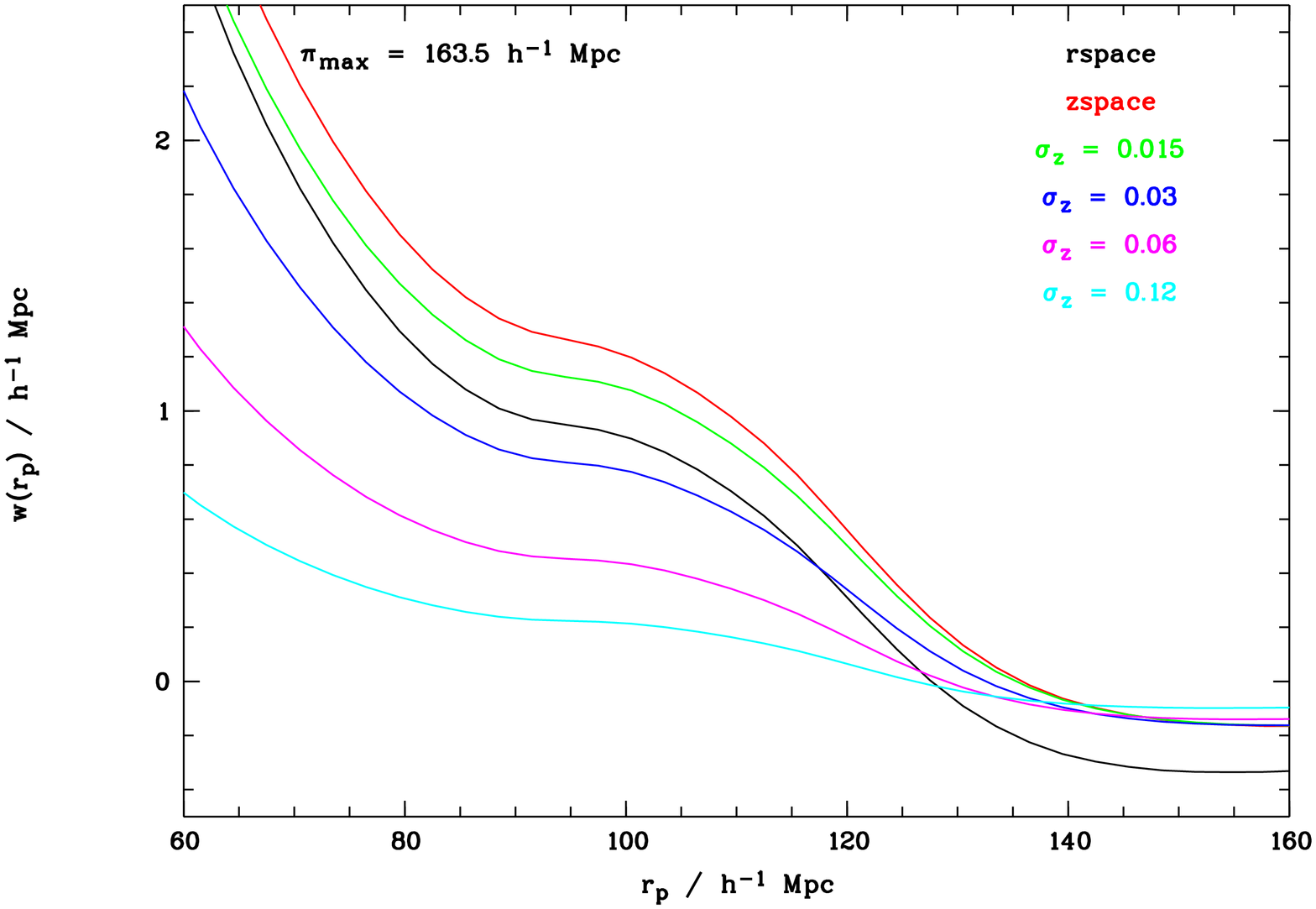,angle=270,clip=t,width=9.cm}}
\caption[ ]{The projected correlation function $w(r_p)$ for real and redshift space (black and red lines, repsectively), and for different widths of the assumed redshift errors, left: for integration limits of $\Delta \pi =298.5 h^{-1}$\,Mpc; right: $\Delta \pi =163.5 h^{-1}$\,Mpc. }\label{wrpplot}
\end{figure*}

For the analysis of the correlation function on small scales often the projected correlation function has been calculated, which is in theory independent of any radial distortions \citealp{1980lssu.book.....P,DavisPeebles83}. For small angles $r^2=r_p^2+\pi^2$. Thus the projected correlation function is defined as
\begin{eqnarray}
w(r_p)&=& \int_{-\infty}^\infty{\xi\left(r_p,\pi\right)~{\mathrm d}\pi}~.\label{projection}
\end{eqnarray}
Note that $w(r_p)$ has dimensions of length. If it were in practice possible to integrate to infinity, it would in principle be possible to recover the three-dimensional real space correlation function $\xi(r)$, and $w(r_p)$ would be far better suited to infer cosmological parameters from the spatial distribution of galaxies than $\xi(r_p, \pi)$, which suffers from redshift space distortions. However, since integrating out even to very large distances without significantly increasing the noise is not feasible (in particular if the signal is smeared out and the amplitude diminished by large redshift errors), the integration limits have to be finite and even rather small, see also \citealp{Norberg09} for an illustration of this. This means that a part of the clustering signal, which depends on the pairwise redshift probability distribution function as well as on the real and the assumed cosmology, can not be recovered. This is illustrated in Figure \ref{wrpplot}, where we show $w(r_p)$ for different widths of the assumed redshift errors ($\sigma_{z}=$0.015, 0.03, 0.06, and 0.12, respectively) and two different choices of the integration limits, $\Delta \pi = 163.5 h^{-1}$\,Mpc and $\Delta \pi = 298.5 h^{-1}$\,Mpc. The resulting shape of $w(r_p)$ depends strongly on both the width of redshift errors and the size of the integration limits: if the integration limits are very large, most of the signal can be recovered and the difference between real and redshift space and the correlation function affected by errors is, although small, still visible. In practice choosing $\Delta \pi = 298.5 h^{-1}$\,Mpc is not advisable, as the measurement will be dominated by noise. On the other hand, if the BAO ring is supposed to be fully included in the integration, the limits can not be much smaller than $\Delta \pi \approx  150h^{-1}$\,Mpc -- in which case the resulting $w(r_p)$ is extremely dependent on the size of the redshift errors (i.e. the fraction of the signal which can be recovered). 

\begin{figure*}
\centerline{\psfig{figure=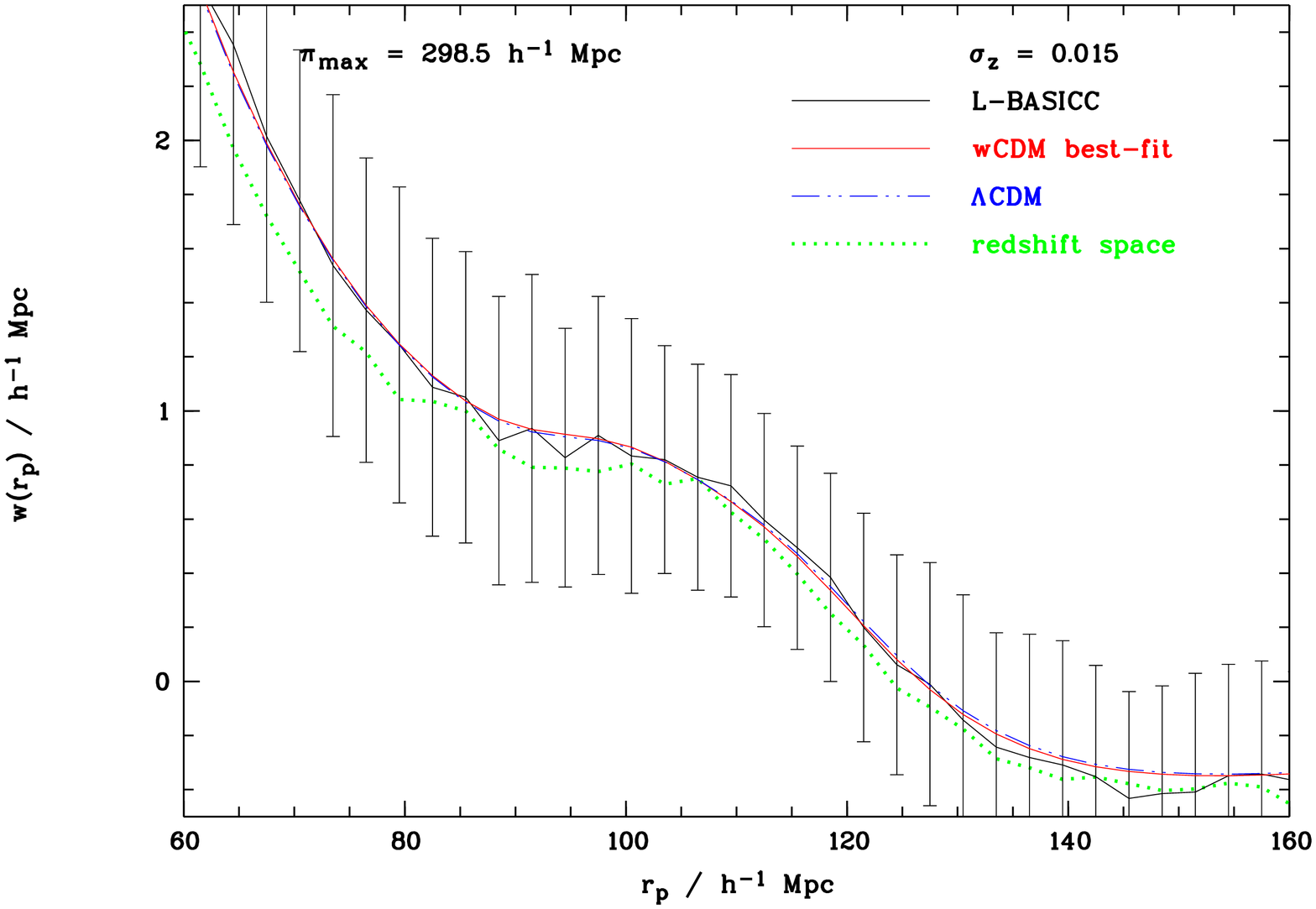,angle=270,clip=t,width=9.cm}
\psfig{figure=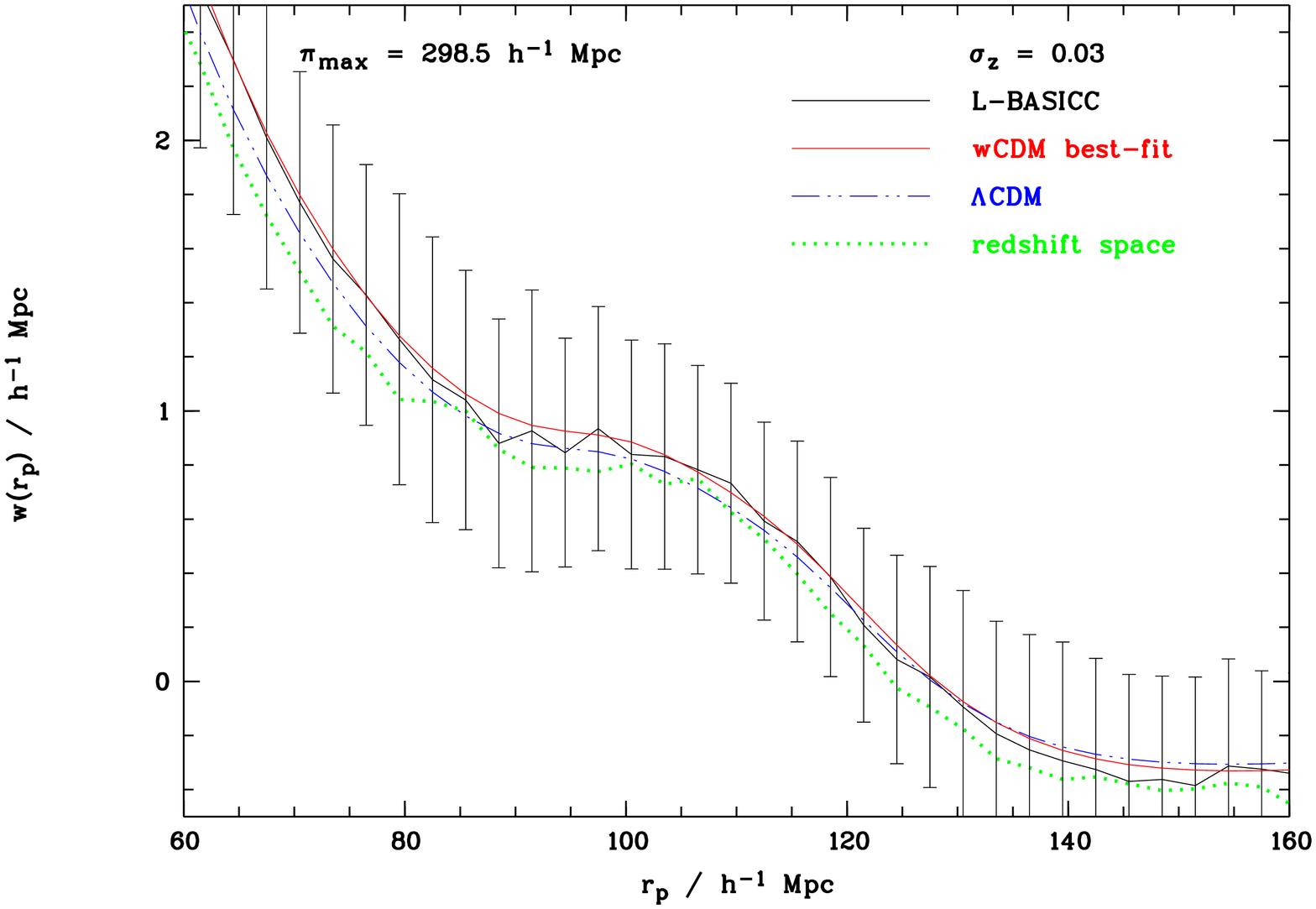,angle=270,clip=t,width=9.cm}}
\centerline{\psfig{figure=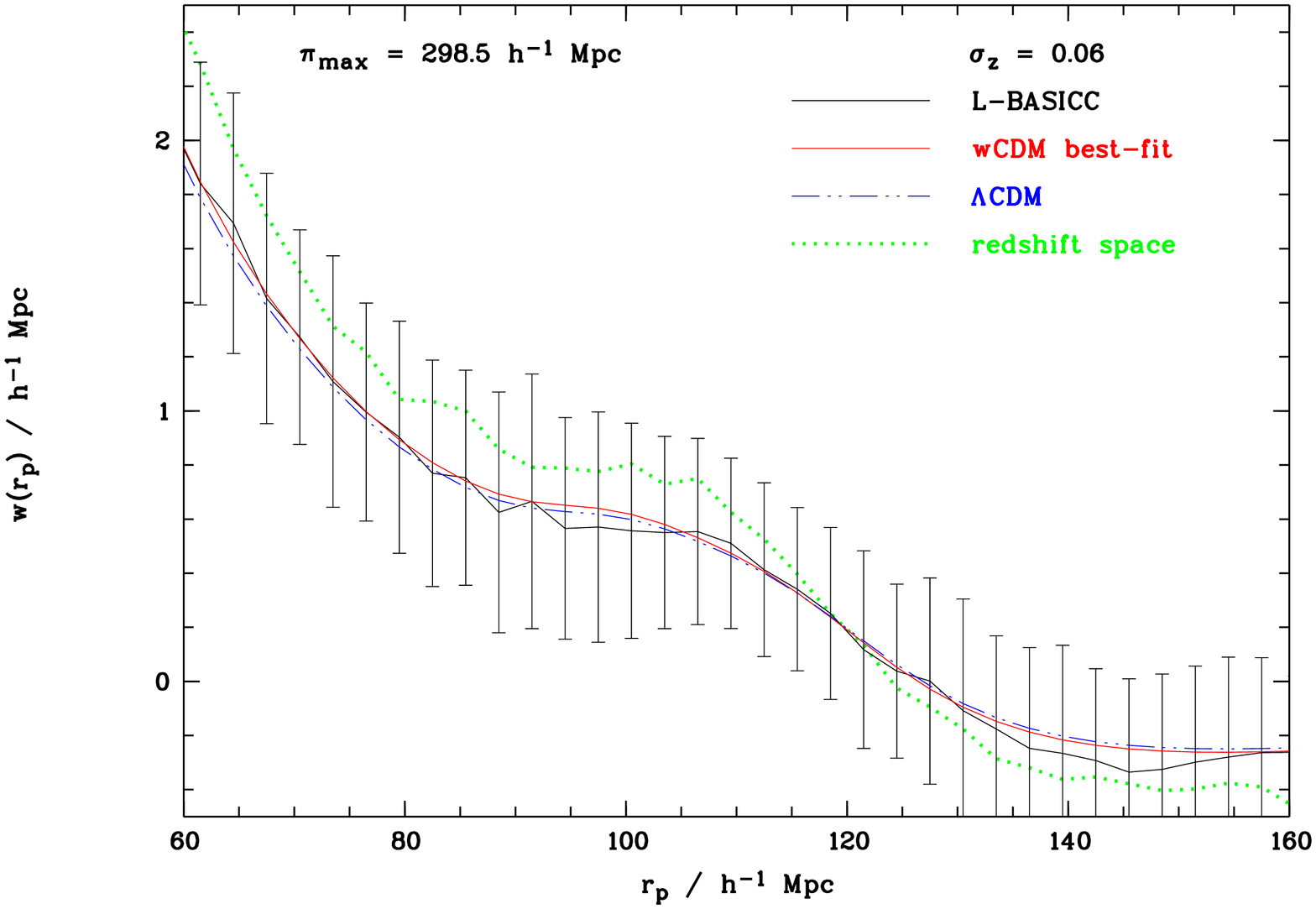,angle=270,clip=t,width=9.cm}
\psfig{figure=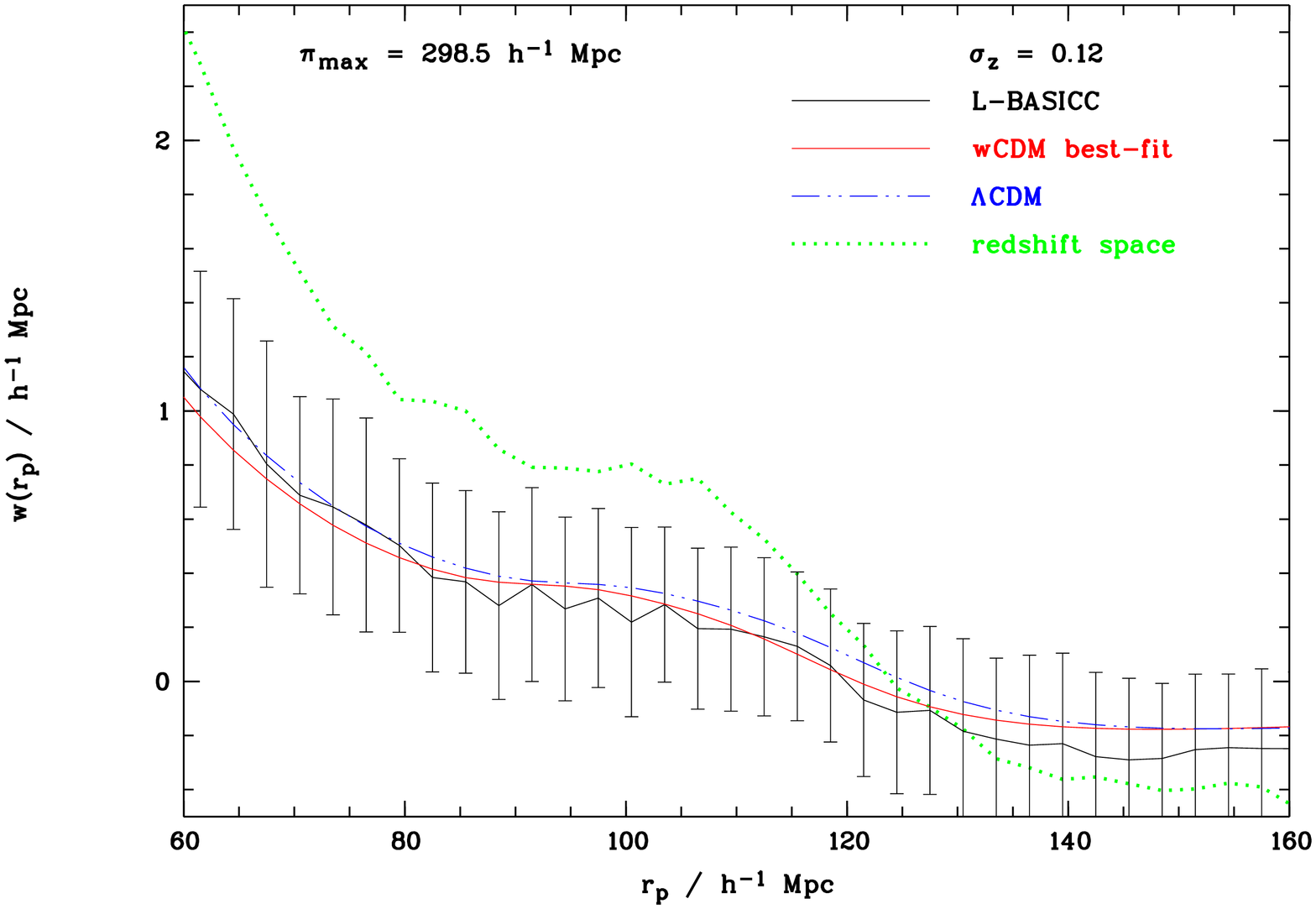,angle=270,clip=t,width=9.cm}}
\caption[ ]{The projected correlation function  $w(r_p)$ of the L-BASICC II dark matter haloes integrated up to $\pi_{max} = 298.5 h^{-1} Mpc$ for redshift errors of $\sigma_z= 0.015$ (top right), $\sigma_z = 0.03$ (top left), $\sigma_z = 0.06$ (bottom right) and $\sigma_z = 0.12$ (bottom left), black solid lines: mean, error bars: $1\sigma$-deviation calculated from the variance of the 50 boxes, red solid line: best-fitting wCDM model, blue dot-dot-dashed line: $\Lambda$CDM case, green dotted line: redshift space ($\sigma_z = 0.00$).}\label{wrpplot_pimax_298.5}
\end{figure*}

\begin{figure*}
\centerline{\psfig{figure=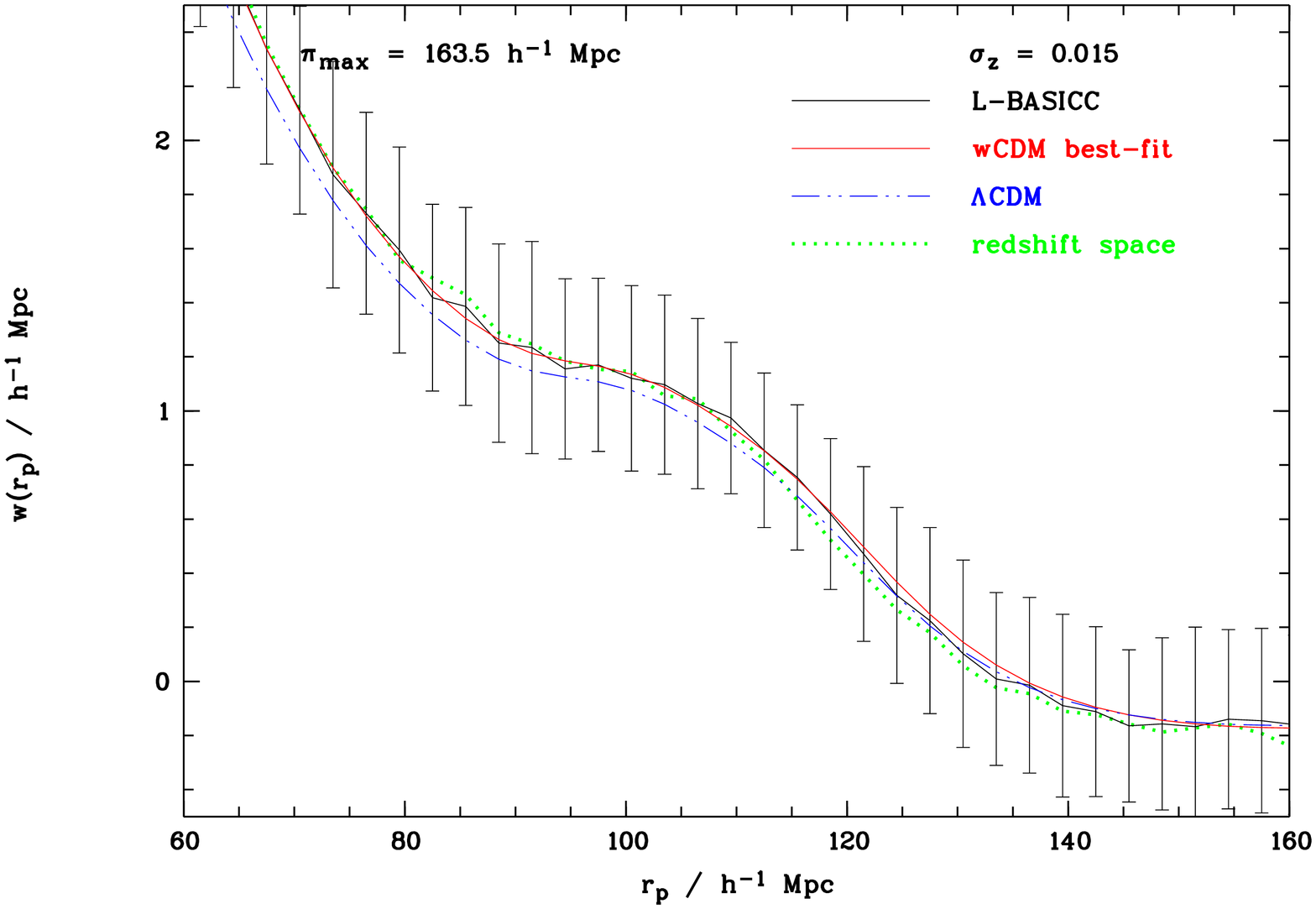,angle=270,clip=t,width=9.cm}
\psfig{figure=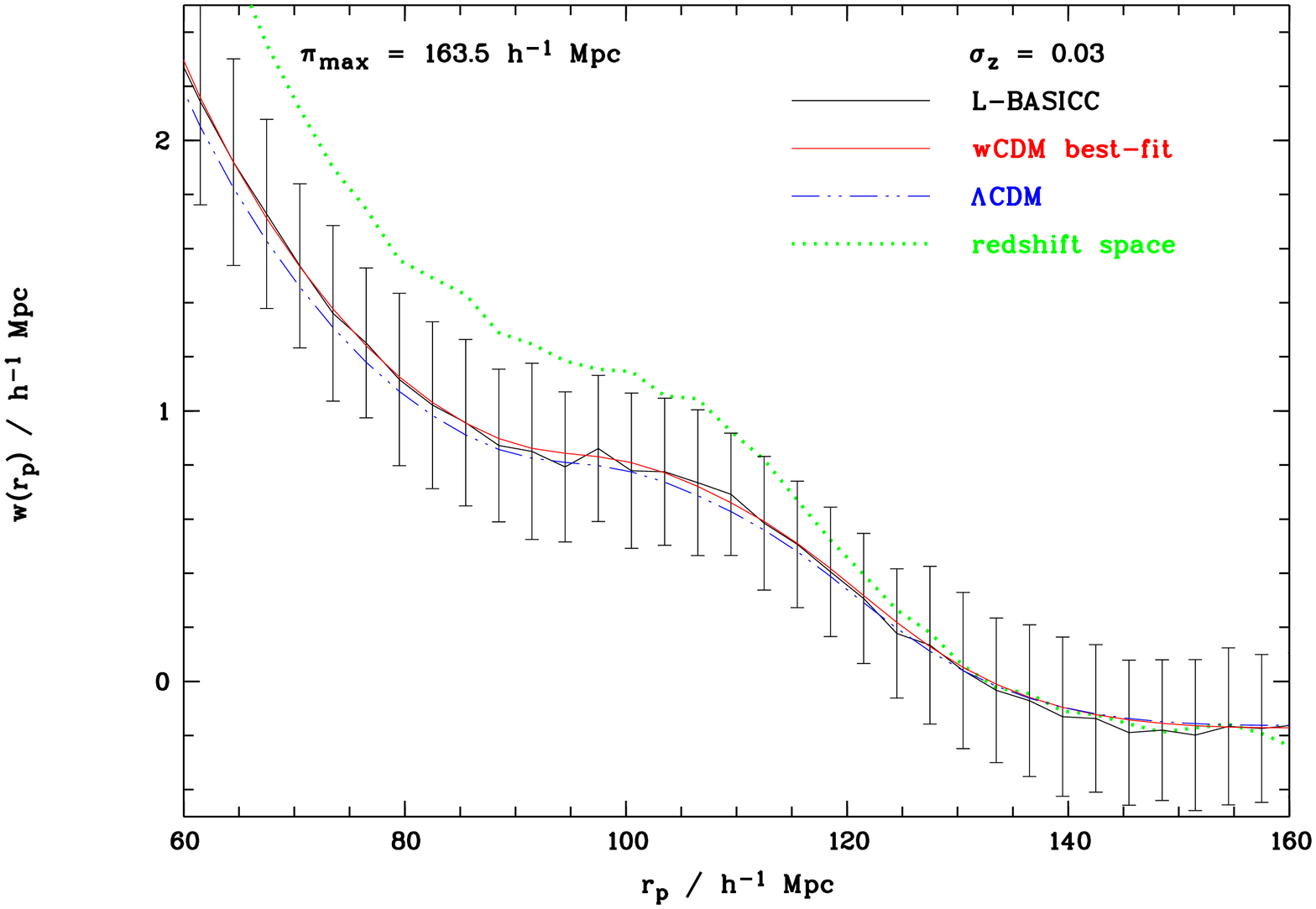,angle=270,clip=t,width=9.cm}}
\centerline{\psfig{figure=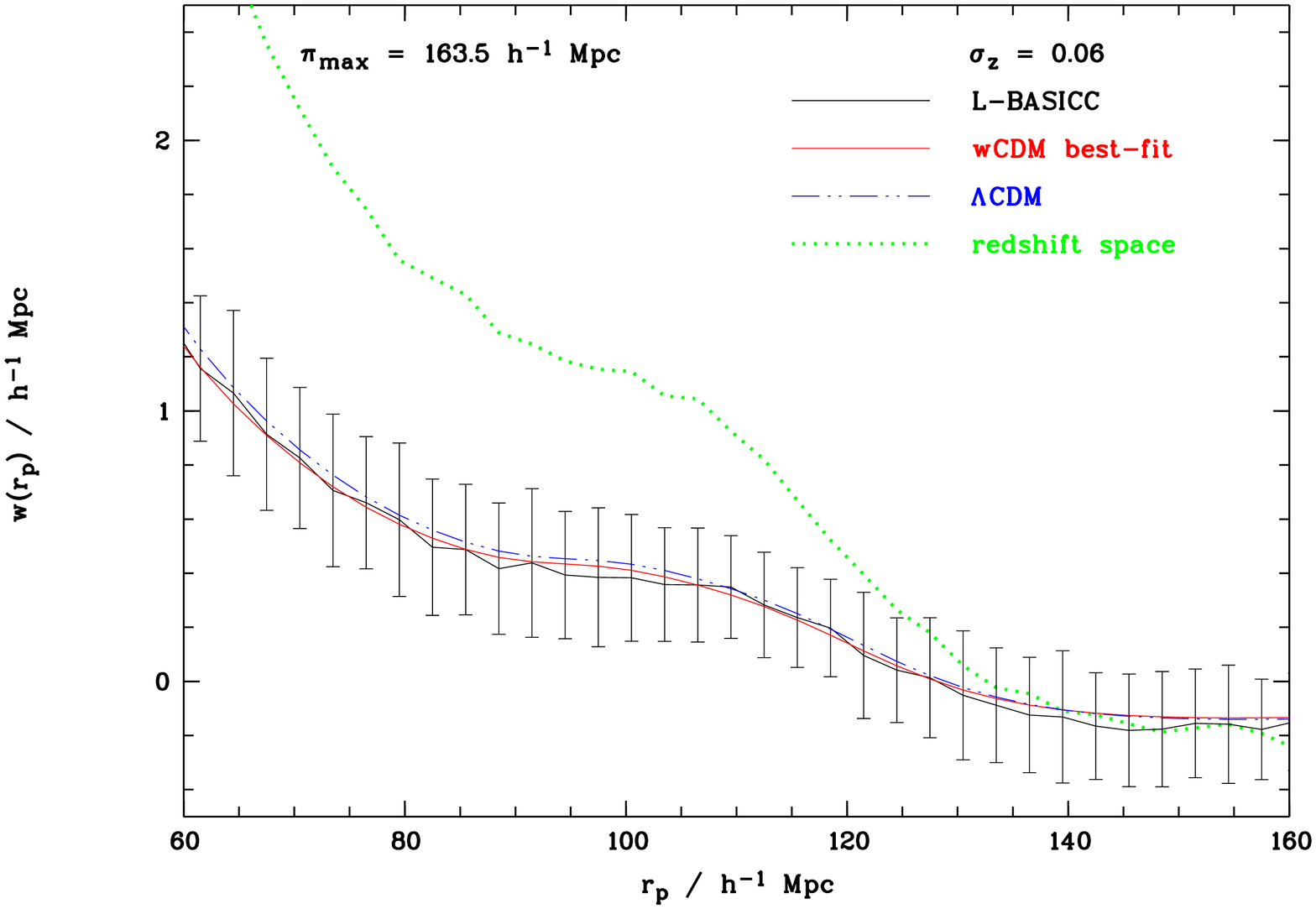,angle=270,clip=t,width=9.cm}
\psfig{figure=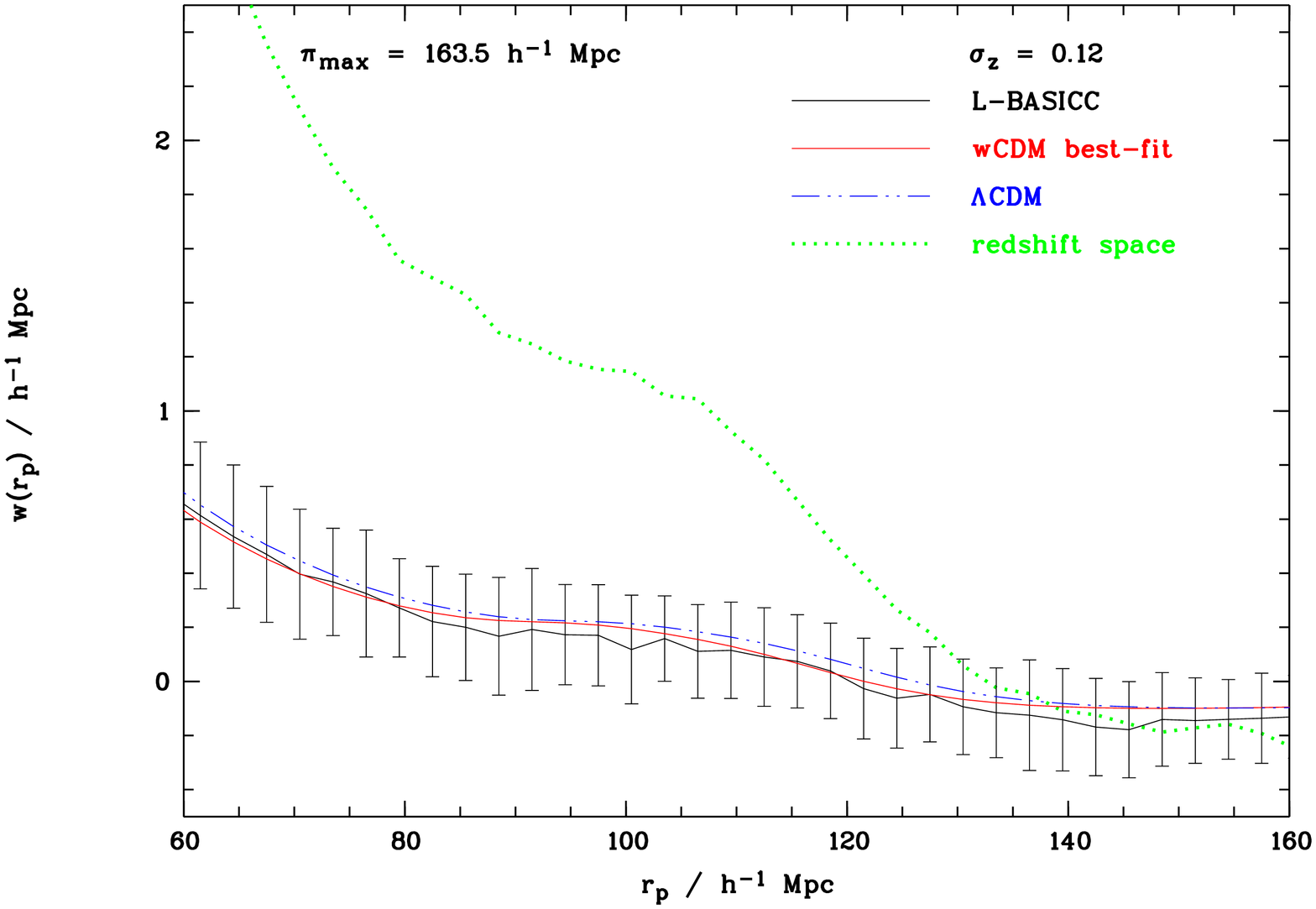,angle=270,clip=t,width=9.cm}}
\caption[ ]{The projected correlation function $w(r_p)$ of the L-BASICC II dark matter haloes integrated up to $\pi_{max} = 163.5 h^{-1} Mpc$ for redshift errors of $\sigma_z= 0.015$ (top right), $\sigma_z = 0.03$ (top left), $\sigma_z = 0.06$ (bottom right) and $\sigma_z = 0.12$ (bottom left), black solid lines: mean, error bars: $1\sigma$-deviation calculated from the variance of the 50 boxes, red solid line: best-fitting wCDM model, blue dot-dot-dashed line: $\Lambda$CDM case, green dotted line: redshift space ($\sigma_z = 0.00$).}\label{wrpplot_pimax_163.5}
\end{figure*}

\begin{figure*}
\centerline{\psfig{figure=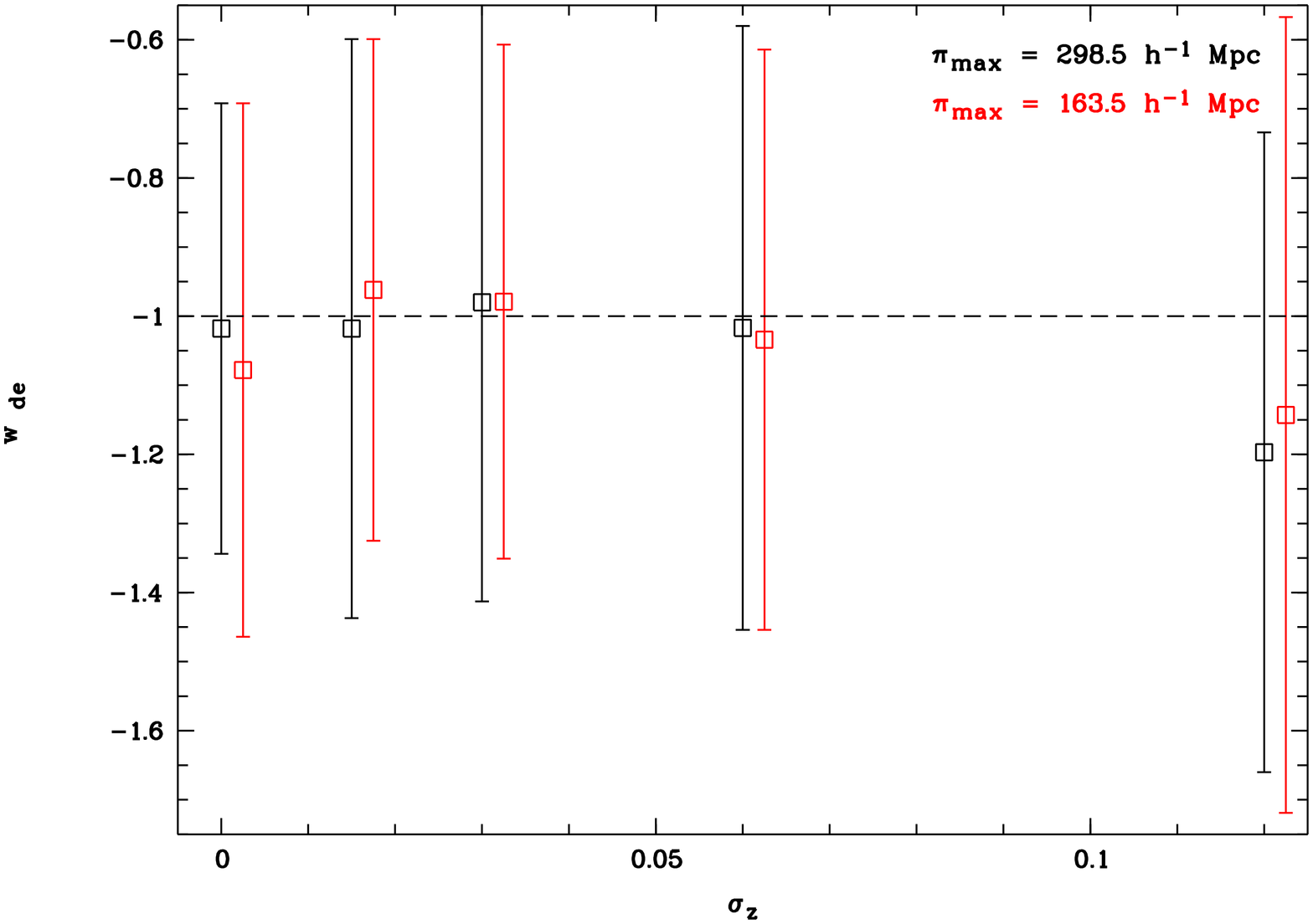,angle=270,clip=t,width=9.cm}
\psfig{figure=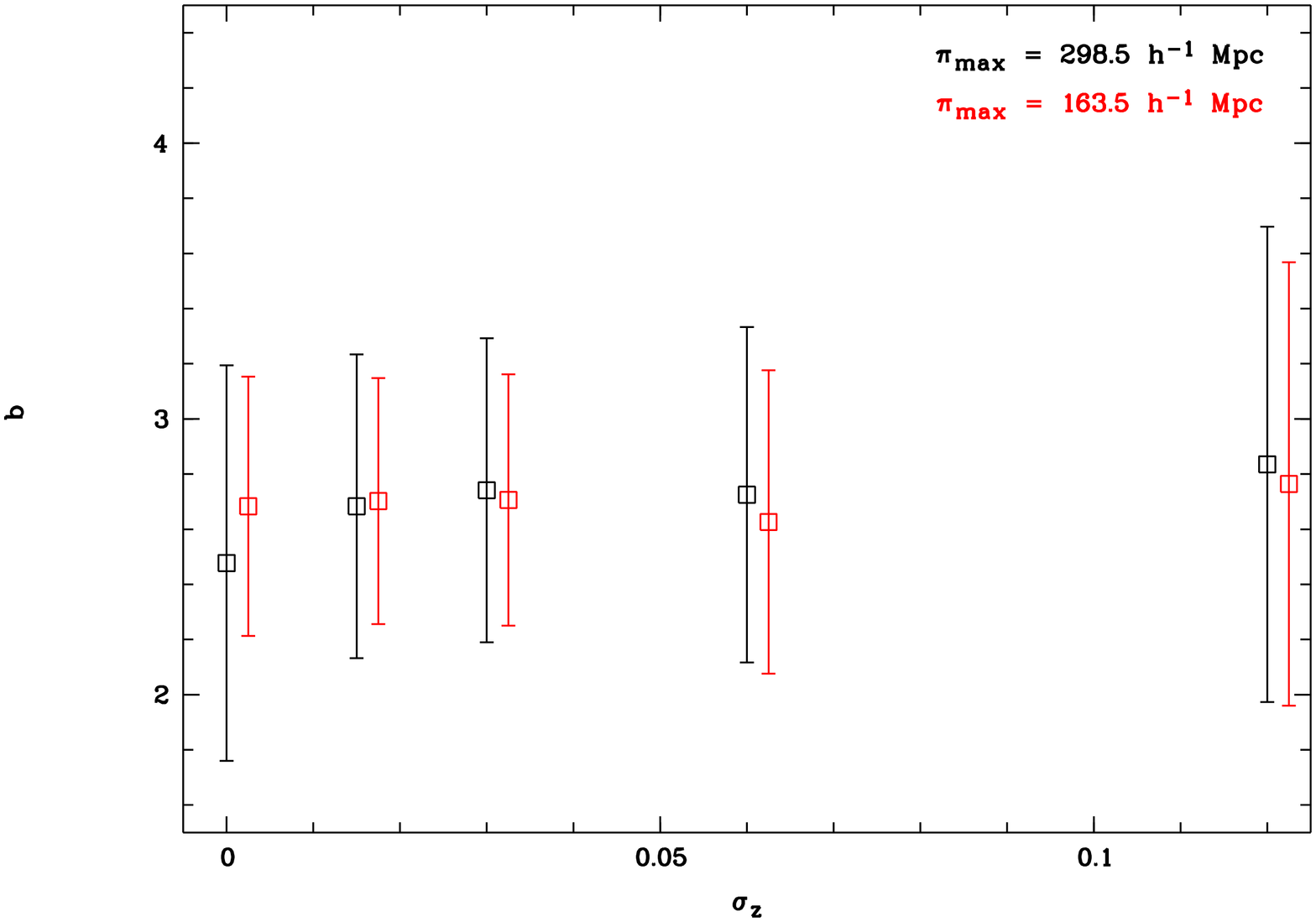,angle=270,clip=t,width=9.cm}}
\caption[ ]{Fitted values of the dark energy equation of state parameter $w_{{\mathrm DE}}$ (left) and the bias $b$  (right) against the width of the redshift errors applied to the L-BASICC II haloes and the model. In black $w(r_p)$ was integrated up to $\pi_{max} = 298.5 h{-1} Mpc$ for the fit, whereas in red $\pi_{max} = 163.5 h^{-1} Mpc$.}\label{zerrwdeandb_wrp}
\end{figure*}

Figures \ref{wrpplot_pimax_298.5} and \ref{wrpplot_pimax_163.5} show the projected correlation function  $w(r_p)$ of the L-BASICC II dark matter haloes integrated up to $\pi_{max} = 298.5 h^{-1}$\,Mpc and $\pi_{max} = 163.5 h^{-1}$Mpc, respectively, for redshift errors of $\sigma_z= 0.015$, $\sigma_z = 0.03$, $\sigma_z = 0.06$, and $\sigma_z = 0.12$, as well as the best fitting model for each case. The amplitude has to be taken into account, otherwise the fit fails: there is not enough information in the shape alone. The corresponding fitted values of the dark energy equation of state parameter $w_{{\mathrm DE}}$ and the bias $b$ and their errors are listed in Table \ref{zerrwrptab}, and shown in Figure \ref{zerrwdeandb_wrp} as a function of $\sigma_z$. Although the fit is not biased for $\sigma_z\la 0.06$, the errors are, as expected, much larger and more quickly increasing with increasing redshift errors than for the corrsponding fits of $\xi(r_p, \pi)$. Hence we conclude that for the analysis of the large-scale correlation function as a means to constrain cosmological parameters from photometric data, $\xi(r_p, \pi)$ is better suited than the projected correlation function $w(r_p)$.

\begin{table}
\begin{center}
\begin{tabular}{l|c|c|}
$\sigma_z$&$w_{{\mathrm DE}}$($\pi_{max}=298.5 h^{-1}$\,Mpc)  & $b$($\pi_{max}=298.5 h^{-1}$\,Mpc)\\
\hline
0.0&$-1.018\pm0.326$&$2.477\pm 0.717$\\
0.015&$-1.018\pm 0.419$&$2.683\pm 0.551$\\
0.03&$-0.980\pm 0.433$&$2.741\pm 0.551$\\
0.06&$-1.017\pm 0.437$&$2.725\pm 0.608$\\
0.12&$-1.197\pm 0.463$&$2.835\pm 0.862$\\
\hline
\hline
$\sigma_z$&$w_{{\mathrm DE}}$($\pi_{max}=163.5 h^{-1}$\,Mpc)  & $b$($\pi_{max}=163.5 h^{-1}$\,Mpc)\\
\hline
0.0&$-1.078\pm 0.386$&$2.683\pm 0.470$\\
0.015&$-0.962\pm 0.363$&$2.702\pm 0.446$\\
0.03&$-0.979\pm 0.372$&$2.706\pm 0.456$\\
0.06&$-1.034\pm 0.420$&$2.626\pm 0.550$\\
0.12&$-1.143\pm 0.576$&$2.764\pm 0.804$\\

\end{tabular}
\end{center}
\caption{The best fitting values of $w_{{\mathrm DE}}$ and $b$, as deduced from the projected correlation function $w(r_p)$, with integration limits of $\pi_{max} = 298.5 h^{-1}$\,Mpc (top rows) and $\pi_{max}=163.5 h^{-1}$\,Mpc (bottom rows), respectively, for gaussian redshift errors with  $\sigma_z= 0.015$, $\sigma_z = 0.03$, $\sigma_z = 0.06$, and $\sigma_z = 0.12$.}
\label{zerrwrptab}
\end{table}

 \subsection{Other work in the literature}

In the last decade substantial effort has been invested into the accurate modelling of the power spectrum or correlation function of galaxies, in order to derive tight constraints on cosmological parameters. While the available and anticipated data sets have become larger, it has become clear that non-linear structure growth, peculiar velocities and galaxy biasing have to be described as precisely as possible. A variety of different methods have been developed and employed to tackle these problems, and the properties of different statistics have been investigated. Since ours is the first systematic investigation of the influence of redshift errors on the measurement of the dark energy equation of state parameter $w_{{\mathrm DE}}$ using $\xi(r_p,\pi)$ (\citet{2009MNRAS.395.1185C} have investigated the impact of photometric redshift errors on the {\it power spectrum}), a direct comparison with other work in the literature is not possible. Instead, we will highlight the differences of ours to existing models in which redshift errors can potentially be included.

\citet{2008arXiv0807.3551G} modeled $\xi(r_p,\pi)$ in order to estimate the position of the radial acoustic peak ($\xi(r_p= 0 h^{-1}$ \,Mpc$,\pi)$) and infer the bias $b$ and $\Omega_M$ from the SDSS DR6 LRG sample \citep{2008ApJS..175..297A}. For the modeling \citet{2008arXiv0807.3551G} use linear perturbation theory and linear redshift space distortions on large scales, whereas our model is non-linear in both the description of structure growth and the Kaiser effect. While we do not include Fingers of God in our model (since they do not occur in a dark matter halo catalogue without substructure, such as we use to compare our model with), they do consider the one dimensional velocity dispersion $\sigma_v$, in order to measure $\beta=\Omega_M^\gamma/b$ from it, as discussed in detail in \citet{2009MNRAS.396.1119C}. 

A second study of $\xi(r_p,\pi)$ was carried out by \citet{2008ApJ...676..889O}. Also in their approach both structure growth and Kaiser effect were treated linearly, but they take the wide angle effect \citep{1998ApJ...498L...1S} and the high-z distortion effect \citep{1996ApJ...470L...1M} into account, which are combined in \citet{2004ApJ...615..573M}. The scales examined were set to $60.0 h^{-1} < s < 160.0 h^{-1}$\,Mpc, in spherical shells like we do in this paper for real and redshift space, similar to our choice. Fitting their model to the anisotropic correlation function of the SDSS DR5 LRG sample \citep{2005ApJ...633..560E} not only $w_{{\mathrm DE}}$ and $b$ were constrained but also several other cosmolgical parameters were determined. It remains to be tested if taking the wide angle effect into account improves the accuracy of our model such that when the redshift errors become larger than the $\sigma_z\ga 0.06$ at which point our fit becomes extremely inaccurate, it is still possible to retrieve reliable constraints, but this analysis is beyond the scope of this paper).

In \citet{2010arXiv1008.4822C} $\xi(r_p,\pi)$ was modeled using halo fit \citep{2003MNRAS.341.1311S} to take the nonlinear structure growth into account. The redshift space distortions are included by the linear Kaiser effect on large scales and by the Fingers of God on small scales. In \citet{2010arXiv1008.4822C} $H(z)$ and $D_A(z)$ was constrained instead of $w_{{\mathrm DE}}$. 

Instead of fitting an analytic model to observed data it is also possible to fit numerical $N$-body simulations (i.e. mocks, where the dark matter haloes are populated with galaxies either using halo occupation modeling or a semi-analytic treatment of galaxy evolution, to which the same selection function, mask and survey geometry has been applied). Then all kinds of clustering statistics can be calculated from the mocks and compared to the observed data. One example for such an attempt is the work of \citet{2010ApJ...710.1444K,2010ApJ...719.1032K}, although they did not measure  $\xi(r_p, \pi)$, but studied the detectability of the BAO peak in $\xi(r)$ as well as of the radial peak in SDSS LRG DR7 sample \citep{2005ApJ...621...22Z}. The advantage of using a mock to compare the clustering statistics with is that nonlinear clustering growth and redshift space distortions do not have to be modeled analytically, but occur naturally in the simulation. Also if many mocks are generated, the covariance matrix can be calculated. However, the big disadvantage is of course that a proper fit which takes differences in the nonlinear clustering growth due to different values of $w_{{\mathrm DE}}$  into account requires to cover the full parameter space with a large number of mocks, which is extremely time consuming.

\citet{2008PhRvD..77l3540P} suggested to utilize the multipole moments for estimating cosmological parameters. Although this requires a smaller number of mock catalogues to calculate the covariance matrix than would be needed for $\xi(r_p,\pi)$, it can not be used in the presence of photometric redshift errors. 

\citet{KSB11} transformed the analysis of the multipole expansion by \citet{2008PhRvD..77l3540P} from Fourier space  to configuration space. They compared the multipole expansion including the hexadecapole with so-called "clustering wedges" ($\xi(\Delta_\mu, s)$, where $\mu = s_{||}/s$ and $s_{||}$ is the radial component of separation $s$) to constrain $H$ and $D_A$, in order to break the degeneracy between these two parameters usually found when only using the monopole. The "clustering wedges" are able to provide constraints at least at the same level of accuracy as the multipole expansion. Their fit is based on $N$-body simulations, but they also argue that for practical use an analytic modeling based on physical principles is needed.

Although our model includes neither the large angle effect, nor Fingers of God (which we do not have in the data we test it against), since we do take non-linear clustering growth and non-linear velocities into account, and we are able to fit the linear bias by making use of the information contained in the redshift space distortions, we believe it is competitive. Also redshift errors can easily be included, which makes it a valuable tool to apply to photometric data, if the redshift errors are small enough ($\sigma_z\la 0.06$).

\section{Conclusions}\label{conclusions}

In this work we developed and tested a model of the anisotropic two point correlation function $\xi(r_p,\pi)$, which we used to investigate the influence of photometric redshift errors on the measurement of the dark energy equation of state parameter $w_{{\mathrm DE}}$. We modeled $\xi(r_p,\pi)$ using third order perturbation theory \citep{1994ApJ...431..495J} to account for the nonlinear nature of the growth of structure and the nonlinear Kaiser effect \citep{2004PhRvD..70h3007S}. Redshift errors can be included in the model by convolving it with the pairwise redshift error distribution, which can easily be computed from the (known) photometric redshift errors.

In order to test the validity of our model, we fit it to the mean correlation function measured from the dark matter haloes in a suite of 50 large-volume, medium-resolution $N$-body simulations (the L-BASICCS II, \citealp{2008MNRAS.387..921A,2008MNRAS.390.1470S}). Both in real and redshift space the fit yields unbiased values of the dark energy equation of state parameter $w_{{\mathrm DE}}$ and the linear bias $b$. With approximately $300\,000$ haloes per box, in real space $w_{{\mathrm DE}}$ and $b$ can be determined with an accuracy of about  $12$\% and $7$\%, respectively. In redshift space these constraints become slightly weaker, $w_{{\mathrm DE}}$ can be measured with an accuracy of approximately  $15$\%, and the relative error of $b$ becomes $\sim 8$\%.

If only the shape information is used to infer $w_{{\mathrm DE}}$ and $b$, the errors on both will increase due to the lack of information contained in the amplitude. The relative error of the bias increases more than the relative error of $w_{{\mathrm DE}}$, since the value of the bias is mainly encoded in the amplitude (and less in the quadrupole and hexadecapole contribution to the redshift space distortion), whereas the equation of state parameter of dark energy influences both shape and amplitude likewise.

In order to investigate the effect of redshift errors on the measurement, we added a small offset to one of the coordinates of the dark matter haloes, which we drew randomly from a gaussian error distribution, and convolved the model with the corresponding pairwise redshift error distribution in the direction along the line-of-sight ($\pi$). Redshift errors smear out the clustering signal and diminish its amplitude; at the same time the convolution leads to a mixing and increase of the noise of the measurement in the single pixels, because intrinsic errors are also distributed along the line-of-sight. The impact of this on the constraints on cosmological parameters is two-fold: Since the signal of the BAOs (as the main feature of the otherwise smooth correlation function) becomes weaker in the observed range of scales, its predictive power decreases -- in the case of very large redshift errors ($\sigma_z\ga 0.06$) the signal is smeared out over such a large range of distances that it completely disappears in the noise. However, since much higher accuracies can be achieved in realistic ongoing or near-future photometric surveys such as e.g. Pan-STARRS (see \citealp{Saglia2012}), this is not a cause for concern. Integrating $\xi(r_p,\pi)$ to obtain $w(r_p)$, as originally proposed by \citet{1980lssu.book.....P} as a means to overcome redshift space distortions, does not help to improve the constraints, as in real space the BAO is a {\it ring} in the $\xi(r_p,\pi)$ plane, and, when integrated, is distributed over $0.\leq r_p\la 120 h^{-1}$\,Mpc. Since it is impossible to integrate $\xi(r_p,\pi)$ to $\pi=\infty$, the resulting amplitude and shape of $w(r_p)$ depends on the choice of integration limits as well as the underlying cosmology, which adds a further complication. Secondly, the noise itself increases in the presence of redshift errors, which creates an additional difficulty. Due to the decreased signal to noise of the correlation function, the accuracy of the constraints on $w_{{\mathrm DE}}$ and $b$ decreases. 

In order to beat down systematics coming from cosmic variance (which is still large, even on BAO scales), it is desireable (and important) to observe the largest volumes possible at one particular redshift. Also, in order to measure a possible variation in the equation of state with lookback time, observations have to be carried out at higher redshifts as well. At this moment in time both is still only feasible with photometric redshifts. The anisotropic correlation function $\xi(r_p,\pi)$, which can be used to infer cosmological parameters like the dark energy equation of state $w_{{\mathrm DE}}$, is well suited to incorporate photometric redshifts. We have developed a model of $\xi(r_p,\pi)$ which will be able to provide {\it unbiased} constraints on $w_{{\mathrm DE}}$ and $b$ for photometric redshift surveys. The maximum redshift error for which this model will work certainly depends on the exact shape of the redshfit error distribution, the volume and number density of the survey to which it is applied.

\section{Acknowledgments}

HAS acknowledges support by the Trans-regional Collaborative Research Centre TR33 'The Dark Universe' of the German Research Foundation (DFG). We thank Raul Angulo for his supportive help by answering our questions about the L-BASICC Simulation II. We also thank Carlton Baugh, Klaus Meisenheimer, and John A. Peacock for their extremely valuable comments, which considerably helped improving the paper. We also thank the anonymous referee for her/his careful reading of our manuscript and his useful comments.

\label{lastpage}

\end{document}